\documentclass[11pt]{article}
\pdfoutput=1
\usepackage{graphicx}
\graphicspath{{./figures/}}
\usepackage{appendix}
\usepackage{latexsym,amsmath,amsfonts,amssymb,booktabs}
\usepackage[font=small]{caption}
\usepackage{slashed,upgreek,amscd,cancel,tensor,color}
\usepackage{adjustbox}
\usepackage[numbers,compress,square]{natbib}
\usepackage{epsfig,latexsym}
\usepackage{amsmath}
\usepackage[pdfencoding=auto]{hyperref} 
\usepackage{url}
\numberwithin{equation}{section}
\usepackage{doi}
\usepackage{subcaption}
\definecolor{MyBlue}{rgb}{0.15,0.15,0.70}
\definecolor{linkblue}{rgb}{0,0,0.8}
\definecolor{linkgreen}{rgb}{0,0.5,0}

\hypersetup{
colorlinks=true,
citecolor=linkgreen,
linkcolor=linkblue,
urlcolor=linkblue
}

\input epsf
\usepackage{amssymb}
\usepackage{amsmath}
\usepackage{amsfonts}
\usepackage{upgreek}
\usepackage{latexsym}
\usepackage{stfloats}
\usepackage{afterpage}

\numberwithin{equation}{section}

\newcommand{\bea}{\begin{eqnarray}}
\newcommand{\eea}{\end{eqnarray}}
\newcommand{\be}{\begin{equation}}
\newcommand{\ee}{\end{equation}}

\newcommand{\cH}{\mathcal{H}}

\newcommand{\knl}{k_{\rm NL}}
\newcommand{\kvec}{\vec{k}}
\newcommand{\qvec}{\vec{q}}
\newcommand{\xvec}{\vec{x}}
\newcommand{\half}{\frac{1}{2}}

\newcommand{\eqn}[1]{Eq.~(\ref{#1})}
\newcommand{\eqns}[2]{Eqs.~(\ref{#1})~-~(\ref{#2})}

\newcommand{\unitsk}{\, h \, { \rm Mpc^{-1}}}

\newcommand{\mpl}{M_{\rm Pl}}
\newcommand{\andd}{\ , \quad \text{and}  \quad}
\newcommand{\Om}{\Omega_{\rm m}}
\newcommand{\secref}[1]{Sec.~\ref{#1}}
\newcommand{\pvec}{\vec{p}}
\newcommand{\appref}[1]{App.~\ref{#1}}
\newcommand{\figref}[1]{Fig.~\ref{#1}}
\newcommand{\tabref}[1]{Tab.~\ref{#1}}

 \definecolor{MattOrange}{rgb}{1.0,0.4,0.2}

\newcommand{\Comment}[1]{{}}

\begin{document}
\def\thefootnote{\fnsymbol{footnote}}


\setcounter{page}{1} \baselineskip=15.5pt \thispagestyle{empty}

{\small \hfill NUHEP-TH/20-07 }

\begin{center}

{\Large \bf  Baryonic effects\\
 in the Effective Field Theory of Large-Scale Structure \\   {and} an analytic recipe for lensing in CMB-S4 \\[0.7cm]}
{\large Diogo~P.~L.~Bragan\c{c}a,$^{1}$ Matthew Lewandowski,${}^{2}$ David Sekera,${}^{3}$ \\ Leonardo Senatore$^{1}$ and Raphael Sgier${}^{1,4}$}
\\[0.7cm]
{\normalsize { \sl $^{1}$ SITP and KIPAC, Department of Physics and SLAC,\\ Stanford University, Stanford, CA 94305}}\\
\vspace{.3cm}

{\normalsize { \sl $^{2}$ Department of Physics and Astronomy,\\ Northwestern University, Evanston, IL 60208}}\\
\vspace{.3cm}

{\normalsize { \sl $^{3}$ \'Ecol\'e Polyt\'echnique F\'ed\'erale de Lausanne, Department of Physics \\}}
\vspace{.3cm}

{\normalsize { \sl $^{4}$ Institute for Particle Physics and Astrophysics, Department of Physics,\\
ETH Zurich, Wolfgang-Pauli-Strasse 27, 8093, Zurich, Switzerland}}\\
\vspace{.3cm}

\end{center}

\vspace{.5cm}

%
%
%

\begin{abstract}\noindent
Upcoming Large-Scale Structure surveys will likely become the next leading sources of cosmological information, making it crucial to have a precise understanding of the influence of baryons on cosmological observables. The Effective Field Theory of Large-Scale Structure (EFTofLSS) provides a consistent way to predict the clustering of dark matter and baryons on large scales, where their leading corrections in perturbation theory are given by {a simple and calculable} functional form even after the onset of baryonic processes. In this paper, we extend the two-fluid-like system up to two-loop order in perturbation theory.  Along the way, we show that a new linear counterterm proportional to the relative velocity of the fluids could generically be present, but we show that its effects are expected to be small in our universe.  Regardless, we show how to consistently perform perturbation theory in the presence of this new term.  We find that the EFTofLSS at two-loop order can accurately account for the details of baryonic processes on large scales.  We compare our results to a hydrodynamical $N$-body simulation at many redshifts and find that the counterterms associated with the leading corrections to dark matter and baryons start to differ between redshifts $z \approx 3$ and $z \approx 2$, signaling the onset of star-formation physics.  We then use these fits to compute the lensing power spectrum, show that the understanding of baryonic processes will be important for analyzing CMB-S4 data, and show that the two-loop EFTofLSS accurately captures these effects for $\ell \lesssim 2000$.  Our results are also potentially of interest for current and future weak lensing surveys. 

\end{abstract}

\newpage
\tableofcontents
\newpage

\def\thefootnote{\arabic{footnote}}
\setcounter{footnote}{0}

%
%
%
%

\section{Introduction}

Upcoming Large-Scale Structure (LSS) surveys may very well become our primary sources of cosmological information, as they will probe the matter distribution with unprecedented accuracy. However, the success of lensing surveys such as CMB-S4 \cite{Abazajian:2016yjj,Abitbol:2017nao,Abazajian:2019eic}, and many LSS surveys in general, crucially depends on the impact of small-scale baryonic processes on the large-scale matter distribution, such as energy feedback mechanisms driven by active galactic nuclei (AGN), supernovae, black hole accretion, and wind mass loading. Despite the {continuous progress} of numerical hydrodynamical simulations, the effect of baryonic processes on the large-scale gas distribution can not be properly resolved by current simulations, restricting their predictability for cosmological observables.

Given the current situation, it is of increasing importance to have an accurate theoretical understanding of how baryons affect LSS formation in the mildly non-linear regime in order to complement hydrodynamical simulations in the non-linear regime. Furthermore, because the amount of information retrievable from large-scale surveys scales as the cube of the largest wavenumber under theoretical control, it is of the utmost importance to have an accurate understanding of the LSS observables at the highest wavenumber possible, so that we can use them to infer cosmological information. This very objective, restricted to the context of analytic predictions, is approached by the research program called the Effective Field Theory of Large-Scale Structure (EFTofLSS)~\cite{Baumann:2010tm, Carrasco:2012cv, Porto:2013qua, Senatore:2014via, Carrasco:2013sva, Carrasco:2013mua, Pajer:2013jj, Carroll:2013oxa, Mercolli:2013bsa, Angulo:2014tfa, Baldauf:2014qfa, Senatore:2014eva, Senatore:2014vja, Lewandowski:2014rca, Mirbabayi:2014zca, Foreman:2015uva, Angulo:2015eqa, McQuinn:2015tva, Assassi:2015jqa, Baldauf:2015tla, Baldauf:2015xfa, Foreman:2015lca, Baldauf:2015aha, Baldauf:2015zga, Bertolini:2015fya, Bertolini:2016bmt, Assassi:2015fma, Lewandowski:2015ziq, Cataneo:2016suz, Bertolini:2016hxg, Fujita:2016dne, Perko:2016puo, Lewandowski:2016yce, Lewandowski:2017kes, delaBella:2017qjy, Senatore:2017hyk}: the idea is to study LSS in the mildly non-linear regime by correctly describing the effect of ultraviolet (UV) modes on long-wavelength observables. Thanks to the inclusion of counterterms to account for the effect of short distances on long distances, the EFTofLSS gives accurate predictions of long-wavelength observables in a perturbative expansion in powers of $k/k_{\text{NL}}$, where  $k_{\text{NL}}$ is the wavenumber associated to the scale where perturbation theory breaks down, {which is expected to be the size of clusters in our universe}. This stands in contrast to standard perturbation theory techniques, where the long-wavelength effects of non-linearities are not considered accurately and thus introduce errors in the perturbative expansion. On distances larger than the non-linear scale, the EFTofLSS can make more and more accurate predictions (up to non-perturbative effects) by taking into account higher-order terms in the perturbative expansion and fitting the arising coefficients of the counterterms (i.e. {the EFT parameters, or coupling constants}) to observations or simulation data.

So far, it has been shown that the EFTofLSS can accurately describe the structure formation of cold dark matter (CDM)~\cite{Foreman:2015lca}, dark energy~\cite{Lewandowski:2016yce,Cusin:2017wjg, Lewandowski:2017kes,Bose:2018orj}, galaxies~\cite{Senatore:2014eva, Angulo:2015eqa}, and massive neutrinos~\cite{Senatore:2017hyk}, in both real and redshift space~\cite{Lewandowski:2015ziq, Perko:2016puo}. In the context of dark matter, the long-wavelength regime is described as a fluid-like system with a non-trivial stress tensor. In~\cite{Lewandowski:2014rca}, it was shown that a system composed of two fluid-like systems, endowed with approximately the same free-streaming scale, can accurately describe a universe filled with dark matter and baryons. This holds true because it is a fact about our universe that baryonic effects involved in star formation processes affect the baryons in a way such that the relative displacement between dark matter and baryons is not larger than the non-linear scale, which is about 10 Mpc. In other words, while star formation physics induces very complicated dynamics on scales within a cluster, it does not {significantly} displace mass beyond the scale of a cluster. This indeed implies that we can describe the system with two fluid-like species characterized by an approximately equal mean free path. In turn, this implies that the functional form of the one- and two-loop corrections to the baryonic and dark matter power spectrum on large scales \textit{is known} up to a number of numerical coefficients. In particular, as first explored in~\cite{Lewandowski:2014rca} and as we will further explore in \secref{theory}, the leading effects of baryonic physics on the power spectrum are fixed:  at one loop, they are proportional to the linear adiabatic power spectrum $ (k / k_{\text{NL}})^2 P_{11}^{A} (k)$ \cite{Lewandowski:2014rca}, and at two loops they have a form that we will derive later in \eqn{candbtwoloop}, which is essentially the same as the two-loop dark-matter power spectrum, but where each fluid has its own EFT coefficients.  In this way, the finite number of numerical coupling constants can be fit to data to extract both the value of these parameters and the cosmological parameters.  {This is particularly compelling since the EFT has been used to analyze BOSS data \cite{DAmico:2019fhj,Ivanov:2019pdj,Colas:2019ret} and has placed tighter constraints on cosmological parameters than traditional methods.}

{In this work, we consider an effect proportional to the long-wavelength relative velocity that was assumed to be small in \cite{Lewandowski:2014rca}, but that could become relevant at the two-loop order that we currently work (although we show that it is not).  This term arises from integrating out UV modes in the theory, in the same way that the other counterterms in the EFT arise.}  Assuming that the small-scale physics obeys conservation of mass of each species separately, conservation of total momentum, and overall {diffeomorphism} invariance (see \eqn{galcoordtrans}), we show in \secref{theory} that a counterterm proportional to the relative velocity that is not derivatively suppressed is allowed in the {effective force \cite{Lewandowski:2014rca}} that appears in the dynamical {equation that controls the relative motion of the two effective fluids}.  Indeed, we show that this term is needed to cancel the cutoff-dependent part of certain one-loop terms.  As we discuss in {\secref{relvel}}, the situation is even more extreme than this.  For a CDM linear power spectrum, this term \emph{diverges} in the UV like $(\log \Lambda_{\rm UV})^3$, where $\Lambda_{\rm UV}$ is the UV cutoff.  This means that the counterterm that we consider is actually \emph{necessary} to have a well-defined mathematical framework. The finite part of this linear counterterm can change the linear equation of motion for the isocurvature mode, {making it act somewhat like a biased tracer.}  We discuss {how this term can be accounted for in perturbation theory} in detail in \secref{linearevo} and \appref{pertwithlinctsec}.  By looking at estimates both in perturbation theory (\secref{sec:realuniverse}) and using a one-dimensional UV model (\secref{sec:sphcoll}), however, we show that the size of the effect of this counterterm on the power spectrum is expected to be small (see \figref{fig:DESI2loop} for example).\footnote{  {This new effect should not be confused with calculable IR effects coming from the relative velocity, which manifest themselves most notably in the violation of the so-called consistency conditions of LSS \cite{Peloso:2013zw, Kehagias:2013yd, Creminelli:2013mca} when there are two fluids \cite{Peloso:2013spa, Bernardeau:2011vy, Bernardeau:2012aq, Creminelli:2013poa, Lewandowski:2014rca} (see \cite{Crisostomi:2019vhj, Lewandowski:2019txi} for example when the extra fluid is dark energy).  This violation is due to the fact that one cannot eliminate the effects of both long-wavelength velocities with a single boost if there is a large-scale relative velocity.  For the case of baryons and dark matter, the effect decays like the relative velocity, proportional to $a^{-1}$ ($a$ is the scale factor of the metric, see \eqn{metric}), and so is negligible at late times, but not necessarily at early times \cite{Tseliakhovich:2010bj}.  }  }  

In order to investigate how well the two-loop EFT with baryons is able to describe baryonic physics on large scales, in \secref{Comparison} we compare our results to non-linear data from the OWLS simulations \cite{vanDaalen:2011xb,Schaye:2009bt}.  In particular, we fit the ratio of power spectra {of two quantities measured} in a simulation which includes baryonic processes to a simulation which only includes dark matter.  In this way, the cosmic variance of the simulation is greatly reduced.  We fit at 18 different redshifts and find an impressive fit to the data.  For example, we are able to fit the ratio until $k \approx 0.8 \unitsk$ at $z = 0$, and $k \approx 3.6 \unitsk$ at $z = 4$ {using three time-dependent parameters per fit}.  We find that the EFT parameters of the baryon and dark-matter fluids (essentially the pressures) start to differ significantly between $z = 3$ and $z = 2$, signaling the onset of star-formation processes.  We also give a suggestion for parametrizing the time dependence of the  EFT parameters.  In this way, we show that the EFT is able to accurately capture baryonic physics in just a few time-dependent parameters, and we view this as a significant improvement in our analytic understanding of the effects of baryons.  { As the functional form of the effect of baryons is known up to a small number of numerical coefficients, this means that one can apply our formalism directly to cosmological data, potentially using simulations to put priors on the parameters. }

\begin{figure}[t]
\centering
\begin{tabular}{cc}
\hspace{-.4in}\includegraphics[width=17cm]{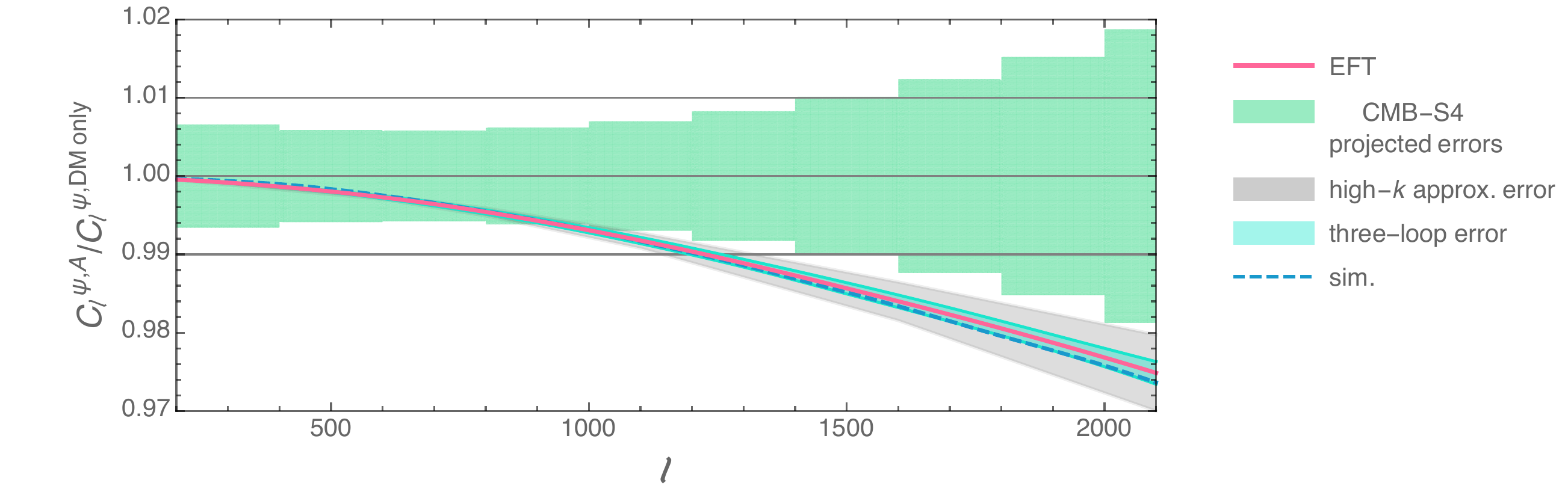} 
\end{tabular} 
\caption{\footnotesize Ratio of the adiabatic and dark matter lensing potential power spectra $C_{\ell}^{\psi, A}/ C_{\ell}^{\psi, \text{DM only}}$ using the two-loop EFT prediction and $0.25 \%$ error on the ratio of the OWLS data.  {The green band is the projected error for CMB-S4 \cite{Abazajian:2016yjj}, the gray band is the estimated error coming from the `high-$k$ approximation' described in \secref{lensing}, and the teal band is the estimated error coming from the three-loop terms in the EFT ratio fits in \secref{Comparison}.}  The dashed blue line is the result of direct numerical integration of the outputs of the simulations, and gives an idea of the systematic theory error in our calculation.  We see that CMB-S4 will be highly sensitive to the effects of baryons on the lensing potential, and that the two-loop EFT can reliably capture these effects up to $\ell \lesssim 2000$, and actually even beyond. }
\label{lensing3}
\end{figure}

Finally, in \secref{lensing}, we apply our results to computing the effect of baryons on CMB lensing.  The CMB-S4 effort will substantially reduce the error bars on the lensing potential to the point that percent-level baryonic effects must be understood to obtain, for example, unbiased neutrino mass constraints \cite{Natarajan:2014xba,Chung:2019bsk}.  {Even if the size of baryonic effects is comparable to or larger than the projected CMB-S4 errors, we show that the EFT is able to reproduce the effect of baryons on the lensing power spectrum up to $\ell \approx 2000$, and actually even beyond. } This is shown in \figref{lensing3} and described in detail in \secref{lensing}.  Thus, the EFT approach offers a compelling recipe for analytically parametrizing baryonic effects on the lensing power spectrum in a systematic way.    Our work here is also applicable to weak lensing of galaxies, as we discuss in \secref{lensing}. 

\paragraph{Notation}

In order to consolidate notation, we present some of the relevant definitions for two fluids here.  We have two species, CDM (denoted with a subscript $c$) and baryons (denoted with a subscript $b$), but we will sometimes find it useful to work in a different basis, with an adiabatic mode (subscript $A$) and an isocurvature mode (subscript $I$).  
For the mass densities $\rho_\sigma$ (here and in the rest of the work, $\sigma \in \{c,b\}$, and multiple appearances of $\sigma$ in an equation are \emph{never} summed over), we define the overdensities $\delta_\sigma$ by
\be
\rho_\sigma \equiv \bar \rho_\sigma ( 1 + \delta_\sigma ) \ ,
\ee
where an overbar denotes the background value.  The adiabatic density is defined by
\be
\rho_A \equiv \rho_c + \rho_b \ , 
\ee
so that for the background, we have $\bar \rho_\sigma = w_\sigma \bar \rho_A$, where $w_\sigma$ are the time-independent matter fractions, and $w_c + w_b = 1$.  They are time independent because $\bar \rho_c$ and $\bar \rho_b$ have the same $a^{-3}$ time dependence.  Then, we define the adiabatic and isocurvature density fluctuations as 
\be \label{basischange}
\delta_A  \equiv w_c \delta_c + w_b \delta_b \ , \quad \text{and} \quad \delta_I \equiv \delta_c - \delta_b \  ,
\ee
so that
\be
\rho_A = \bar \rho_A ( 1 + \delta_A )  \ , \quad \text{and} \quad \rho_I = \frac{\rho_c}{w_c} - \frac{\rho_b}{w_b}  = \bar \rho_A   \delta_I  \ .
\ee
We use the same definitions for the velocity,
\be \label{aiveldefs}
v^i_A \equiv w_c v^i_c + w_b v^i_b \ , \quad  \text{and}  \quad v^i_I \equiv v_c^i - v^i_b  \ .  
\ee
Additionally, we will sometimes use the momentum densities $\pi^i_\sigma \equiv \rho_\sigma v^i_\sigma$
and define the adiabatic and isocurvture momentum densities as
\be \label{adiaisomom}
\pi^i_A \equiv \pi^i_c + \pi^i_b   \ , \quad \text{and} \quad \pi^i_I \equiv \bar \rho_A \left(  \frac{\pi^i_c}{\rho_c} - \frac{\pi^i_b}{\rho_b} \right)  \ . 
\ee

Our Fourier conventions are
\be
f(t , \xvec ) = \int_{\kvec} f(t , \kvec) e^{i \kvec \cdot \xvec} \ , \quad \text{with} \quad \int_{\kvec} \equiv \int \frac{d^3 k}{( 2 \pi)^3} \ .
\ee
Furthermore, in terms of the variables in the metric \eqn{metric}, we denote $\dot f \equiv \partial f / \partial t$, $f '  \equiv \partial f / \partial a$, the Hubble parameter $H$ is defined by $H \equiv \dot a / a$, and we use the notation $\cH \equiv a H$.  

%
%
%
%

\section{The EFTofLSS with cold dark matter and baryons}
\label{theory}

The EFTofLSS describing dark matter and baryons was developed in~\cite{Lewandowski:2014rca} {(see also {\cite{Shoji:2009gg,Somogyi:2009mh,Bernardeau:2012aq}} for work including baryons in the standard perturbation theory (SPT) context)}. Extending  the EFTofLSS for dark matter to incorporate baryons is based on the following idea: dark and baryonic matter can be described by the same equations of motion from the time of recombination to the formation of the first stars because the dynamics do not depend on their different initial conditions. Then, even after the onset of star formation processes, the baryons are still described by an effective fluid-like system with mean free path of order of the non-linear scale (as for dark matter), with the only difference being now that the numerical coefficients of the counterterms are no longer equal to those of dark matter (although we predict them to be within the same order of magnitude). Additionally, since now the two fluids can move differently, they can exchange momentum. {This makes the counterterms appear within an effective force, rather than just an effective stress tensor.  Crucially, incorporating baryons in the EFT allows us to  describe baryons accurately at long wavelengths, at the cost of adding  free coefficients that are both necessary and sufficient for the perturbative approach to converge to the true answer.
}

%
%
%
\subsection{Gravitationally coupled systems} \label{coupledsec}

In this subsection, we derive the effective equations governing two non-relativistic fluid-like systems coupled through gravity.  When only dark matter is involved, the standard way to obtain the effective equations is to smooth the Boltzmann hierarchy for the non-interacting dark-matter particles \cite{Baumann:2010tm, Carrasco:2012cv}.  However, when baryons are involved, we do not know the exact equations in the UV and so this procedure is not as rigorous.  Therefore, we choose to take the generic EFT approach of starting with the low energy degrees of freedom (i.e. the long-wavelength fields), which are the mass density $\rho$ and the momentum density $\pi^i$ (or equivalently the velocity $v^i$), and writing down the most general equations consistent with the symmetries of the system, in an expansion in derivatives and powers of the perturbations (see \cite{Mercolli:2013bsa} for a related discussion with one fluid).  {As with all EFT constructions, this one as well relies on a separation of scales.  The reason that we can describe the system as fluid-like (i.e. using the mass density and momentum density as our fundamental low energy degrees of freedom) is because dark matter and baryons do not move too much in the history of the universe.}  This means that there is a non-linear scale $\knl$, which is related to the typical distance that the particles have traveled by $\knl^{-1} \sim v \cH^{-1}$ \cite{Baumann:2010tm}, and that our results rely on the hierarchy  $k / \knl \lesssim 1$.\footnote{{Of course, a small fraction of baryons at late times are expelled in explosions and travel farther than the non-linear scale.  The EFT approach does not take these effects into account, but the amount of baryons traveling outside of the non-linear scale is expected to be tiny.  This does, however, represent a tiny uncalculable systematic error in our calculations. } }
{In this section we take a more bottom-up approach, while in \appref{GRsec} we take a somewhat more top-down approach.  Our results agree with the results of \cite{Baumann:2010tm} for one fluid, and of \cite{Lewandowski:2014rca} for two fluids.}  

The symmetries relevant for our discussion here are conservation of the total number of dark matter and baryonic particles separately, conservation of the total momentum, and overall Galilean invariance, which is the result of a residual large gauge symmetry (or equivalently, residual large diffeomorphisms) {\cite{Creminelli:2012ed, Hinterbichler:2012nm, Hinterbichler:2013dpa, Creminelli:2013mca, Horn:2014rta}} of the (scalar part of the) Newtonian-gauge metric\footnote{In the absence of anisotropic stress, the Einstein equations imply that $\Psi = \Phi$, and we assume this throughout our work unless otherwise stated.  { Notice that, although anisotropic stress can be generated in the EFTofLSS, for what concerns its effect on the metric, it is a relativistic correction.}}
 \be \label{metric}
 ds^2 = - ( 1 + 2 \Phi ) dt^2 + a(t)^2 ( 1 - 2 \Psi ) d \xvec^2 \ ,
 \ee
which we use throughout this work.

As shown in \appref{GRsec}, the equations of motion for matter in the Newtonian limit  take the form of the divergence of a pseudo tensor 
\be \label{pseudocons1}
a^{-3} \partial_\mu ( a^3 t^\mu{}_\nu ) = 0 \ , 
\ee
where $t^{\mu }{}_\nu$ is a symmetric stress-energy pseudo-tensor, which involves the individual stress tensors of CDM and baryons as well as gravitational non-linearities.  This form of the equations suggests that we should be able to write a system of first order differential equations for the CDM and baryon mass densities $\rho_c$ and $\rho_b$, and the CDM and baryon momentum densities $\tilde \pi^i_c$ and $ \tilde \pi^i_b$ (we use the tilde here for notational convenience to simplify later expressions), all of which are components of the CDM and baryon stress tensors (see \appref{GRsec} for details).

We start with the equations for $\rho_\sigma$.  {Because we assume the separate conservation of dark matter and baryons, we start by writing separate equations for each $\rho_\sigma$.}  In $\Lambda$CDM, both $\rho_c$ and $\rho_b$ have time-dependent background values $\bar \rho_c$ and $\bar \rho_b$, which satisfy $\dot{\bar \rho}_\sigma = - 3 H \bar \rho_\sigma$.  This means that, in order to ensure that we expand around the correct background, our continuity equations must start with
\be
\text{Continuity:} \quad \dot \rho_\sigma + 3 H \rho_\sigma \ ,
\ee
where, without loss of generality, we have assumed that the coefficient of $\dot \rho_\sigma$ is unity.
{Next, we impose diffeomorphism invariance.  The subset of diffeomorphisms that  keep us in Newtonian gauge and that is relevant for the Newtonian limit are the so-called Galilean transformations} 
\be \label{galcoordtrans}
t \rightarrow t+a^2 n^i(t) x^i \  , \quad \text{and} \quad x^i \rightarrow x^i + n^i ( t ) \ , 
\ee
 which act on the terms in the equations of motion at leading order in a relativistic expansion, as
\begin{align} \label{fluidreps}
\begin{split}
\partial_i & \rightarrow \partial_i \ , \quad  \partial_t   \rightarrow \partial_t - \dot n^i ( t )   \partial_i \ , \quad  \rho_\sigma  \rightarrow \rho_\sigma    \ , \quad \tilde \pi^i_\sigma \rightarrow \tilde \pi^i_\sigma + \rho_\sigma \, a \dot n^i(t) \ ,  \\ 
\Phi  &  \rightarrow \Phi - a^2 (\ddot n^i ( t ) + 2 H \dot n^i ( t )   ) x^i    \  ,
\end{split}
\end{align}
so that, by construction, $v^i_I$ and $\pi^i_I$ are Galilean scalars.  {The transformation of the momentum density above follows directly from the transformation of the velocity.}
This means that a Galilean invariant combination is 
\be
\text{Continuity:} \quad \dot \rho_\sigma + 3 H \rho_\sigma + a^{-1} \partial_i \left( \tilde  \pi^i_\sigma + \frac{b_\sigma \rho_\sigma}{\bar \rho_A}  \tilde \pi^i_I \right) +a^{-1}  \frac{\tilde b_\sigma \rho_\sigma}{\bar \rho_A}  \partial_i\tilde \pi^i_I+\ldots \ ,
\ee
for any time-dependent functions $b_\sigma$ and $\tilde b_\sigma$, {where the ellipsis $\dots$ represents higher order terms}.  Finally, we impose conservation of the number of dark-matter particles and baryon particles separately 
\be
\partial_t \int d^3 x \,  a^3 \rho_\sigma = 0 \  ,
\ee
which implies that $\partial_t ( a^3 \rho_\sigma)$ is a total spatial derivative.  This means that the most general equations for $\dot \rho_\sigma$ that expand around the correct background, that are Galilean invariant, and that conserve mass are
\be \label{contpif}
\text{Continuity:} \quad \dot \rho_\sigma + 3 H \rho_\sigma + a^{-1} \partial_i \left( \tilde  \pi^i_\sigma  + F_\sigma^i \right) = 0  \ ,
\ee
where $F^i_\sigma$ contains the previous term $ b_\sigma \rho_\sigma \tilde \pi^i_I / \bar \rho_A$, and also any other Galilean invariant terms, including non-linear and higher derivative terms.

Next, we move to the momentum equations for $\tilde \pi^i_\sigma$.  Before we start, there is an important subtlety that we should briefly mention.  The question is, given the form of \eqn{contpif}, whether we should write the momentum equations starting with $\partial_t \tilde \pi^i_\sigma$, or with $\partial_t ( \tilde \pi^i_\sigma + F^i_\sigma)$.  As is well known, the stress-energy pseudo-tensor $t^{\mu}{}_\nu$ in \eqn{pseudocons1} is symmetric (which is guaranteed because the system is coupled to gravity through the symmetric metric).  As we show explicitly in \appref{GRsec}, this implies that whatever appears under the $\partial_i$ in \eqn{contpif} should appear under the $\partial_t$ in the momentum equations.  This means that we should start to construct the momentum equation with $\partial_t ( \tilde \pi^i_\sigma + F^i_\sigma) + \dots$.  To do this, we define the combination\footnote{Since $F^i_\sigma$ is a Galilean scalar, $\pi^i_\sigma$ has the same transformation properties as $\tilde \pi^i_\sigma$ under \eqn{galcoordtrans}.}
\be
 \pi^i_\sigma \equiv \tilde  \pi^i_\sigma + F^i_\sigma \ , 
\ee
and then write everything in terms of $ \pi^i_\sigma$.  Thus the final form of the continuity equations are 
\be \label{contreal}
\text{Continuity:} \quad \dot \rho_\sigma + 3 H \rho_\sigma + a^{-1} \partial_i  \pi^i_\sigma = 0  \ .
\ee

Next, we start constructing the momentum equations with $\dot \pi^i_\sigma$, which transforms like
\be \label{dotpitransf}
\dot \pi^i_\sigma \rightarrow \dot \pi^i_\sigma + \dot \rho_\sigma a \dot n^i + \rho_\sigma \partial_t ( a \dot n^i ) - \dot n^j \partial_j \pi^i_\sigma -a  \dot n^i \dot n^j \partial_j \rho_\sigma \ . 
\ee
Then, adding terms to make a Galilean invariant combination, we have
\be \label{momentum1}
\text{Momentum:} \quad \dot \pi^i_\sigma - \frac{\dot \rho_\sigma}{\rho_\sigma} \pi^i_\sigma + a^{-1} \pi_\sigma^j \partial_j \left( \frac{\pi^i_\sigma}{\rho_\sigma} \right) + H \pi^i_\sigma  + a^{-1} \rho_\sigma \partial_i \Phi +a^{-1} G_\sigma^i  = 0  \  , 
\ee
where $G_\sigma^i$ must be Galilean invariant but otherwise can be a sum of any non-linear and higher derivative terms, and, without loss of generality, we have assumed that the coefficient of $\dot \pi^i_\sigma$ is unity.  The form of these equations is unique for pressureless fluids up to the freedom just mentioned in $G^i_\sigma$.\footnote{One can obtain different equations if there is a background pressure, as in clustering quintessence (see for example \cite{Creminelli:2008wc, Creminelli:2009mu, Sefusatti:2011cm} and \cite{Lewandowski:2016yce} within the EFTofLSS context), or if there are other fields in the low-energy spectrum, as with dark energy (see for example \cite{Bartolo_2013, Bloomfield_2013, Gleyzes_2013, Takushima_2014, Bellini:2014fua, Gleyzes:2014rba, Cusin:2017wjg, Hirano:2018uar, Crisostomi:2019vhj, Lewandowski:2019txi}).  In particular, to obtain the equations for clustering quintessence in the limit of vanishing speed of sound (i.e. including background pressure $\bar p$), we first realize that the background Einstein equations imply $\dot{\bar \rho} = - 3 H ( \bar \rho + \bar p )$, and that the momentum density transforms as $\pi^i \rightarrow \pi^i+ ( \rho + \bar p ) a \dot n^i $ under a Galilean transformation.   This means that the continuity equation that expands around the correct background and is Galilean invariant is 
\be
\dot \rho + 3 H ( \rho  + \bar p) + a^{-1} \partial_i \pi^i = 0 \ . 
\ee
Then, using the same construction that lead to \eqn{momentum1} (but now for a single fluid), and taking into account the new transformation of $\pi^i$, we see that the momentum equation that is Galilean invariant is simply the same as \eqn{momentum1}, but with $\rho$ replaced by $\rho + \bar p$.  Finally, defining the velocity by $\pi^i = ( \rho + \bar p ) v^i$, one obtains the equations for clustering quintessence (see for example \cite{Sefusatti:2011cm}), which have the same Euler equation as dark matter, but a different continuity equation that reflects the different background. }  In particular, we see that the coupling to gravity through $\partial_i \Phi$ is forced because it is the only field with a transformation that depends on $\ddot n^i$, which is needed to cancel the term proportional to $\ddot n^i$ in \eqn{dotpitransf}. This is at the level of equations of motion the same phenomenon that happens in the Lagrangian where diffeomorphism invariance forces a minimal coupling to gravity. Notice furthermore that since baryons and dark matter are two independent degrees of freedom (as evident from the very early universe dynamics), we have a diffeomorphism-invariant equation of motion for the momentum of each species. 

Next, we use the continuity equations \eqn{contreal} to write \eqn{momentum1} as 
\be \label{momentum2}
\text{Momentum:} \quad \dot \pi^i_\sigma  + 4 H \pi^i_\sigma + a^{-1}  \partial_j \left( \frac{\pi^i_\sigma \pi_\sigma^j }{\rho_\sigma} \right)  + a^{-1} \rho_\sigma \partial_i \Phi +a^{-1} G_\sigma^i  = 0  \  .
\ee
Finally, we must impose total momentum conservation in the form\footnote{ {The factor of  $a^4$  can be easily understood working in Fermi coordinates and then going back to FRW coordinates as done in \cite{Baumann:2010tm}: one should keep in mind that only the obvious factors of $a$ need to be included in these transformations, and that the comoving velocity is related to the proper velocity by another factor of $a$. } }
\be
\partial_t \int d^3 x  \, a^4 ( \pi^i_c + \pi^i_b ) = 0 \ ,
\ee
which means that $\partial_t ( a^4 ( \pi^i_c + \pi^i_b))$ must be a total spatial derivative.  Adding together the two equations in \eqn{momentum2}, we have a term of the form
\be \label{rhoaterm}
(\rho_c + \rho_b) \partial_i \Phi = \bar \rho_A ( 1 + \delta_A ) \partial_i \Phi \ ,
\ee
which we would like to write as a total derivative.  To do that, we use the Poisson equation
\be
a^{-2} \partial^2 \Phi = \frac{3}{2} \Om ( t ) H(t)^2 \delta_A\ , 
\ee
where $\Om (t)$ is the time-dependent total-matter fraction,\footnote{This is defined by $\Om ( t ) \equiv \bar \rho_A ( t ) / (3 \mpl^2 H(t)^2 )$, where $\mpl$ is the Planck mass, which is related to the Newton constant $G_N$ by $\mpl^2 = 1 / (8 \pi G_N)$.  In terms of the scale factor, this is given by $\Om ( a ) = \Omega_{{\rm m},0} (H_0 / H(a) )^2 (a / a_0)^{-3}$, where the subscript ``$0$'' means the present value.  In $\Lambda$CDM, we parametrize Hubble by $H(a)^2 / H_0^2 = \Omega_{{\rm m},0} \left( a / a_0  \right)^{-3} + (1 - \Omega_{{\rm m},0}) $ . } 
to write 
\be
\delta_A \partial_i \Phi = \frac{2 a^{-2}}{3 \Om ( t ) H(t)^2 } \partial_j \left( \partial_i \Phi \partial_j \Phi - \half \delta_{ij} (\partial \Phi)^2 \right)  \ , 
\ee  
which shows that \eqn{rhoaterm} is a total derivative.  All in all, this means that $G^i_c + G^i_b$ must be a total derivative, so we can have
\begin{align}
\begin{split}
G^i_c = -  \gamma^i + \partial_j \tau^{ij}_c \andd  G^i_b = + \gamma^i + \partial_j \tau^{ij}_b \  .
\end{split}
\end{align}
Here, $\gamma^i$ is a Galilean scalar, and the results of \appref{GRsec} show that $\tau^{ij}_c$ and $\tau^{ij}_b$ are both Galilean scalars.  

All in all, this means that the most general equations for $\dot \rho_\sigma$ and $\dot \pi^i_\sigma$ that expand around the correct $\Lambda$CDM background, are Galilean invariant, satisfy conservation of the number of dark-matter and baryon particles separately, and satisfy total momentum conservation, are \cite{Lewandowski:2014rca}
\begin{align}
 \text{Continuity:} &\quad \dot \rho_\sigma + 3 H \rho_\sigma + a^{-1} \partial_i  \pi^i_\sigma = 0  \label{finalcont1}  \ , \\
 \text{Momentum:} &\quad   \dot \pi^i_c   + 4 H \pi^i_c  + a^{-1} \partial_j \left( \frac{\pi^i_c \pi^j_c}{\rho_c} \right) + a^{-1} \rho_c \partial_i \Phi = + a^{-1} \gamma^i - a^{-1} \partial_j \tau^{ij}_c  \label{momeqc1}\ ,  \\
& \quad   \dot \pi^i_b   + 4 H \pi^i_b  + a^{-1} \partial_j \left( \frac{\pi^i_b \pi^j_b}{\rho_b} \right) +  a^{-1} \rho_b \partial_i \Phi =  -  a^{-1} \gamma^i - a^{-1} \partial_j \tau^{ij}_b \  .  \label{momeqb1}
\end{align}
The important new possibility is a term $\gamma^i$, which is allowed by the symmetries, and is in fact generically needed to cancel UV divergences in the one-loop power spectrum, as we show in \secref{linrelvelsec}.  

As always, the velocity fields, defined by $v_\sigma^i \equiv \pi^i_\sigma / \rho_\sigma$, are contact operators, and so one will have in general that 
\be \label{longwlvel}
v^i_\sigma \neq \left[ \frac{\pi^i_{\sigma,s}}{\rho_{\sigma,s}} \right]_{L}  \ , 
\ee
where the subscript $s$ stands for short modes, and the brackets $[\dots ]_L$ mean smoothing to form a long-wavelength field.  This means that when one computes correlation functions involving the velocity, one needs to use the renormalized fields \cite{Carrasco:2013mua}.  {Now, for the case of two fluids, one must keep in mind that the relative velocity $v^i_I$ is allowed in the counterterms of the two renormalized velocities because it is a Galilean scalar and will not affect the transformation properties of the velocity fields.  }

%
%
%

%
%
%
\subsection{The effective force and stress tensors for two loops} \label{forcesandsts}

In this section, we explicitly construct the effective force and stress tensors that are relevant for the two-loop calculation.  First, we write the equations of motion in terms of the velocities $v_\sigma^i \equiv \pi^i_\sigma / \rho_\sigma$, and then take the divergence of the velocity equations to obtain \cite{Lewandowski:2014rca}
\begin{equation}
\label{BARYONS:EOMgeneral}
\begin{split}
& a^{-2} \partial^2 \Phi = \frac{3}{2} \Om (a ) H^2 ( w_c \delta_c + w_b \delta_b)   \ ,  \\
& \dot{\delta}_c  +  a^{-1} \partial_i((1 + \delta_c) v_c^i) = 0 \ , \quad \dot{\delta}_b  +  a^{-1}  \partial_i((1 + \delta_b) v_b^i)  = 0  \ , \\
&\partial_i\dot{v}_c^i + H \partial_iv_c^i  + a^{-1}\partial^2 \Phi    + a^{-1}\partial_i(v_c^j \partial_j v_c^i)   = - a^{-1} \partial_i \left( \partial \tau_{\rho}\right)_c^i +  a^{-1}\partial_i(\gamma)_c^i  \ ,  \\
&\partial_i\dot{v}_b^i + H \partial_iv_b^i    + a^{-1}\partial^2 \Phi   + a^{-1}\partial_i(v_b^j \partial_j v_b^i)  = - a^{-1} \partial_i \left( \partial \tau_{\rho}\right)_b^i+ a^{-1} \partial_i(\gamma)_b^i \ ,
\end{split}
\end{equation}
where 
\begin{equation}
(\gamma)_c^i = \frac{1}{\rho_c}\gamma^i, \quad (\gamma)_b^i = -  \frac{1}{\rho_b}\gamma^i\,, \quad \text{and} \quad \left( \partial \tau_{\rho}\right)_{\sigma}^i = \frac{1}{\rho_{\sigma}}\partial_j \tau^{ij}_{\sigma}\,.
\end{equation}

Next, we expand the force and the stress tensors in powers and derivatives of the long-wavelength fields.  
Since at this order we can approximate the stress tensors as being local in time\footnote{ Non-locality in time of the EFTofLSS tells us that the coefficients of the counterterms are integrals in time of some kernel of an expansion in powers and derivatives of $\partial_i \partial_j \Phi$ and $\partial_i v_{\sigma}^{j}$. To lowest order we have
\begin{align}
& (\partial \tau_{\rho})_{\sigma}^{i}\left(a,\vec{x}\right) - \left( \gamma\right)_{\sigma}^{i}\left(a,\vec{x}\right)= \int da' \Big[\kappa_{\sigma}^{(1)}\left(a,a'\right)\partial^{i}\partial^2\Phi\left(a';\vec{x}_{\text{fl}}\left(\vec{x};a,a'\right)\right)  +\kappa_{\sigma}^{(2)}\left(a,a'\right)\frac{1}{H(a')}\partial^{i}\partial_{j}v_{\sigma}^{j}\left(a';\vec{x}_{\text{fl}}\left(\vec{x};a,a'\right)\right) \nonumber  \\ 
&\hspace{2in}  {+ \kappa^{(3)}_\sigma ( a , a') H(a') v^i_I \left(a';\vec{x}_{\text{fl}}\left(\vec{x};a,a'\right)\right)   } + \ldots   \Big]\, .
\label{kappa}
\end{align}
The linear evolution of the modes is scale independent. In this way, the complication associated to the non-local time kernels is reduced greatly. In fact, the perturbative solutions schematically have the structure
\begin{eqnarray}
\delta^{(n)} (a, \xvec) = D(a)^n \delta^{(n)}(\xvec) \, ,
\end{eqnarray}
where $D(a)$ represents any of the growth factors in \appref{linearandedsapp}.  Therefore, schematically, we have
\begin{align}
{ \partial \tau_{\rho}   - \gamma  \sim \int d a' \left(  \kappa (a, a') \partial \delta (a', \xvec) +  K(a , a') \delta (a', \xvec) \right) = \sum_n \left(  \tilde \kappa_n(a) \partial \delta^{(n)}( a, \xvec) + \tilde K_n ( a ) \delta^{(n)}( a, \xvec) \right)  \ ,  }
\end{align}
where
\begin{align}
{\tilde \kappa_n(a) = \int da' \kappa(a,a') \left( \frac{D(a')}{D(a)} \right)^n \andd    \tilde K_n(a) = \int da' K(a,a') \left( \frac{D(a')}{D(a)} \right)^n  \ , }
\end{align}
(the actual expression is slightly more complicated due to the flow terms (see~\cite{Carrasco:2013mua})). In other words, the non-locality in time is reduced to having a set of local counterterms, with a different coefficient for each order in perturbation theory. If one uses the counterterms at leading order, the non-locality in time is degenerate with a local in time counterterm. At higher order, this is mathematically not the case anymore. In our case, we will be interested in the two-loop power spectrum. In this scenario, there are several quadratic counterterms that are evaluated at leading order, together with the linear counterterms evaluated at higher order. The functional form of those terms is quite degenerate, so that, as has been shown in~\cite{Foreman:2015lca}, one can just include the tree level terms, and therefore for our computation, limit ourself to treat the counterterms as if they were local in time.} (see~\cite{Carrasco:2013mua} for a more detailed discussion), the effective stress tensor for dark matter relevant for the two-loop power spectrum has the form
\begin{align}
\begin{split}\label{stress1}
&  \partial_i (\partial \tau_{\rho})_{c}^{i}  - \partial_i(\gamma)^i_c = -  g\, w_b  \, a H \,\partial_i v_I^i + 9 ( 2 \pi) H^2 \Big\{   \frac{ c_{c,g}^2}{k^2_{\text{NL}}}  \left ( w_c  \partial^2\delta_c +w_b  \partial^2\delta_b \right )+ \,\frac{  c_{c,v}^2 }{k^2_{\text{NL}}} \, \partial^2  \delta_c \\  
&\hspace{2in} + \,\frac{1}{ k^2_{\text{NL}}} \left( c_{1c}^{cc} \partial^2 \delta_{c}^2 +  c_{1c}^{cb} \partial^2 \left(\delta_{c} \delta_{b}\right) + c_{1c}^{bb} \partial^2 \delta_{b}^2  \right)\,   \\  
& \hspace{2in}   + \, \frac{ c_{4c,g}^2}{a^2 k^4_{\text{NL}}}  \left ( w_c  \partial^4\delta_c +w_b  \partial^4\delta_b \right )+ \, \frac{  c_{4c,v}^2}{a^2 k^4_{\text{NL}}} \, \partial^4  \delta_c  \Big\}   + \dots \ , 
\end{split}
\end{align}
 and the effective stress tensor for baryons relevant for the two-loop power spectrum has the form
\begin{align}
\begin{split} \label{stress2}
&  \partial_i (\partial \tau_{\rho})_{b}^{i} - \partial_i(\gamma)^i_b = + g  \, w_c \,  a H \,\partial_i v_I^i  + 9 ( 2 \pi) H^2 \Big\{   \frac{ c_{b,g}^2 }{k^2_{\text{NL}}}  \left ( w_c  \partial^2\delta_c +w_b  \partial^2\delta_b \right ) +  \frac{  c_{b,v}^2+c_{\star(1)}^2   }{k^2_{\text{NL}}} \, \partial^2  \delta_b \,  \\
&\hspace{2in}   + \frac{1}{ k^2_{\text{NL}}} \left( c_{1b}^{cc}  \partial^2 \delta_{c}^2 +  c_{1b}^{cb}  \partial^2 \left(\delta_{c} \delta_{b}\right) + c_{1b}^{bb}  \partial^2 \delta_{b}^2  \right)\,   \\  
& \hspace{2in}  +   \frac{ c_{4b,g}^2}{a^2 k^4_{\text{NL}}}  \left ( w_c  \partial^4\delta_c +w_b  \partial^4\delta_b \right )+ \frac{    c_{4b,v}^2+c_{4\star}^2 }{a^2 k^4_{\text{NL}}} \, \partial^4  \delta_b \Big\}  + \dots \ ,  \\ 
\end{split}
\end{align}
where $w_c$, $w_b$, and $\knl$ are constants, the fields $\delta_\sigma$ and $v^i_\sigma$ depend on $(a , \xvec)$, and the rest of the coefficients depend on $a$,  $v_I^i \equiv v_c^i - v_b^i$ is the relative velocity, and quadratic terms $\sim \partial^2 \delta^2$ and higher derivative terms $\sim \partial^4 \delta$ have been included.  The ellipsis $\dots$ represents higher-order or higher-derivative terms.  Notice that we did not include any cubic counterterms, since they would contribute terms proportional to $k^2 P_{11} ( k )$, and so are degenerate with other terms that we have included for the two-loop power spectrum \cite{Foreman:2015lca}.  Notice also the presence of a term $\propto g \,\partial_i v^i_I$ from expanding the effective force.  {In \eqn{stress1} and \eqn{stress2}, we considered this counterterm to be local-in-time, while in \secref{linearevo} and \appref{pertwithlinctsec} we consider the possibility of non-locality in time. }

Recall that the effective force and stress tensors of the two fluids has the form $- \frac{1}{\rho_\sigma} \partial_{j} \tau_{\sigma}^{ij} \pm \frac{1}{\rho_{\sigma}} \gamma^{i}$. Therefore by symmetry, further quadratic terms, which are not total derivatives (for example  $\delta_b \partial_i \delta_c$ and $\delta_c \partial_i \delta_b$) can be added as well. 
However, as we will see, these terms seem not to be needed for the fitting procedure at the level of precision at which the analysis is performed.  Additionally, we have only presented the subset of quadratic counterterms that contribute non-degenerately to the final form of the two-loop power spectrum that we present below in \eqn{candbtwoloop}.  See \cite{Foreman:2015lca} for a study of other possible combinations in the dark-matter case, and \cite{Angulo:2014tfa} for the specific forms of other quadratic counterterms.

The above equations for the effective stress tensors and force \eqn{stress1} and \eqn{stress2} introduce, suggestively,  the EFT coefficients induced by gravity $\{g\}$, $\{c_{c,g} , c_{b,g} \}$, $\{c_{1c}^{(cc)} , c_{1c}^{cb}, c_{1c}^{bb}, c_{1b}^{cc}, c_{1b}^{cb}, c_{1b}^{bb} \}$, $\{c_{4c,g} , c_{4b,g} \}$ and by gradients of the velocity fields $\{c_{c,v} , c_{b,v} \}$, $\{c_{4c,v} , c_{4b,v} \}$ for dark matter and baryons. The parameters induced by star-formation physics are described by the coefficients $\{c_{\star (1)}^2, c_{4\star}^2 \}$, which are assumed to be the main difference between the two species.

In terms of the derivatives with respect to the scale factor $a$ and the velocity divergences 
\be
\Theta_\sigma \equiv - \partial_i v^i_\sigma  / \cH \ ,
\ee
the non-linear evolution equations \eqn{BARYONS:EOMgeneral} in Fourier space now become
\begin{align}
\begin{split} \label{ceomwithst}
&a  \delta_{c}' ( a , \kvec )  -  \Theta_c ( a , \kvec ) =  \alpha_{cc} ( a , \kvec)  \ ,    \\
& a  \Theta_{c} ' ( a , \kvec )  + \left( 1 + \frac{a \cH'}{\cH} \right) \Theta_{c} ( a , \kvec )  - \frac{3 \Om(a) }{2} ( w_c \delta_c ( a , \kvec )  + w_b \delta_b ( a , \kvec ) ) =  \\
& \hspace{2.5in}  + \beta_{cc}   ( a , \kvec )  + \cH^{-2} \left(  [\partial_i (\partial \tau_{\rho})_{c}^{i}(a)]_{\kvec}  - [\partial_i(\gamma)^i_c (a )  ]_{\kvec}  \right)  \ , 
\end{split}
\end{align}
\begin{align}
\begin{split} \label{beomwithst}
& a  \delta_{b}' ( a , \kvec )   - \Theta_b ( a , \kvec ) =  \alpha_{bb} ( a , \kvec ) \ ,  \\
& a \Theta_{b}' ( a , \kvec )  + \left( 1 + \frac{a \cH'}{\cH} \right) \Theta_{b} ( a , \kvec)   - \frac{3 \Om ( a ) }{2} ( w_c \delta_c ( a , \kvec ) + w_b \delta_b ( a , \kvec) )  =  \\
& \hspace{2.5in} + \beta_{bb} ( a , \kvec )  + \cH^{-2} \left(  [\partial_i (\partial \tau_{\rho})_{b}^{i}(a)]_{\kvec}  - [\partial_i(\gamma)^i_b (a )  ]_{\kvec}  \right)  \ , 
\end{split}
\end{align}
where the non-linear terms are defined by 
\begin{align}
\begin{split} \label{alphabetadefs}
& \alpha_{\sigma \sigma'} ( a , \kvec ) \equiv \int_{\qvec} \alpha ( \kvec - \qvec , \qvec ) \delta_{\sigma} ( a , \kvec - \qvec) \Theta_{\sigma'} ( a , \qvec ) \ , \\
& \beta_{\sigma \sigma'} ( a , \kvec ) \equiv \int_{\qvec} \beta(\kvec - \qvec , \qvec ) \Theta_\sigma ( a , \kvec - \qvec ) \Theta_{\sigma'} ( a , \qvec )  \ ,
\end{split}
\end{align}
with
\be
\alpha ( \kvec_1 , \kvec_2 ) = 1 + \frac{\kvec_1 \cdot \kvec_2}{k_2^2} \ , \quad \text{and} \quad  \beta( \kvec_1 , \kvec_2 ) = \frac{|\kvec_1 + \kvec_2|^2 \kvec_1 \cdot \kvec_2}{2 k_1^2 k_2^2 } \ ,
\ee
and we have used the shorthand notation $[\cdots ]_{\kvec}$ to mean the Fourier transform evaluated at momentum $\kvec$.


\subsection{Two-loop solution}
\label{oneloop}

Using the equations of motion including counterterms \eqn{ceomwithst} and \eqn{beomwithst}, we can now compute the two-loop power spectra.  As we will show in \secref{linrelvelsec}, the size of the contribution from the linear counterterm proportional to $ g \,   v^i_I$ in \eqn{ceomwithst} and \eqn{beomwithst} is expected to be small.  The success of our fits in \secref{Comparison}, which do not include this counterterm, also supports this.  Thus, we take $g = 0$ in this section.  For completeness, we discuss how one can include this term in perturbation theory in \secref{linrelvelsec} and \appref{pertwithlinctsec}.  

Defining the normalized adiabatic growth factor $D_1 ( a ) \equiv D_{A_+}(a) / D_{A_+}(a_0)$, the final expressions are,  
\begin{eqnarray} \label{candbtwoloop}
&&P_{\text{EFT-2-loop}}^{\sigma }(a, k)= P^{\sigma}_{\text{EFT-1-loop}}(a, k) +  [D_{1}(a)]^{6} P^{ A}_{\text{2-loop}}(k) - 2 (2 \pi) c_{\sigma (2)}^2(a) \frac{k^2}{k_{\text{NL}}^2} P^{A}_{11}(k) \nonumber \\
&& \qquad + (2 \pi) c_{\sigma(1)}^2 (a) [D_{1}(a)]^{4} P_{\text{1-loop}}^{A,(c_s)}(k) + (2 \pi)^2 \left(1 + \frac{\xi_\sigma + \frac{5}{2}}{2(\xi_\sigma + \frac{5}{4})}\right) [c_{\sigma (1)}^2 (a)]^2 [D_{1}(a)]^{2} \frac{k^4}{k_{\text{NL}}^4} P^{A}_{11}(k) \, \nonumber \\
&&\qquad  + (2 \pi) c^2_{1\sigma }(a) [D_{1}(a)]^4 P_{\text{1-loop}}^{A\text{,(quad,1)}} (k) + 2 (2 \pi)^2 c^2_{4\sigma }(a) [D_{1}(a)]^2 \frac{k^4}{k_{\text{NL}}^4} P^{A}_{11}(k) \ , \label{UVimprovedA}
\end{eqnarray}
where the one-loop expressions are given by  
\be
P^\sigma_{\text{EFT-1-loop}} ( a , k ) =  P^\sigma_{11} (a ,  k ) + [D_1 ( a ) ]^4 P_{\text{1-loop}}^A  ( k )  - 2 ( 2 \pi ) c_{\sigma (1)}^2 ( a ) [D_1(a)]^2 \frac{k^2}{\knl^2} P_{11}^A ( k  )  \ ,
\ee
for $\sigma \in \{ A, c , b \}$, denoting the adiabatic, CDM, and baryon power spectra, respectively.  The linear power spectra $P_{11}^\sigma ( a , k) $ are defined by
\be
\langle \delta_{\sigma}^{(1)} ( a , \kvec ) \delta_{\sigma}^{(1)} ( a , \kvec' ) \rangle = (2 \pi)^3 \delta_D ( \kvec + \kvec') P_{11}^\sigma ( a , k) \ ,
\ee
where $\delta_D$ is the Dirac delta function, and we have written $P_{11}^A ( k ) \equiv P_{11}^A ( a_0 , k)$ for convenience.  The time dependence of the linear adiabatic power spectrum is given by 
\be
P^A_{11} ( a , k ) = [D_1 ( a )]^2 P_{11}^A ( k ) \ , 
\ee
whereas the time dependence of the CDM and baryon linear power spectra can be computed using the definitions for the CDM and baryon overdensities in terms of the adiabatic and isocurvature modes \eqn{basischange}, and then using the linear solutions for the adiabatic and isocurvature modes in \eqn{adiabaticisolinearsols}.   {Neglecting the isocurvature mode}, the time dependence of CDM and baryons is simply $[D_1(a)]^2$, but there are subleading corrections, particularly important at early times, which generally make the CDM and baryons evolve differently on linear scales if an isocurvature mode is present.  As always, though, one can simply obtain the linear power spectra for CDM and baryons directly from CAMB at each redshift.

  The adiabatic EFT parameters are defined in terms of the CDM and baryon parameters by 
\begin{align}
\begin{split} \label{cavaluedefs}
& c_{A(1)}^2 ( a  ) \equiv w_c c_{c(1)}^2(a) + w_b c_{b(1)}^2 (a) \ , \quad c_{A(2)}^2 (a) \equiv w_c c_{c(2)}^2(a) + w_b c_{b(2)}^2(a) \ , \\
& c^2_{1A}(a) \equiv w_c c^2_{1c} (a) + w_b c^2_{1b}  (a)  \andd   c^2_{4A}(a) \equiv w_c c^2_{4c} (a) + w_b c^2_{4b}  (a) \ . 
\end{split}
\end{align}
Finally, we note that $c_{\sigma(2)}^2$ is not a free parameter; it is determined in terms of $c_{\sigma(1)}^2$ using the fitting procedure in \cite{Carrasco:2013mua, Foreman:2015uva, Foreman:2015lca}.  In this work, we use the UV-improved two-loop power spectrum, so we set $c_{\sigma (2)}^2 = 0$ from now on \cite{Foreman:2015lca}.  

Let us unpack and explain the above expressions.  First, notice that, besides the linear power spectra, all of the higher order terms are computed with the adiabatic mode, which makes these expressions very similar to those of the pure dark-matter case \cite{Carrasco:2013mua, Foreman:2015uva, Foreman:2015lca};\footnote{In particular, the SPT contributions $P^A_{\text{1-loop}}$ and $P^A_{\text{2-loop}}$ (see for example \cite{Bernardeau:2001qr}) are implemented with the IR-safe integrand \cite{Carrasco:2013sva,Carrasco:2013mua}.  } the main differences are in the values of the counterterms for each fluid. This approximation is justified because the inclusion of an isocurvature mode in the one-loop power spectra is down by a factor of approximately $ 5 \times 10^{-3}$ at $z = 0$ from the one-loop adiabatic power spectrum, which is subleading to the two-loop adiabatic contribution (see \figref{fig:DESI2loop}) .  

Next, we quickly review where the various terms come from.  In each power spectrum, there are three different contributions going like $k^4 P_{11}^A$.  The first, proportional to $[c_{\sigma(1)}^2]^2$, comes from contracting two of the one-loop counterterms which go like $ c_{\sigma(1)}^2 k^2 \delta^{(1)}$ \cite{Carrasco:2013mua}.  The second comes from plugging the one-loop counterterm back into itself in the stress tensor.  This term is also proportional to $[c_{\sigma(1)}^2]^2$ but can have a different time dependence from the previous counterterm due to the nested Green's functions; capturing this different time dependence is the role of the term involving $\xi_\sigma$ \cite{Foreman:2015uva}.  The final term, proportional to $c_{4\sigma}$ comes from explicitly adding higher derivative terms to the stress tensors \cite{Foreman:2015lca}.  The term $P^{A,(c_s)}_{\text{1-loop}}$ comes from expanding the fluid line element $\xvec_{\rm fl.}$ in the argument of the linear counterterm $c_{\sigma(1)}^2 k^2 \delta^{(1)}$ (and so is also proportional to $c^2_{\sigma(1)}$ but is a two-loop term) \cite{Carrasco:2013mua},\footnote{In particular, we use   $P^{A,(c_s,p)}_{\text{1-loop}}$ for $p\rightarrow \infty$, which is the local in time limit \cite{Carrasco:2013mua,Foreman:2015uva}.  Also, see \cite{Carrasco:2013mua} and references therin for the definition of $\xvec_{\rm fl.}$ and how and why it enters.} while the $P^{A,(\text{quad,1})}_{\text{1-loop}}$ term comes from explicit quadratic terms added to the stress tensors \cite{Foreman:2015lca}.\footnote{In particular, $P^{A,(\text{quad,1})}_{\text{1-loop}}$ is a contraction of the quadratic counterterm proportional to $K_1$ in the appendix of \cite{Angulo:2014tfa} with $\delta^{(2)}$ from SPT, divided by $\knl^2$.}

We have also dealt with non-locality in time in the same way as in previous dark-matter studies \cite{Carrasco:2013mua, Foreman:2015uva, Foreman:2015lca}.  In this work, though, there is the additional complication that CDM and baryons have slightly different Green's functions, so that various different combinations appear when computing the loops.  This is relevant for the $P^{A,(c_s)}_{\text{1-loop}}$ and $P^{A,\text{(quad,1)}}_{\text{1-loop}}$ terms, for example. However, we expect this difference to be small, since it is proportional to the isocurvature Green's function, and we do not find any evidence of needing to include it in this study at the two-loop order that we work.  For similar reasons, and for simplicity, we have used the same value of $\xi_\sigma$ for all power spectra.  Specifically, we have chosen $\xi_\sigma =3$ as in \cite{Foreman:2015uva}.  In \appref{coefficientsapp}, we relate the coefficients in the stress tensors \eqn{stress1} and \eqn{stress2} to the coefficients appearing in the power spectra \eqn{candbtwoloop}, for the case of EdS scaling, i.e. when $D_1(a) = a / a_0$.

In the case of two gravitationally coupled fluids, the effect of large bulk flows becomes relevant. Advection leading to bulk motion in LSS is due to the large relative velocity $\left< v_{bc}^2(\vec{x})\right>$ between dark matter and baryons at recombination \cite{Tseliakhovich:2010bj}. This has carefully been analyzed in \cite{Senatore:2014via} and applied to the present two fluid-like system in \cite{Lewandowski:2014rca} to correctly reproduce the baryonic acoustic oscillations (BAO). We use the same formalism as in \cite{Lewandowski:2014rca} to perform the IR-resummation on our two-loop prediction given by \eqn{UVimprovedA}.

%
%
%

%
%
%
\section{Linear relative-velocity counterterm} \label{linrelvelsec}

%

\subsection{Generation at one loop}
\label{relvel}

In this section we discuss the importance of the term proportional to $g  \, v_I^i$ in Eqs.~\eqref{stress1} and \eqref{stress2}. Perhaps intuitively, this term can be thought of as an effective dynamical friction between CDM and baryons. We first show that the counterterm is necessary to cancel new UV divergences at one loop in perturbation theory.  To see this, it is easiest to work in the adiabatic-isocurvature basis, where the equations of motion with counterterms set to zero are given by 
\begin{align}
a \delta'_A - \Theta_A &=  \alpha_{AA} + \alpha_{II} w_b w_c \,,\label{eq:contA} \\
a \delta'_I - \Theta_I &= \alpha_{AI} + \alpha_{IA} + \alpha_{II} (w_b - w_c)  \,,\label{eq:contI}\\
a  \Theta'_A +\left(1 +  \frac{a \mathcal{H}'}{\mathcal{H}}\right)\Theta_A - \frac{3}{2} \Om \delta_A &= \beta_{AA} +\beta_{II} w_b w_c \,,  \label{eq:eulerA}\\
a \Theta'_I + \left(1 +  \frac{ a \mathcal{H}'}{\mathcal{H}}\right)\Theta_I &= 2 \beta_{AI} + \beta_{II} (w_b - w_c) \label{eq:eulerI} \,,
\end{align}
where the $a$ and $\kvec$ arguments were suppressed for clarity, and the $\alpha$ and $\beta$ functions are defined by \eqn{alphabetadefs} but with $\sigma$ and $\sigma'$ allowed to be $A$ and $I$, and $\Theta_\Upsilon \equiv - \partial_i v^i_\Upsilon  / \cH$ where $\Upsilon \in \{ A , I\}$. 
Since isocurvature modes are suppressed by $\sim 5 \times 10^{-3}$ relative to adiabatic modes {at $ z = 0$}, we can safely neglect loops that have two isocurvature modes. {We describe the full perturbative solutions to the above equations explicitly in \appref{solsec}, but most relevant for our discussion now are the linear isocurvature solutions 
\be \label{linearsolsiso}
{\delta_I^{(1)} ( a , \kvec ) =   \epsilon^2  \delta_{I_+}^{(1)} ( \kvec)  +   \epsilon^3 \frac{D_{I_-} ( a ) }{D_{I_-} ( a_0 )}  \delta_{I_-}^{(1)} ( \kvec)  \andd  \Theta_I^{(1)} ( a , \kvec ) =  \epsilon^3 \frac{a D_{I_-}' ( a ) }{D_{I_-}(a_0)}  \delta_{I_-}^{(1)} ( \kvec) \ , }
\ee
where $\epsilon^2 \equiv a_{\rm in} / a_0 \approx 5 \times 10^{-3}$ is approximately the relative size of adiabatic and isocurvature fluctuations at the current time, i.e. $\epsilon^2 \approx \delta_{I}^{(1)} (a_0) / \delta_A^{(1)} ( a_0)$. Here and elsewhere, we use ``$+$'' to denote the growing (in this case constant) mode and ``$-$'' to denote the decaying mode.}

{We also introduce the following notation for the power spectra of the adiabatic and isocurvature modes.  For the linear fields (see \eqn{linearsols}), we define the power spectra of the growing and decaying parts of the linear power spectra by
\be \label{psdefs}
\langle \delta^{(1)}_{\Upsilon_\vartheta} ( \kvec) \delta^{(1)}_{\Upsilon'_{\vartheta'}} ( \kvec ') \rangle  = ( 2 \pi)^3 \delta_D ( \kvec + \kvec ' ) P^{\Upsilon \Upsilon'}_{\vartheta \vartheta'} ( k ) \ ,
\ee
where $\Upsilon$ and $\Upsilon'$ can be either $A$ or $I$, and $\vartheta$ and $\vartheta'$ can be either $+$ or $-$.  
For the rest of the equal-time power spectra, we use the notation
\be
\langle \delta_\Upsilon ( a , \kvec ) \delta_{\Upsilon'} ( a , \kvec') \rangle = ( 2 \pi)^3 \delta_D ( \kvec + \kvec' ) P^{\Upsilon \Upsilon'} ( a , k)  \ ,
\ee
for the full power spectra, and
\begin{align}
\begin{split}
\langle \delta^{(2)}_\Upsilon ( a , \kvec ) \delta^{(2)}_{\Upsilon'} ( a , \kvec') \rangle & = ( 2 \pi)^3 \delta_D ( \kvec + \kvec' ) P^{\Upsilon \Upsilon'}_{22} ( a , k)  \ ,\\
\langle \delta^{(1)}_\Upsilon ( a , \kvec ) \delta^{(3)}_{\Upsilon'} ( a , \kvec') \rangle  & = ( 2 \pi)^3 \delta_D ( \kvec + \kvec' ) P^{\Upsilon \Upsilon'}_{13} ( a , k) \ , \\
 \langle \delta^{(3)}_\Upsilon ( a , \kvec ) \delta^{(1)}_{\Upsilon'} ( a , \kvec') \rangle & = ( 2 \pi)^3 \delta_D ( \kvec + \kvec' ) P^{\Upsilon \Upsilon'}_{31} ( a , k) \ ,
\end{split}
\end{align}
for the one-loop corrections, where again, $\Upsilon$ and $\Upsilon'$ can be either $A$ or $I$.

As shown in detail in \appref{uvlimapp}, the adiabatic power spectrum behaves in the standard way when the loop momentum becomes large, i.e.
\be
P^{AA}_{13} ( a, k ) \sim k^2 P_{++}^{A A } (k) \ , \quad \text{and} \quad P^{A A}_{22} ( a, k ) \sim k^4 \ ,
\ee
but correlations involving the isocurvature mode $\delta_I$ have a different UV behavior which is less derivatively suppressed, for example,
\be
P^{AI}_{13} ( a, k ) \sim k^0 P_{+-}^{A I} ( k ) \ , \quad \text{and} \quad P_{22}^{II} ( a, k ) \sim k^2 \ . 
\ee
Specifically,}
neglecting loops with two isocurvature modes, and using the EdS approximation {(see \eqn{epsilon3termsedsapprox})} for simplicity, we find that there is a strong UV divergence in $P^{AI}_{13}$, that is not suppressed by derivatives, namely
\begin{equation}
\label{eq:UVdiv}
P^{AI}_{13}(a,k)\rightarrow   \epsilon^3 \frac{D_{I_-}(a) D_{A_+}(a)^3}{D_{I_-}(a_0) D_{A_+}(a_0)^3} P^{AI}_{+-}(k) \left[\frac{2}{3}\int^{\Lambda} \frac{dq }{(2 \pi)^2} q^2P^{AA}_{++}(q) \right] \ ,
\end{equation}
for $q \gg k$, where $\Lambda \gg k$ is a UV-cutoff.
Since this term does not go like $k^2$ as $k \to 0$, and the counterterms available in the single-fluid EFTofLSS go like $k^2$ or higher powers of $k$ as $k \to 0$, we need an extra counterterm which is proportional to $\Theta_I$
and that does not have any derivative in front of it, exactly like terms in Eqs.~\eqref{stress1} and \eqref{stress2} that are proportional to $g \, \partial_i v^i_I$.\footnote{Equivalently, we could have chosen a counterterm proportional to $\delta'_I$, which is also proportional to the decaying mode. However, they are degenerate at first order.}

{From \eqn{eq:UVdiv}, we see that the UV divergent term is proportional to $P^{AI}_{+-}$, i.e. it involves the decaying isocurvature mode $\delta^{(1)}_{I_-}$, and not the constant mode $\delta^{(1)}_{I_+}$.  Thus, it is proportional to $\epsilon^3$ instead of $\epsilon^2$.  This was to be expected from our discussion in \secref{theory} where we showed that the new counterterm is proportional to $g \, \Theta_I$, which from \eqn{linearsolsiso} we see is proportional to the decaying isocurvature mode.     }

Let us comment on two additional aspects. First, this counterterm appears at linear order and is not suppressed by derivatives, which means that it is as important as the leading terms in the equations of motion. Importantly, we find in \secref{sec:realuniverse} and \secref{sec:sphcoll} that the finite part of this counterterm is not expected to be much larger than the other linear terms. This still non-trivially complicates perturbation theory for the isocurvature modes, and we discuss in \secref{linearevo} and \appref{pertwithlinctsec} how to treat this term consistently. {Second, the integral in \eqn{eq:UVdiv} actually \emph{diverges} for $\Lambda \rightarrow \infty$ for a realistic {power spectrum $P^{AA}_{++}$ with CDM}.  To see this, we note that on small scales, the linear power spectrum behaves like $P_{++}^{AA} ( q ) \sim (\log q)^2 / q^3$  \cite{Bardeen:1985tr, Goroff:1986ep}.  Plugging this into \eqn{eq:UVdiv}, we find
\be \label{uvdivergencereal}
P^{AI}_{13} ( k ) \sim P^{AI}_{+-} ( k ) (\log \Lambda )^3 \ ,
\ee
as $\Lambda \rightarrow \infty$.  This means that for perturbation theory to be a well-defined mathematical framework for the real universe, the counterterm proportional to $\Theta_I$ is actually \emph{necessary}.   {This is to be contrasted with the loops in the EFTofLSS with a single fluid, where the loops are always UV convergent (since there is an extra factor of $1/q^2$ inside of the loop integral for large $q$), and so the counterterms are needed not to give mathematical consistency, but simply to correct the mistake associated to the finite (but incorrect) UV dependence of the loops.}}  {In the next two subsections, we explicitly show how the counterterm proportional to $g \, v^i_I$ cancels this contribution, and we estimate the expected size of the finite part of the counterterm.}

%
%
\subsection{Estimate of the linear counterterm in perturbation theory}
\label{sec:realuniverse}

In this section, we will assume that the counterterm  $g\, v_I^i$ can be treated perturbatively.
Neglecting EFTofLSS counterterms except for the linear one, proportional to $g\, v_I^i$, we obtain the following linear equations for the $g$-dependent contributions to the isocurvature fields (from Eqs.~\eqref{eq:contI} and~\eqref{eq:eulerI}): 
\begin{align}
\label{eq:geq}
a \delta'_{I,g} - \Theta_{I,g} = 0 \andd
a \Theta'_{I,g} + \left( 1 + \frac{ a \cH'}{\cH} \right) \Theta_{I,g} =  g\, \Theta_I^{(1)}  \ ,
\end{align}
where $\delta_{I,g}$ and $\Theta_{I,g}$ are the counterterm-dependent contributions to $\delta_I$ and $\Theta_I$.  As is the case with the other EFTofLSS counterterms,
the counterterm $g$ has a cutoff dependent part $g_{\rm UV}$, that is needed to cancel the one induced by the loop in Eq.~\eqref{eq:UVdiv}, and a finite part $g_{f}$, i.e., 
\be
g(a) \equiv g_f(a) + g_{\rm UV} ( a ) \ . 
\ee
Similarly, we also break up the solution to \eqn{eq:geq} as
\be
\delta_{I,g} \equiv  \delta_I^{f} + \delta_{I}^{\rm UV}  \ , 
\ee
and analogously for $\Theta_{I,g}$.  

 We first give the UV dependent contribution and later focus on the finite part.  From \eqn{eq:UVdiv}, it is clear that the counterterm contribution needed to cancel the UV divergence (when contracted with $\delta_{A}^{(1)} ( a , \kvec')$) is
 \be \label{uvctsolution}
 \delta_{I}^{\rm UV} ( a , \kvec ) = - \epsilon^3  \frac{ D_{I_-} ( a ) D_{A_+}(a)^2}{D_{I_-} ( a_0 ) D_{A_+}(a_0)^2} \left[\frac{2}{3}\int^{\Lambda} \frac{dq }{(2 \pi)^2} q^2P^{AA}_{++}(q) \right] \delta_{I_-}^{(1)} ( \kvec )  \ .
 \ee
 Next, we find the $g_{\rm UV} ( a ) $ that is needed in \eqn{eq:geq} to produce \eqn{uvctsolution} as a solution.  Using the EdS approximation \eqn{epsilon3termsedsapprox}, we obtain 
 \be
 g_{\rm UV} ( a ) = -12 \frac{a D_{I_-}'(a) }{D_{I_-} ( a ) } \frac{D_{A_+}(a)^2}{D_{A_+} ( a_0 )^2}\left[\frac{2}{3}\int^{\Lambda} \frac{dq }{(2 \pi)^2} q^2P^{AA}_{++}(q) \right]  \ . 
 \ee
 
 \begin{figure}[t!]
\centering
\includegraphics[height=5.5cm]{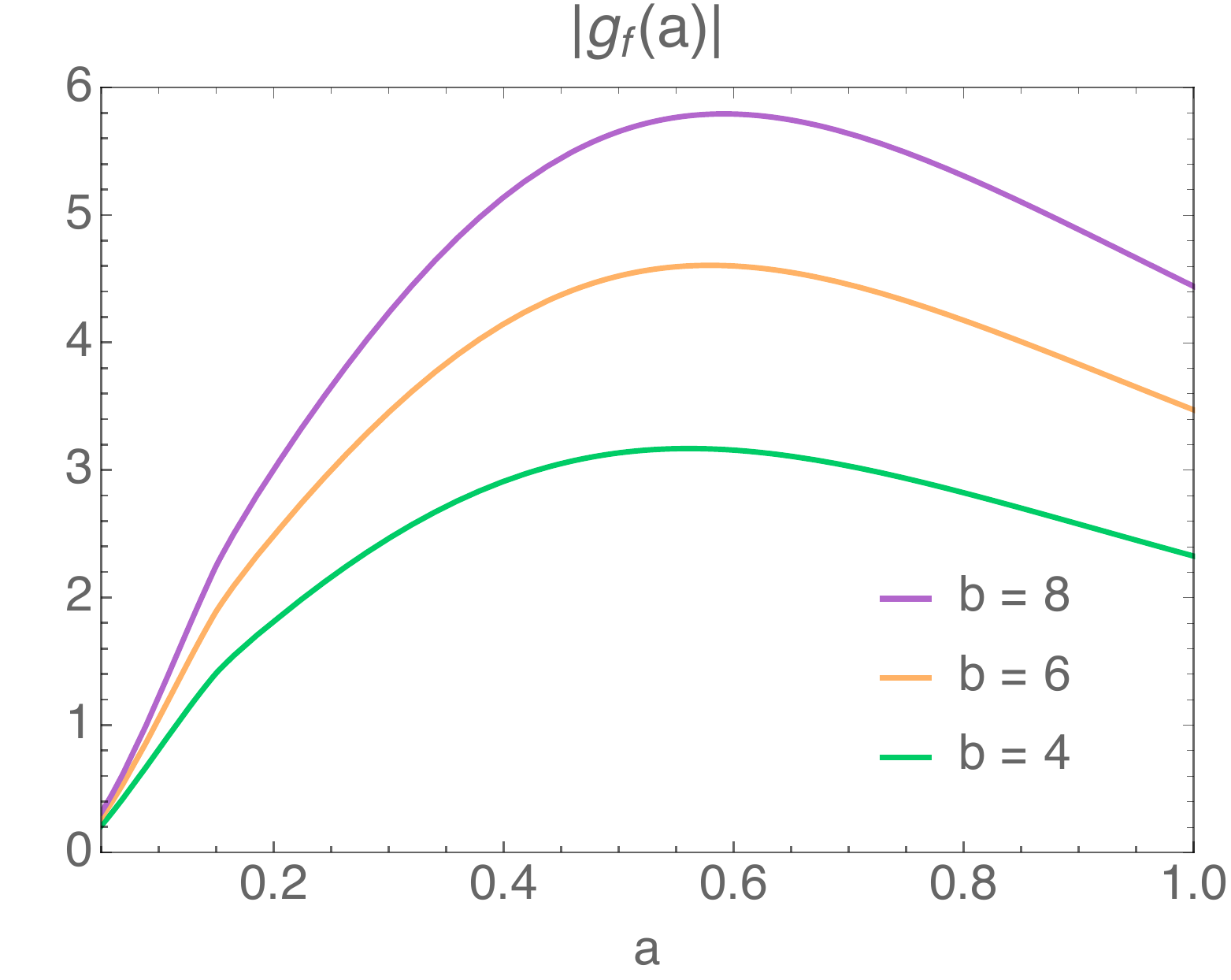} \hspace{.4in}
\includegraphics[height=5.5cm]{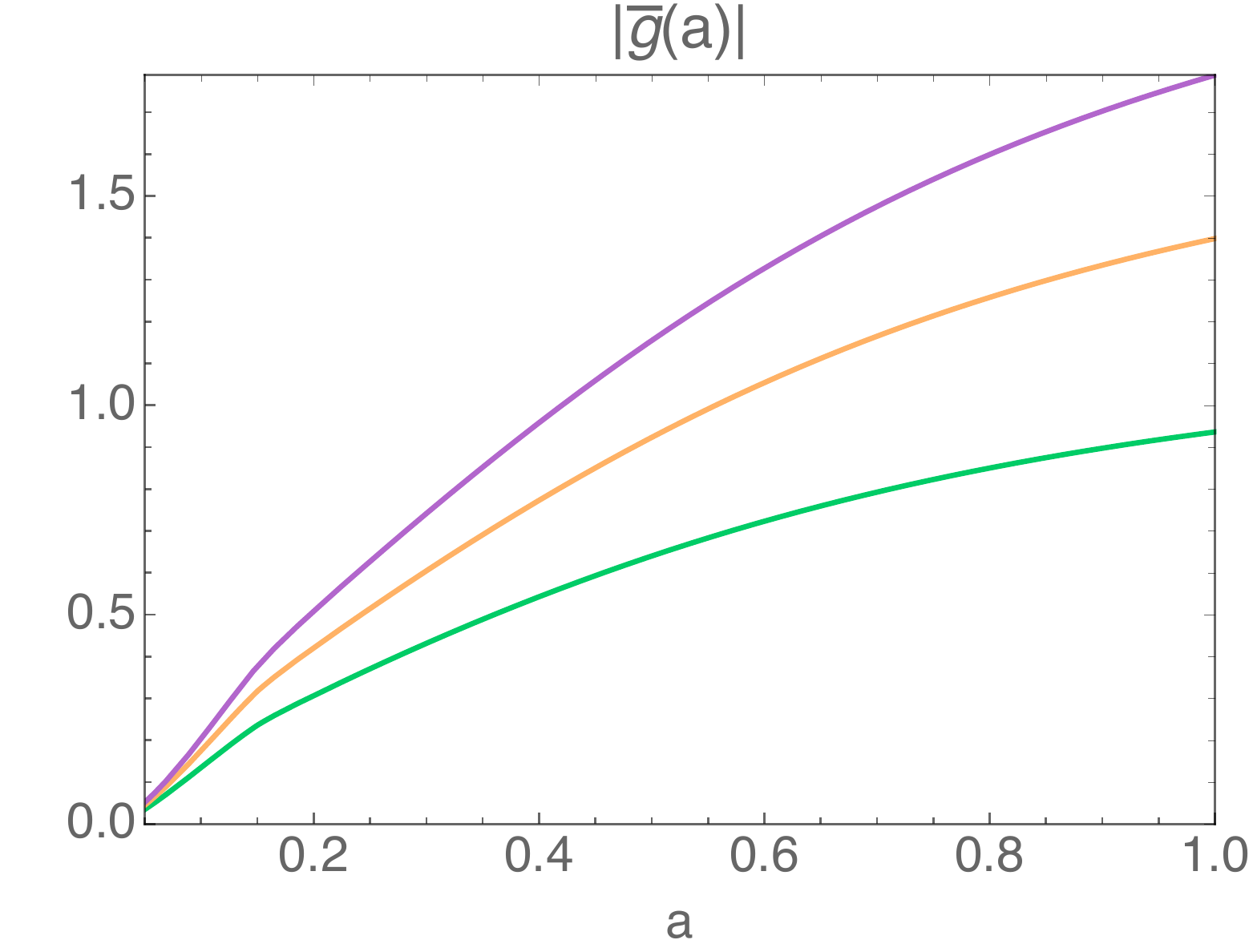}
\caption{\footnotesize Evolution of $g_f(a)$ and $\bar{g}(a)$ using \eqn{gfdef} and \eqn{bargdef}, respectively, with $b=\{4,6,8\}$ for our universe assuming WMAP3 cosmological parameters, given at the start of \secref{Comparison}.} \label{fig:g}
\end{figure}
 
 Now we can estimate the finite part $g_f ( a ) $.  To do this, we assume the same time dependence and form as the UV-dependent contribution above, with one important difference.  Since the integral over $q$ in \eqn{uvctsolution} is divergent (see \eqn{uvdivergencereal}), we assume that the finite piece receives contributions only from scales near the non-linear scale.  Thus, defining\footnote{We define $\knl ( a )$ from 
 \be
 \int_0^{k_{\rm NL}(a)} \frac{dq}{2 \pi^2} q^2 \frac{D_{A_+}(a)^2}{D_{A_+}(a_0)^2} P^{AA}_{++}(q) = 1\,.
 \ee}
 \be
 \sigma_b^2 ( a ) \equiv \frac{2}{3}\int^{b \, \knl ( a ) }_{\knl(a)} \frac{dq }{(2 \pi)^2} q^2P^{AA}_{++}(q)  \ , 
 \ee
 where $b$ is a constant greater than one, we approximate
 \be
 \delta_{I}^f ( a , \kvec) \approx \pm \epsilon^3 \frac{ D_{I_-} ( a ) D_{A_+}(a)^2}{D_{I_-} ( a_0 ) D_{A_+}(a_0)^2}  \sigma_b^2 ( a ) \delta_{I_-}^{(1)} ( \kvec ) \ ,
 \ee
 and
 \be \label{gfdef}
 g_f(a) \approx  \pm 12 \frac{a D_{I_-}'(a) }{D_{I_-} ( a ) } \frac{D_{A_+}(a)^2}{D_{A_+} ( a_0 )^2} \sigma^2_b ( a ) \ ,
 \ee
where we have written $\pm$ above because we cannot in general predict the sign, and we will consider a range of values of $b$ in this section.  Finally, defining the counterterm power spectrum $P_{(\rm ct)}^{AI}$ from 
 \be
 \langle \delta_{A}^{(1)} ( a , \kvec ) \delta_I^f ( a , \kvec ') \rangle = (2 \pi)^3 \delta_D ( \kvec + \kvec') P_{(\rm ct)}^{AI} ( a , k )  \ , 
 \ee
we have
\be \label{newcteapprox}
P_{(\rm ct)}^{AI} ( a , k ) \approx    \epsilon^3 \bar g ( a ) \frac{D_{I_-} ( a ) D_{A_+}(a)}{D_{I_-} ( a_0 ) D_{A_+}(a_0)} P^{AI}_{+-} ( k ) \ , 
\ee
where 
\be \label{bargdef}
\bar g(a) \equiv   \frac{g_f(a)}{12} \frac{D_{I_-} ( a ) }{a D_{I_-}' ( a ) } \ . 
\ee
In \figref{fig:g}, we plot the time dependence of $g(a)$ and $\bar g(a)$ for $b = \{ 4,6,8\}$, where we see that the effect on the decaying isocurvature part of the power spectrum \eqn{newcteapprox}, $\bar g(a)$, is order one around the present time.

\begin{figure}[t!]
\centering
\includegraphics[height=6cm]{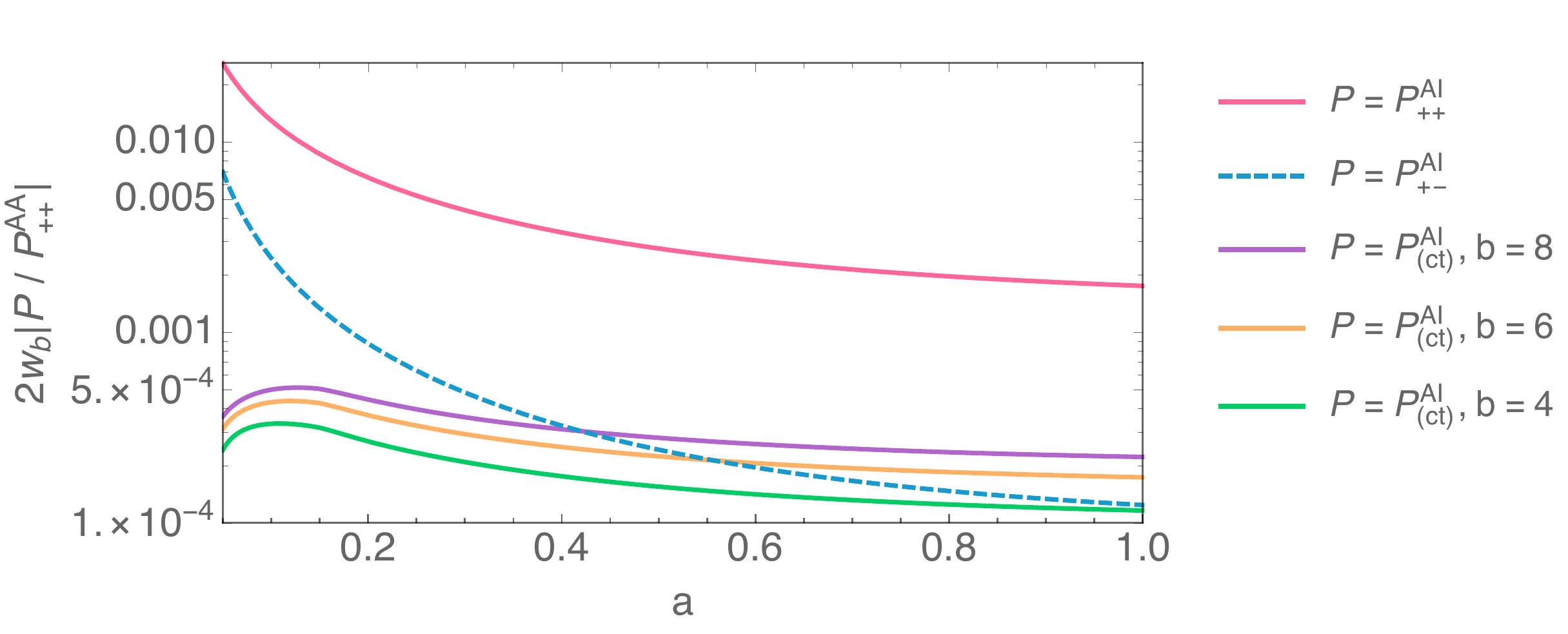} 
\caption{\footnotesize Contribution to $P^{cc}$ of the new counterterm \eqn{newcteapprox} for $b=\{ 4,6,8\}$ as well as $P^{AI}_{++}$, and $P^{AI}_{+-}$.  We see that all terms with a decaying mode remain subleading to the power spectrum containing the constant isocurvature mode.}
\label{fig:wbPAPI} 
\end{figure}

Next, we would like to compare the counterterm contribution \eqn{newcteapprox} to other contributions to the power spectrum.  For example, consider the dark-matter power spectrum  $P^{cc}$ given by
\begin{equation}
P^{cc}=P^{AA} + 2 w_b P^{AI} + w_b^2 P^{II} \ .
\end{equation}
Here, we see that the most relevant contribution from the isocurvature fluctuations is given by $2 w_b P^{AI}$. In \figref{fig:wbPAPI}, we plot the various contributions to $P^{cc}$, including from $P^{AI}_{(\rm ct)}$, $P^{AI}_{++}$, and $P^{AI}_{+-}$, as a function of $a$.  We see that $P^{AI}_{++}$ gives the dominant contribution, while $P^{AI}_{+-}$ and $P^{AI}_{(\rm ct)}$ give subdominant contributions, with the ratio  $2 w_b P^{AI}_{(\rm ct)}/P_{++}^{AA}$  always less than $5 \times 10^{-4}$ for the values of $b$ that we consider. 
Furthermore, we indeed see that, for $a \gtrsim 0.4$, the effect of the counterterm is of the same order as the contribution from the decaying isocurvature mode $P^{AI}_{+-}$, which is suppressed with respect to the leading isocurvature mode by a factor of $\sqrt{a_{\rm in}/a_0}\sim 0.07$.

\begin{figure}[t!]
\centering
\includegraphics[width=13cm]{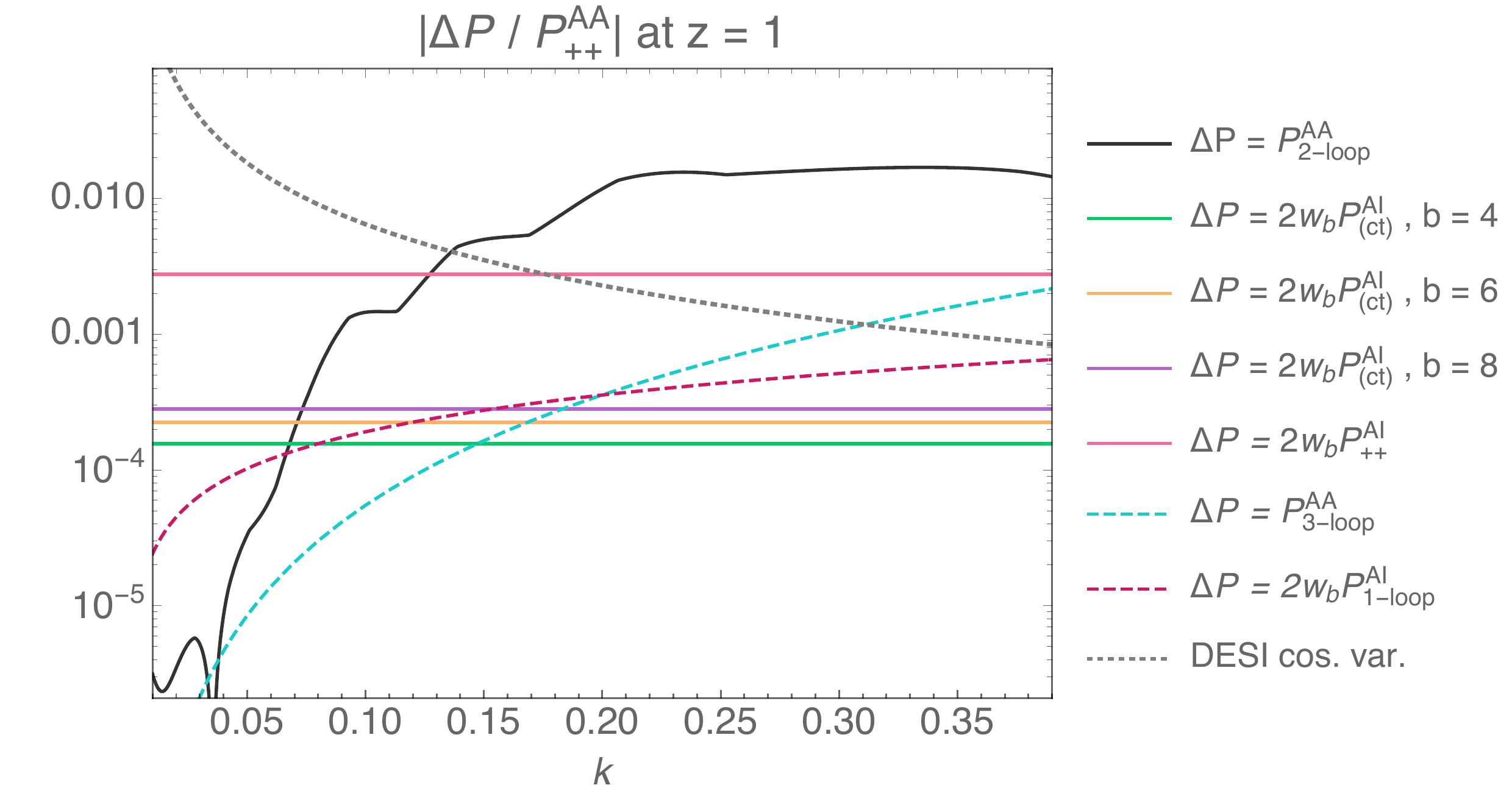}
\caption{\footnotesize  DESI cosmic variance (gray dotted), two-loop contribution to the adiabatic power spectrum used in this paper (black solid), estimate of the three-loop contribution to the adiabatic power spectrum (teal dashed), linear contribution of the adiabatic-isocurvature power spectrum (pink), estimate of the one-loop adiabatic-isocurvature power spectrum (fuchsia dashed), and contribution generated by the counterterm for $b=4$ (green), $b=6$ (orange) and $b=8$ (purple), assuming the WMAP3 cosmological parameters at $z =1$. We see that the effect of the counterterm is safely smaller than the DESI cosmic variance and also, depending on the scale, than many of the purely adiabatic loops.  Dashed curves were estimated using the power-law decomposition of the power spectrum, see \eqn{powerlaw} and below.} \label{fig:DESI2loop}
\end{figure}

It is important to compare the counterterm contribution to $P^{cc}$ with other potential contributions  in order to know which are the next largest corrections after the adiabatic one-loop term.  To do that, in \figref{fig:DESI2loop} we plot $P_{\text{2-loop}}^{AA}$ used in this paper, an estimate of the three-loop contribution $P_{\text{3-loop}}^{AA}$, and an estimate of the adiabatic-isocurvature power spectrum, along with the new counterterm contributions with $b = \{4,6,8\}$.\footnote{To estimate these contributions, we describe the terms at a desired loop order by parametrizing them as being in a scaling universe, which allows us to approximate their behavior through dimensional analysis \cite{Carrasco:2013mua}. The linear adiabatic power spectrum can thus be expressed as a piecewise power law \cite{Carrasco:2013mua, Pajer:2013jj}
\begin{equation}
P_{++, \text{pl}}^{AA}(k)= (2 \pi)^3 \left \{ \begin{array}{rcl}
\frac{1}{k_{\text{NL}}^3} \left( \frac{k}{k_{\text{NL}}}\right)^n & , & k > k_{\text{tr}} \\
\frac{1}{\tilde{k}_{\text{NL}}^3} \left( \frac{k}{\tilde{k}_{\text{NL}}}\right)^{\tilde{n}} & , & k < k_{\text{tr}}
\end{array}\right. \, ,
\label{powerlaw}
\end{equation}
where $k_{\text{tr}}$ denotes the transition scale between the two power laws. 
Here we use the parameters $k_{\text{NL}}=5.50 h\, \text{Mpc}^{-1}$, $\tilde{k}_{\text{NL}}=2.68 h\, \text{Mpc}^{-1}$, $k_{\text{tr}}=0.24 h\, \text{Mpc}^{-1}$, $n=-2.1$ and $\tilde{n}=-1.83$ derived in \cite{Lewandowski:2014rca} by fitting the power laws \eqn{powerlaw} to the linear power spectrum of non-linear simulation data with WMAP3 cosmological parameters.
For a given loop order $L$, the estimate for the corresponding adiabatic loop correction scales as $P^{AA}_{L\text{-loop}}/P_{++}^{AA} \sim {(2 \pi)^L} (k/k_{\text{NL}})^{(3+n)L } / L! $ \cite{Carrasco:2013mua}. {We also use $P^{AI}_{1\text{-loop}} ( a , k)  \sim \epsilon^2 D_1 ( a)^{-1 }P^{AA}_{1\text{-loop}} (a,k)$.  }
}   
We also plot the cosmic variance of DESI, a representative of the leading experiments in large-scale structure. 
We plot at redshift $z=1$ because the linear counterterm contribution is larger at early times and the DESI survey will measure luminous red galaxies up to $z \sim 1$ {\cite{Aghamousa:2016zmz}}.\footnote{{We note that $b = 8$ gives roughly the value of $g( z = 1)$ that we find in \secref{sec:sphcoll}.}}   

We see that when the adiabatic two-loop and three-loop contributions become larger than the cosmic variance (around $k \approx 0.14 \unitsk$ and $k \approx 0.3 \unitsk$ respectively), these terms dominate over the linear counterterm contribution.
At low wavenumbers the linear counterterm is larger than the loops, but cosmic variance dominates there.
 {All in all, we find that the linear adiabatic-isocurvature contribution $2 w_b P^{AI}_{++}$ is the most likely one to be comparable to the three-loop adiabatic term on the scales of interest, but this term can be easily included in the calculation by simply using the correct linear power spectrum from CAMB (in this case $P^{cc}_{11}$).  All of the remaining terms are much smaller than the adiabatic two-loop term, which is the order to which we work in this paper, justifying the use of the expressions given in \secref{oneloop}.}
Furthermore, the comparison with DESI cosmic variance strongly suggests that we can safely neglect the linear counterterm contribution in our perturbative expansion for all practical purposes.

In the next section, we use a UV complete model to support our estimate in this section, and confirm that the EFT coefficient $g$ indeed leads to a small effect on the power spectrum.

%
%
\subsection{Estimate of the linear counterterm with UV model}
\label{sec:sphcoll}

In this section we use a one-dimensional UV model with two species, CDM and baryons, to estimate in a more direct way the size of the new counterterm proportional to $g \, v^i_I$.  {We will do this estimate by first solving a specific UV model, given by \eqn{BARYONS:EOMgeneral} in one dimension with the counterterms set to zero (i.e. the perfect fluid model), and then smoothing the solutions.  The smoothed fields, of course, will not satisfy the same perfect fluid equations as the UV fields: this is the essence of an EFT.  Specifically, we will show that the smoothed fields, if they are smoothed on a scale large enough to diminish the effects of higher derivative counterterms, satisfy \eqn{BARYONS:EOMgeneral} but with the linear velocity counterterms (proportional to $g(a)\,   v^i_I$) in \eqn{stress1} and \eqn{stress2}.  This then allows us to measure the size of $g(a)$ in this specific UV model, and thus estimate its size in the true universe, since we expect  the two to differ by just order one.}

{To this end, we start with \eqn{BARYONS:EOMgeneral} in one dimension, neglecting the right-hand sides because we start with a perfect fluid.\footnote{A very interesting study in one dimension for the adiabatic mode has been already performed in~\cite{McQuinn:2015tva}.}}  Letting $x$ be the one-dimensional spatial coordinate, for each species $\sigma =c,b$ we define the velocity as
\begin{equation}
v_\sigma(a,x) \equiv v_{\sigma}^x ( a , x) \ ,
\end{equation}
which gives the equations of motion,
\begin{align}
\begin{split} \label{sphericaleom}
&a \mathcal{H} \frac{\partial \delta_\sigma}{\partial a} + \frac{\partial}{\partial x }  \left( ( 1 + \delta_\sigma ) v_\sigma \right) = 0  \ ,  \\
&   a \cH \frac{\partial}{\partial a} \frac{\partial v_\sigma}{\partial x}  + \cH \frac{\partial v_\sigma}{\partial x} + \frac{3}{2} \Om  \cH^2 \delta_A  + \frac{\partial }{ \partial x} \left( v_\sigma \frac{\partial v_\sigma}{\partial x} \right)  = 0 \ ,
\end{split}
\end{align}
where $\delta_\sigma$ and  $v_\sigma$ are functions of $a$ and $x$.

\begin{figure}[t!]
\centering
\includegraphics[height=6cm]{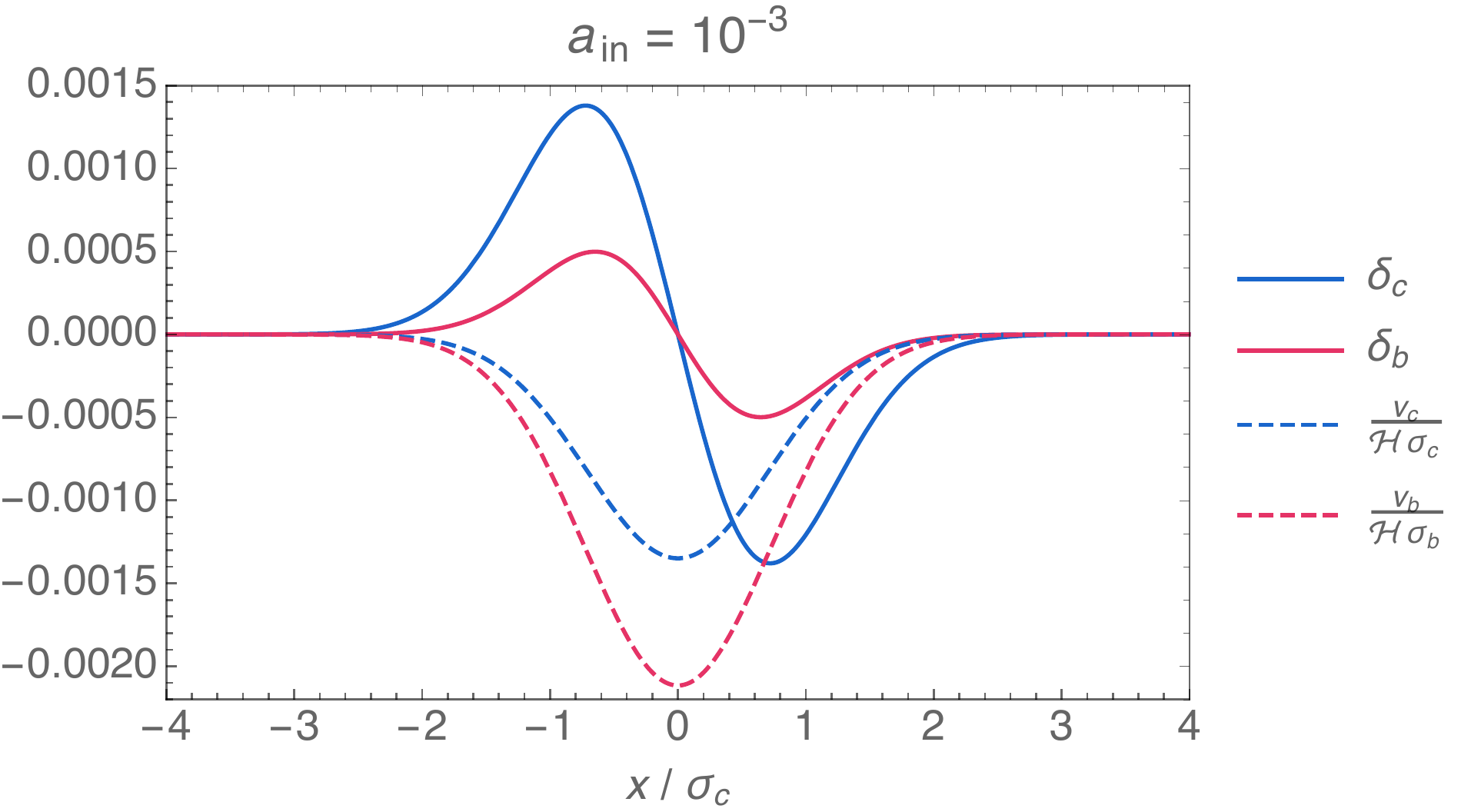}
\caption{\footnotesize Initial CDM and baryon overdensities  and velocities at $a_{\rm in}= 10^{-3}$.  The parameters $\sigma_c$ and $\sigma_b$ are the variances of the CDM and baryon initial gaussian distributions, respectively, in \eqn{sphericalic}.} \label{fig:deltain}
\end{figure}

{Next, we set up an initial configuration. For the initial overdensities, for each species we choose an antisymmetric sum of two gaussians }
\begin{align}
\begin{split} \label{sphericalic}
& \delta_c ( a_{\rm in} , x ) = ( \delta_{A_+ , \text{in}} + w_b \delta_{I_-,\text{in}})\left[ \exp \left\{ - \left( \frac{x}{\sigma_c} + \frac{1}{4} \right)^2 \right\} - \exp \left\{ - \left( \frac{x}{\sigma_c} - \frac{1}{4} \right)^2 \right\}    \right] \ ,  \\
& \delta_b ( a_{\rm in} , x ) = ( \delta_{A_+ , \text{in}} - w_c \delta_{I_-,\text{in}})\left[ \exp \left\{ - \left( \frac{x}{\sigma_b} + \frac{2}{9} \right)^2 \right\} - \exp \left\{ - \left( \frac{x}{\sigma_b} - \frac{2}{9} \right)^2 \right\}    \right] \ ,
\end{split}
\end{align}
where $\sigma_b = 9 \sigma_c / 10$, $\delta_{A_+,\text{in}} = 3 a_{\rm in}$,  $\delta_{I_-,\text{in}} = 2 a_{\rm in}$, $\delta_{A_-,\text{in}} = \delta_{I_+,\text{in}} = 0$, $a_{\rm in} = 10^{-3}$, and we use the subscript ``${\rm in}$'' to label the initial values of the various quantities.    Setting $\delta_{I_+,\text{in}} =0$ ensures that we are focusing on the decaying isocurvature mode, which is relevant for the relative-velocity counterterm.   For the initial velocities, we use the linear part of the first equation in \eqn{sphericaleom} and integrate in $x$, assuming linear time evolution for $\delta_c$ and $\delta_b$.  The fact that the integral in $x$ of the overdensities vanishes implies that the velocity goes to zero at the boundaries.    Furthermore, we take $w_c = 0.824$ and $w_b = 0.176 $, in accordance with the cosmology used in \secref{Comparison}.  In \figref{fig:deltain}, we plot the initial configuration.

\begin{figure}[t!]
\centering
\includegraphics[height=5.5cm]{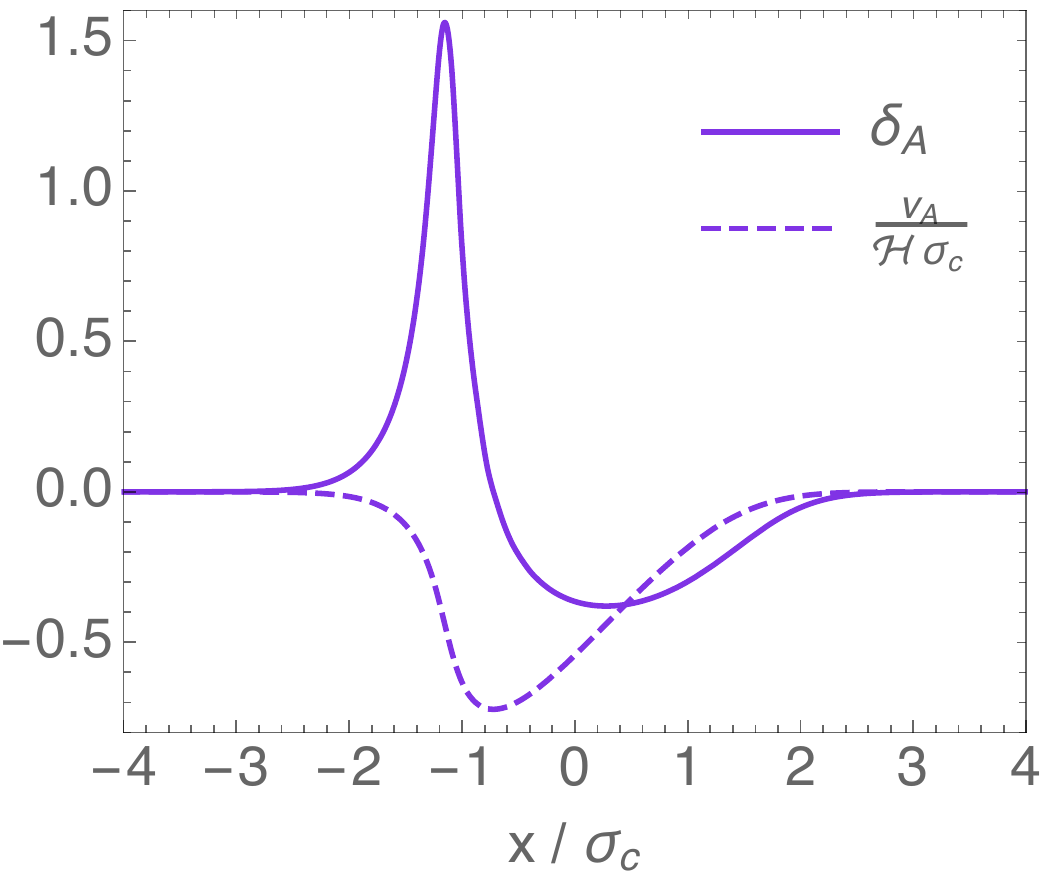} \hspace{.4in}
\includegraphics[height=5.5cm]{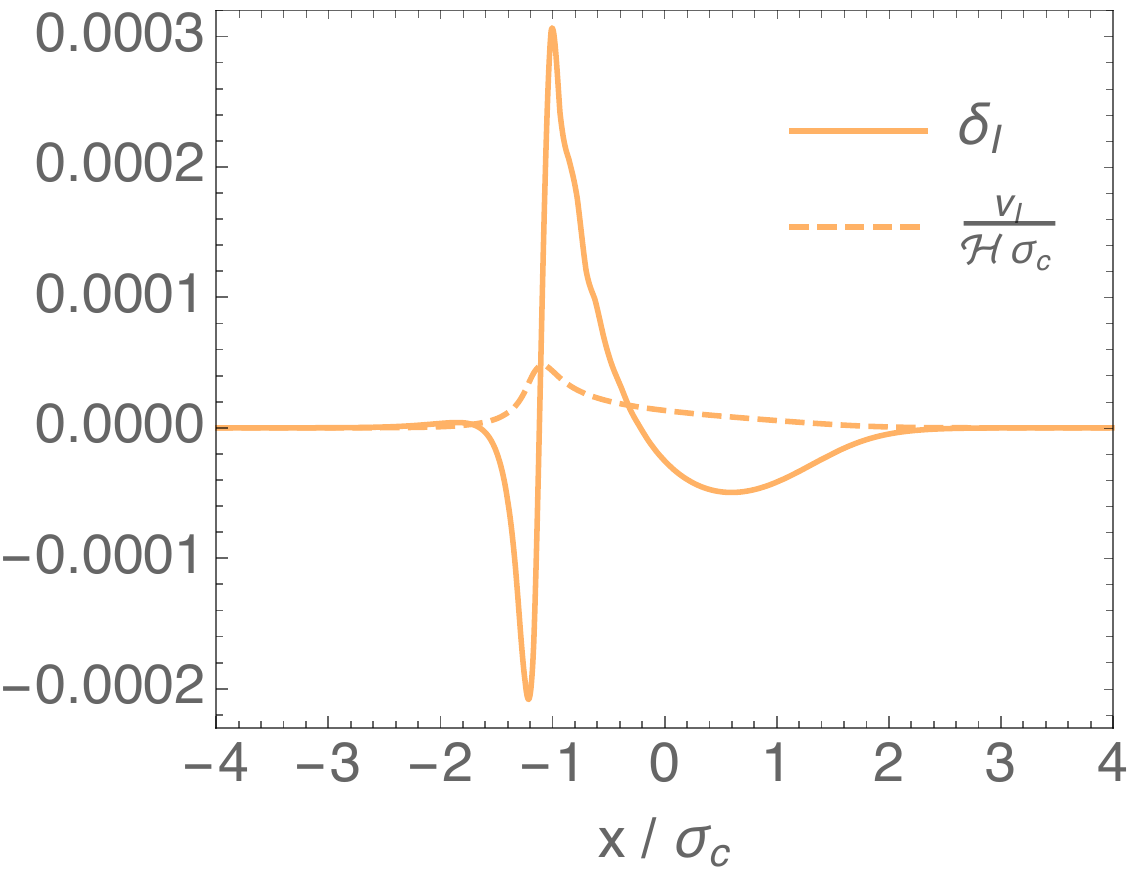} \\
\vspace{.1in}
\includegraphics[height=5.5cm]{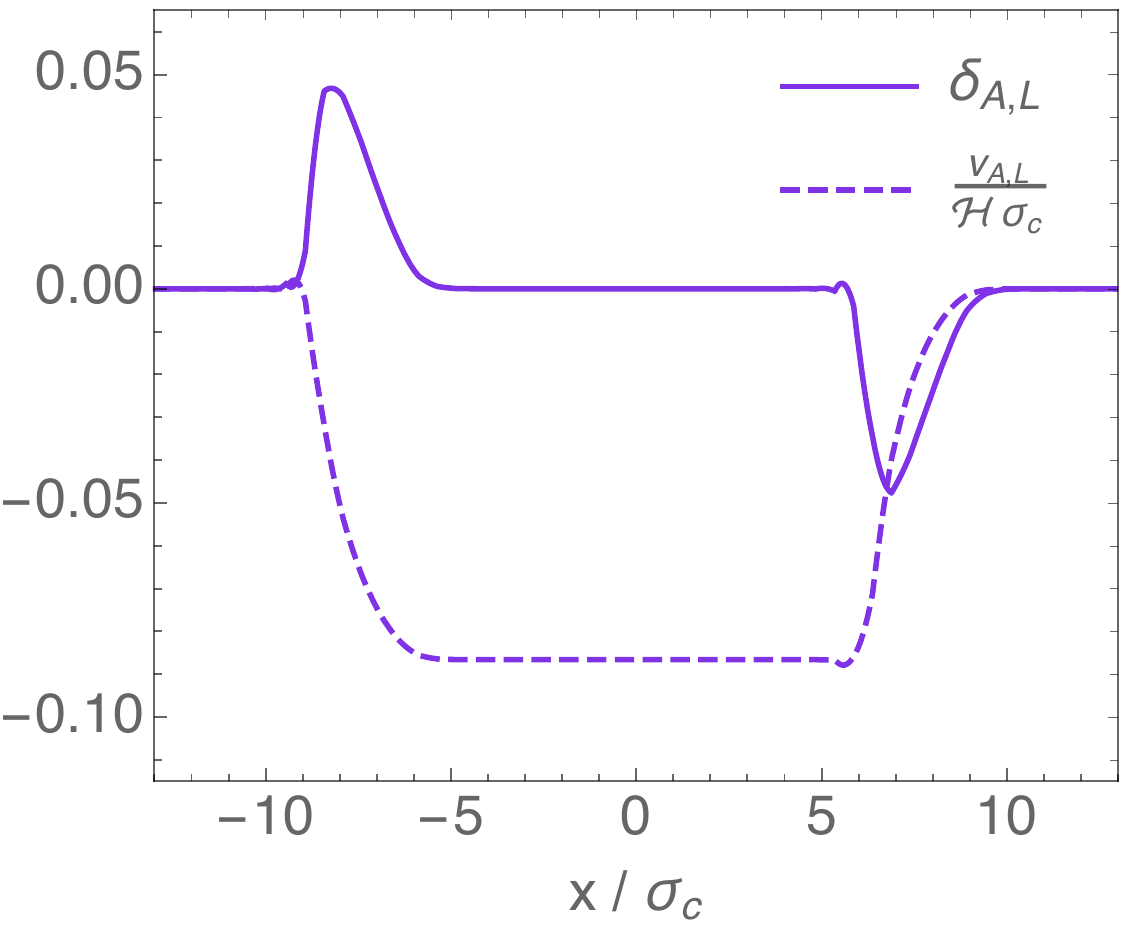} \hspace{.4in}
\includegraphics[height=5.5cm]{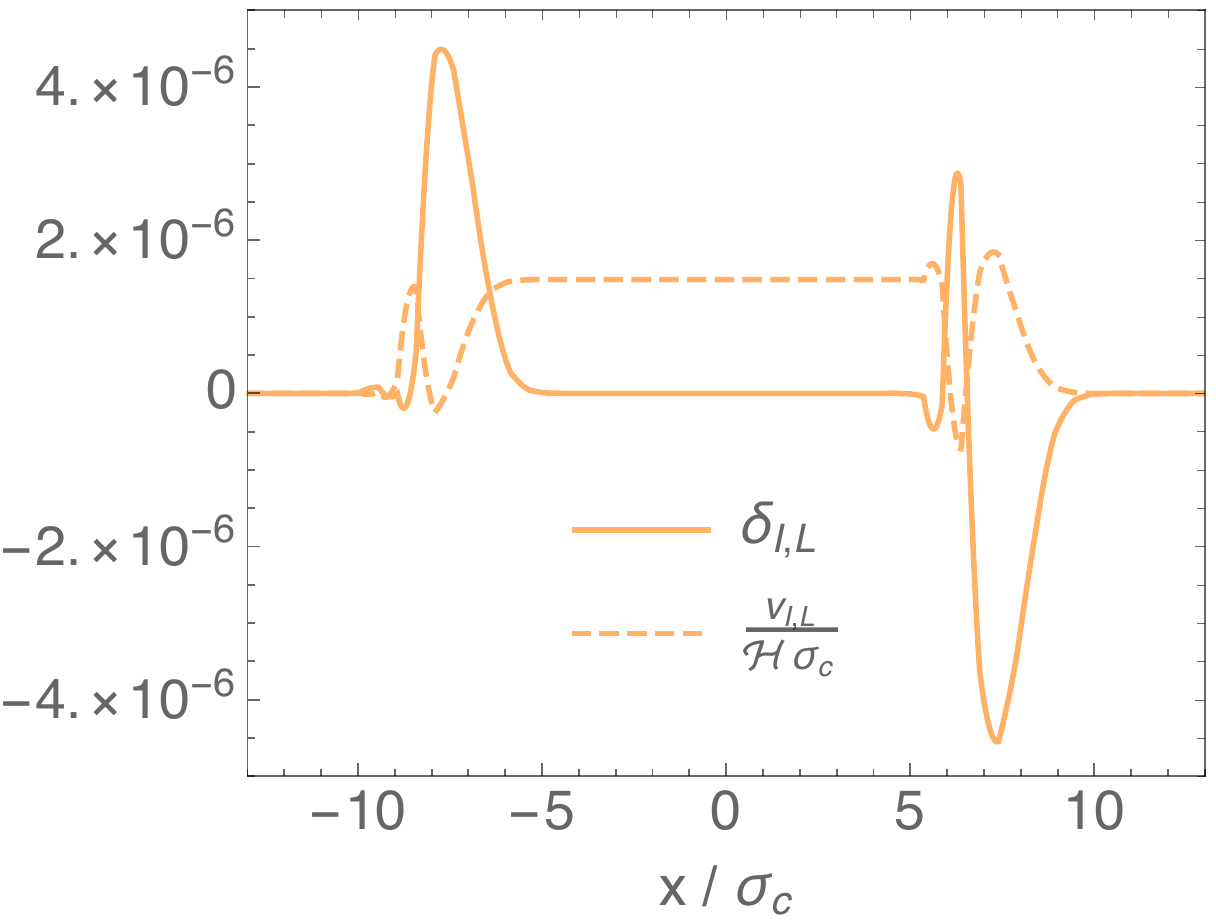}
\caption{\footnotesize  In the top two plots, profiles of the non-linear solutions to \eqn{sphericaleom} at $a = 0.5$, where the system is clearly in the non-linear regime.  We can see matter moving into the overdense region, increasing the density there.  In the bottom two plots, solutions smoothed with $L = 15 \sigma_c$.  } \label{nlsolutions}
\end{figure}

We then numerically solve the equations of motion \eqn{sphericaleom} with the initial conditions \eqn{sphericalic}, assuming an EdS universe for simplicity.   Once we have the UV solutions, we then smooth the fields with a normalized top-hat $W_L$ defined by $W_L ( x ) = L^{-1}$ for $| x | \leq L/2$ and $W_L(x) = 0$ for $|x | > L/2$.  As always, we smooth the overdensities $\delta_\sigma$ and momentum densities $\pi_\sigma \equiv \rho_\sigma v_\sigma$ directly, 
\begin{align}
\delta_{\sigma,L} (a , x ) \equiv \int d x' W_L ( x - x') \delta_\sigma ( a, x') \andd \pi_{\sigma,L} (a , x ) \equiv \int d x' W_L ( x - x') \pi_\sigma ( a, x') \ ,
\end{align}
then define the smoothed velocity as the ratio of the smoothed momentum and smoothed density:
\be \label{longvel}
v_{\sigma,L}(a , x) \equiv \frac{\pi_{\sigma, L} ( a , x)}{\bar \rho_\sigma(a) ( 1 + \delta_{\sigma,L} ( a , x) )} \  ,
\ee
{and finally define $\delta_{A,L}$, $\delta_{I,L}$, $v_{A,L}$ and $v_{I,L}$ analogously to \eqn{basischange} and \eqn{aiveldefs}.\footnote{Recall that \eqn{longvel} is \emph{not} the same as smoothing the UV velocities directly, see \eqn{longwlvel}.  In this sense, one can think of $v_{\sigma,L}$ in \eqn{longvel} as an auxiliary variable which makes the equations of motion for the long-wavelength fields simple.  If one were to smooth the UV velocities directly (which in any case we never do in this section), then that field would be related to $v_{\sigma,L}$ by a series of counterterms which renormalize the velocity. }}
We should choose $L$ large enough so that the smoothed fields satisfy the linear equations ({see \figref{nlcheckplot}}), and in particular in this work we consider the two smoothing scales $L = L_1 \equiv 12 \sigma_c$ and $L= L_2 \equiv 15 \sigma_c$.  In \figref{nlsolutions}, we show examples of fully non-linear and smoothed adiabatic and isocurvature modes at $a = 0.5$. We see that even though $\delta_A $ reaches $\approx 1.5$, signaling that we are in the non-linear regime, the smoothed fields remain small and perturbative.

\begin{figure}[t!]
\centering
\includegraphics[height=5.8cm]{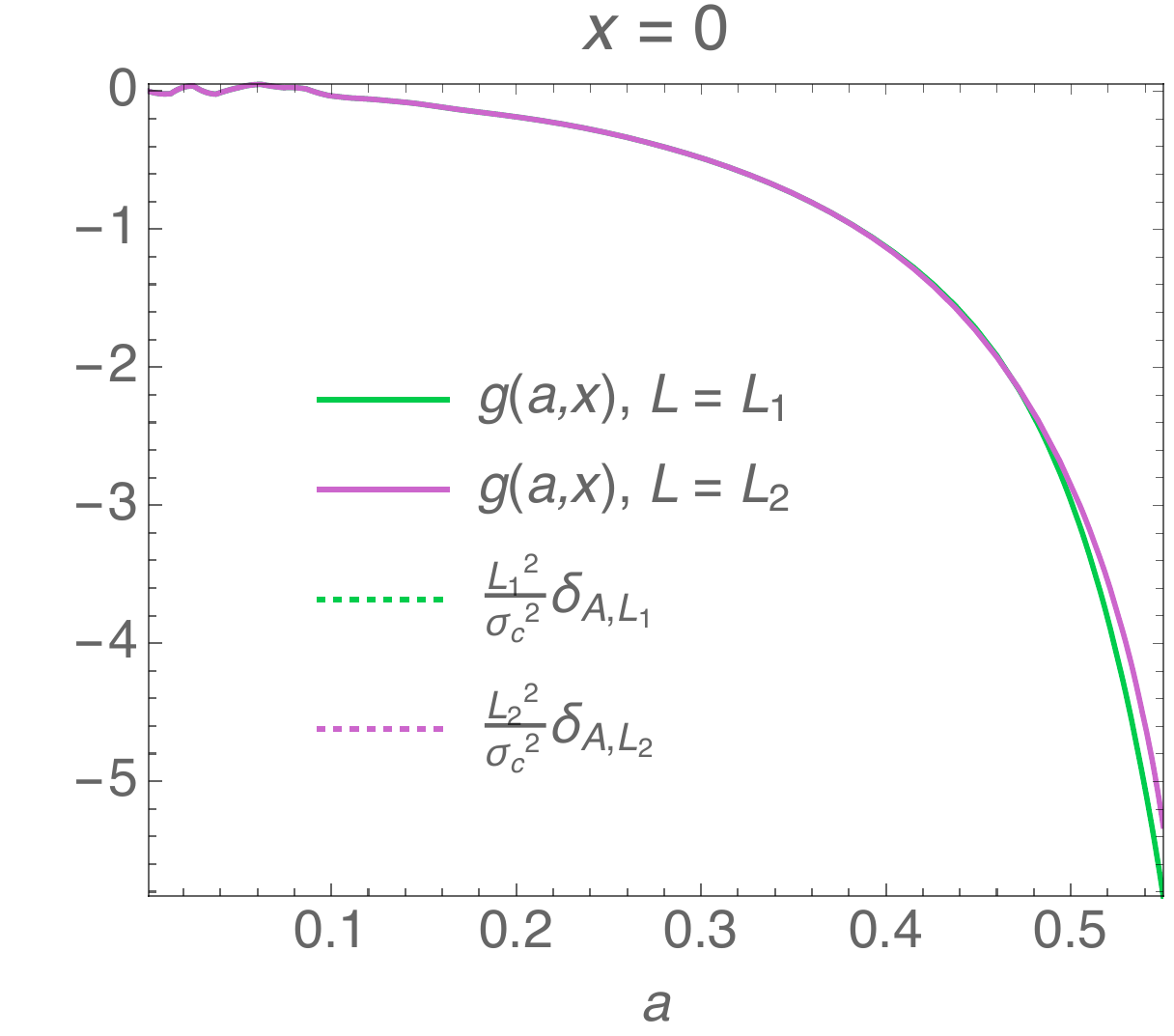} \hspace{.4in}
\includegraphics[height=5.8cm]{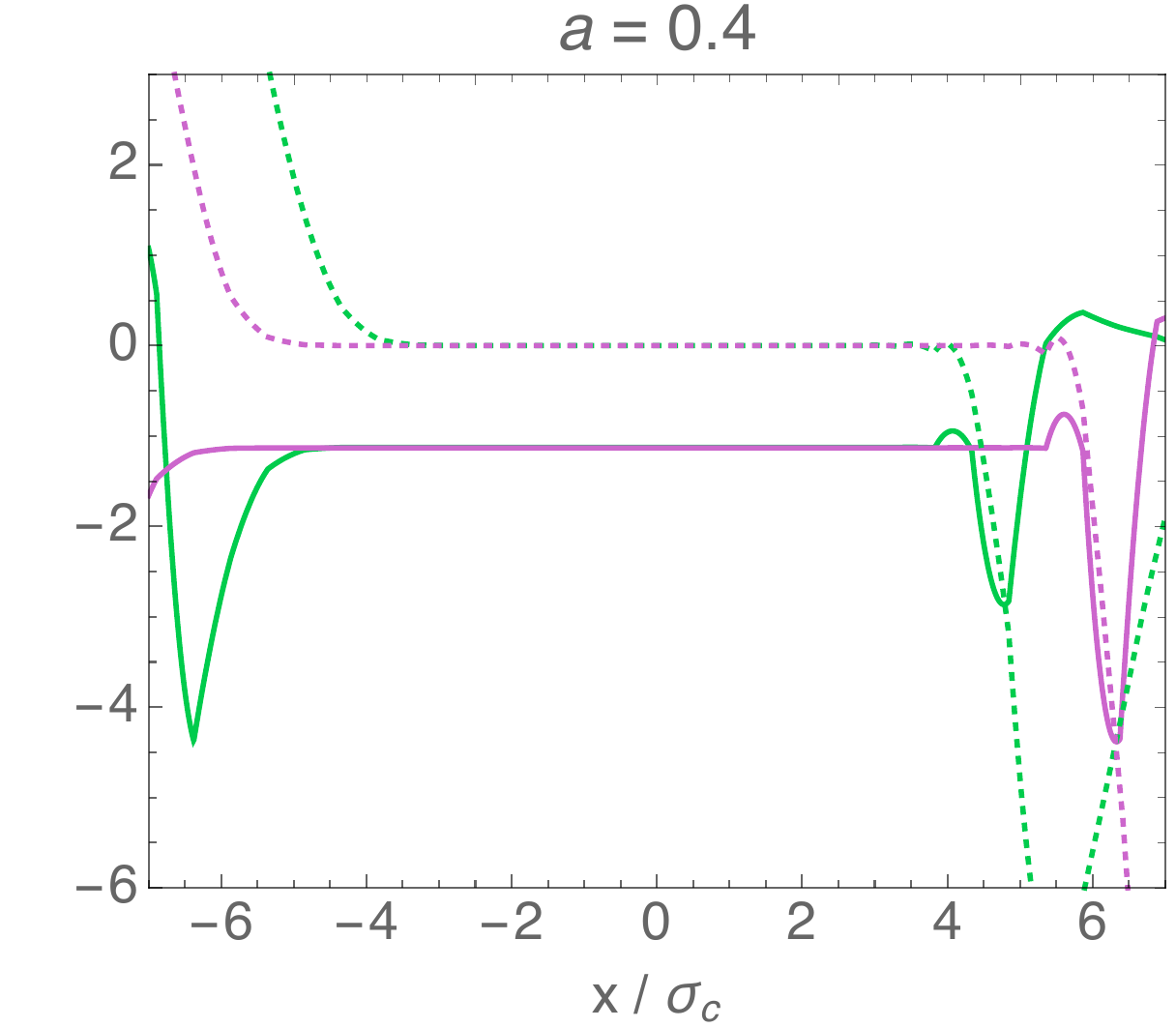} \\
\caption{\footnotesize Left: Evolution of $g$ calculated from the smoothed quantities at the origin using \eqn{gdef}. We see that $g$ remains $\sim \mathcal{O}(1) - \mathcal{O}(5)$ during the non-linear evolution.  Right: Spatial profile of $g$ at $a=0.4$. The profile remains constant within the smoothing scales around the origin, and starts to deviate when higher derivative terms (green and purple dotted lines) become large, as expected from the EFTofLSS.  In this plot and elsewhere, we use $L_1 = 12 \sigma_c$ and $L_2 = 15 \sigma_c$.} \label{plotsofg}
\end{figure}

\begin{figure}[t!]
\centering
\includegraphics[height=5.8cm]{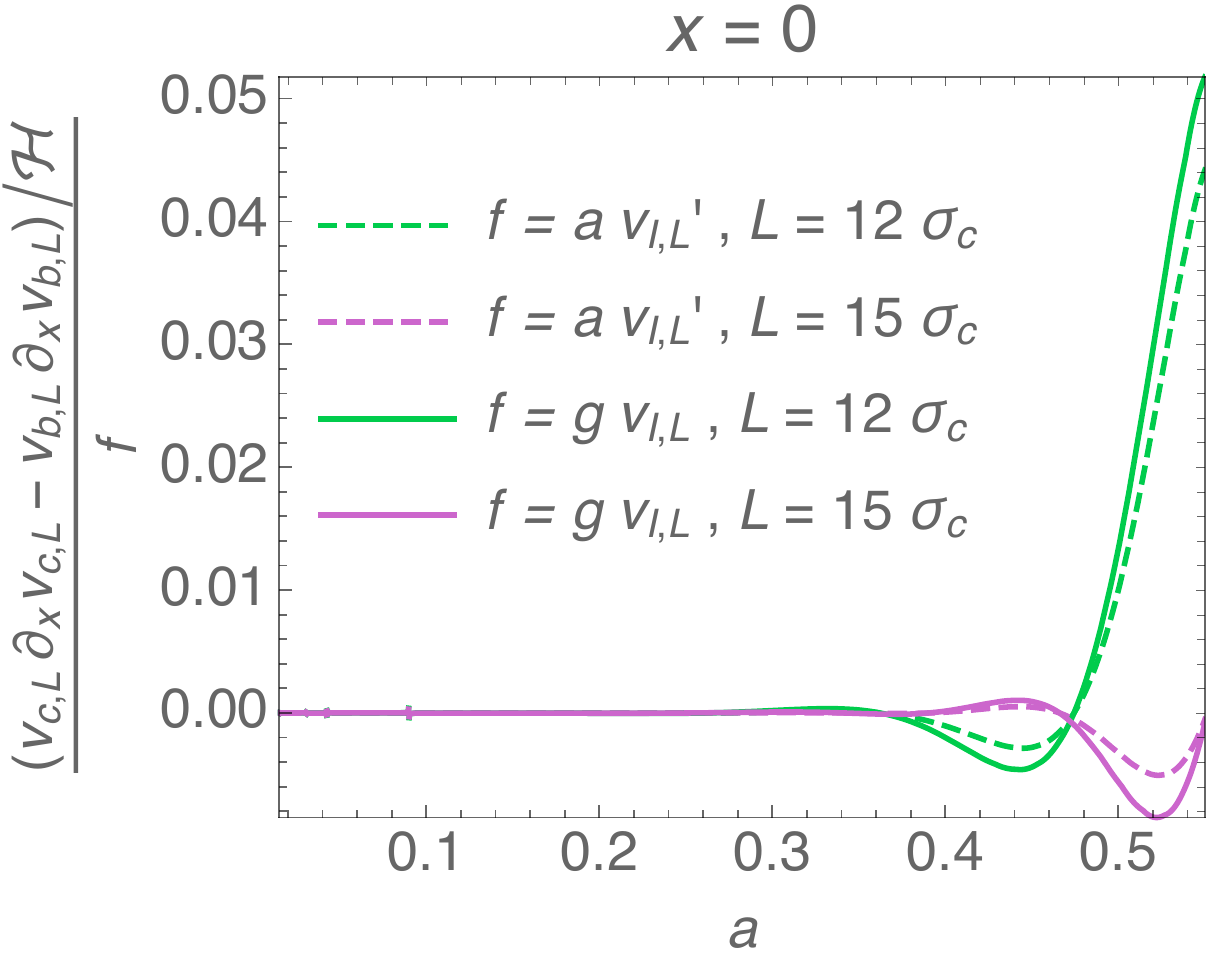} \hspace{.2in}
\includegraphics[height=5.8cm]{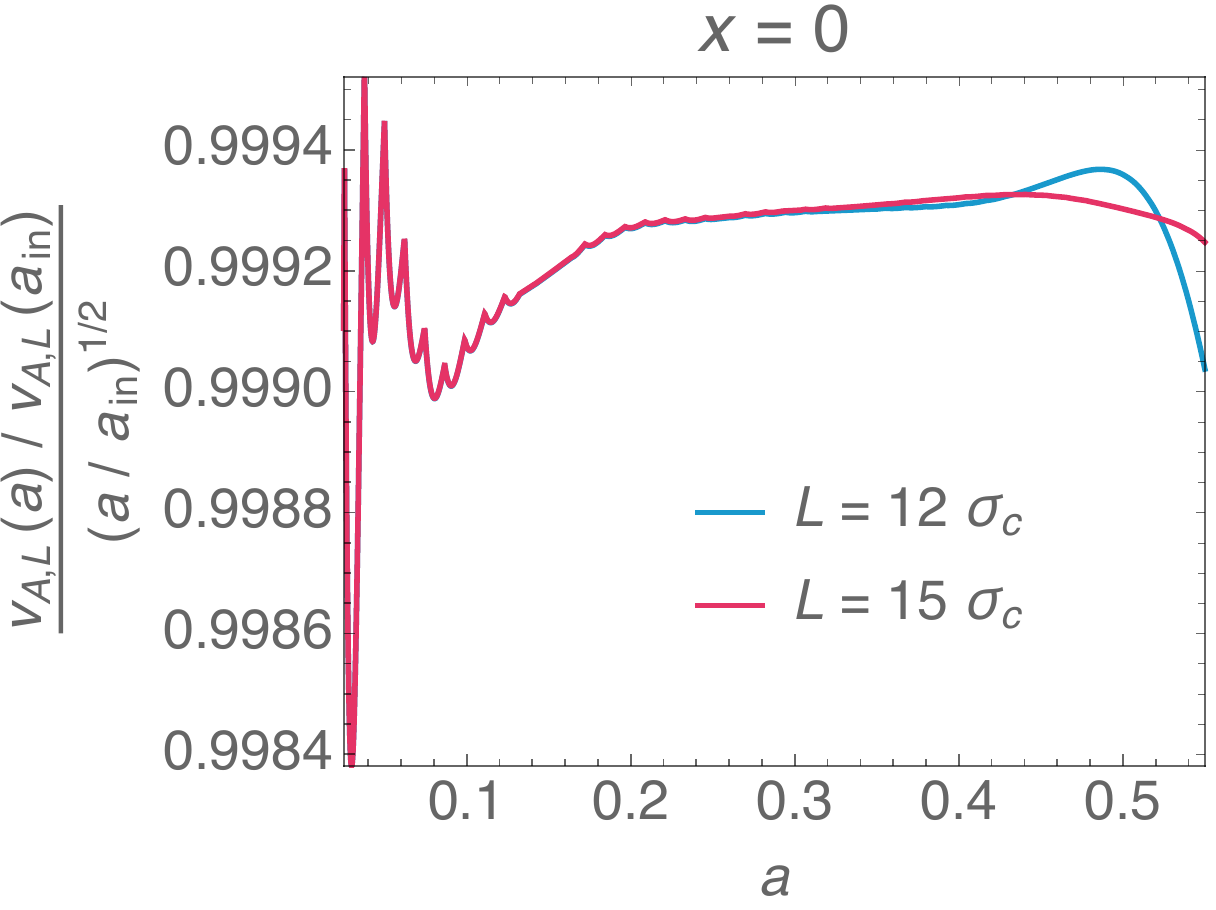} \\
\caption{\footnotesize Left:  The non-linear term that would appear in \eqn{gdef} compared to some linear terms there.  We see that with the smoothing scales that we have used, and the values of $g$ that we have found, the assumption of using the linear equation \eqn{gdef} to determine $g$ is justified.  We also see that using a larger smoothing scale suppresses the non-linear terms more, as expected.  Right:  The time evolution of $v_{A,L}$, which is shown to be within $10^{-4}$ of the linear evolution.  This again confirms that we are indeed in the linear regime after smoothing.} \label{nlcheckplot}
\end{figure}

Next, in order to estimate the size of the relative-velocity counterterm, we determine the equations satisfied by the smoothed fields.  To do that, we focus on the velocity equation for the isocurvature mode, but include the counterterm $g \, v_{I,L}$, 
\be \label{gdef}
a v_{I,L}' ( a , x) + v_{I,L}(a,x) = g(a,x) v_{I,L}(a,x ) + \dots \ ,  
\ee
where the ellipsis $\dots$ represents non-linear terms and higher-derivative terms in the effective stress tensor $\partial_j \tau^{ij}_\sigma$ and in the effective force $\gamma^i$. 
Now, as long as the higher derivative and second order terms are negligible, we can plug the smoothed quantities into \eqn{gdef} and simply solve for $g$.  In \figref{plotsofg}, we give our results, which show that the size of $g(a)$ goes from zero at early times to $\approx -6$ at  $a \simeq 0.55$, which is the latest time where there are significant observations.  {In the EFTofLSS, $g$ is only a function of time (see \secref{forcesandsts}), but if we solve for $g$ with \eqn{gdef}, it is generically a function of $a$ and $x$, as translation are broken in this example.  However, from \figref{plotsofg}, we see that $g$ is independent of $x$ within the smoothing scales around the origin, and that $g$ only starts getting an $x$-dependence when higher derivative terms become important.}  This is exactly what one expects from an EFT.  

Now that we have found $g$, in \figref{nlcheckplot} we confirm that the smoothing that we have performed is such that the non-linear terms that would appear in \eqn{gdef} are in fact negligible.  In \figref{nlcheckplot}, we also show that the time evolution of $v_{A,L}$ is within $10^{-4}$ of the linear evolution, again confirming that our smoothing has put us in the linear regime.  

\begin{figure}[t!]
\centering
\includegraphics[height=5.8cm]{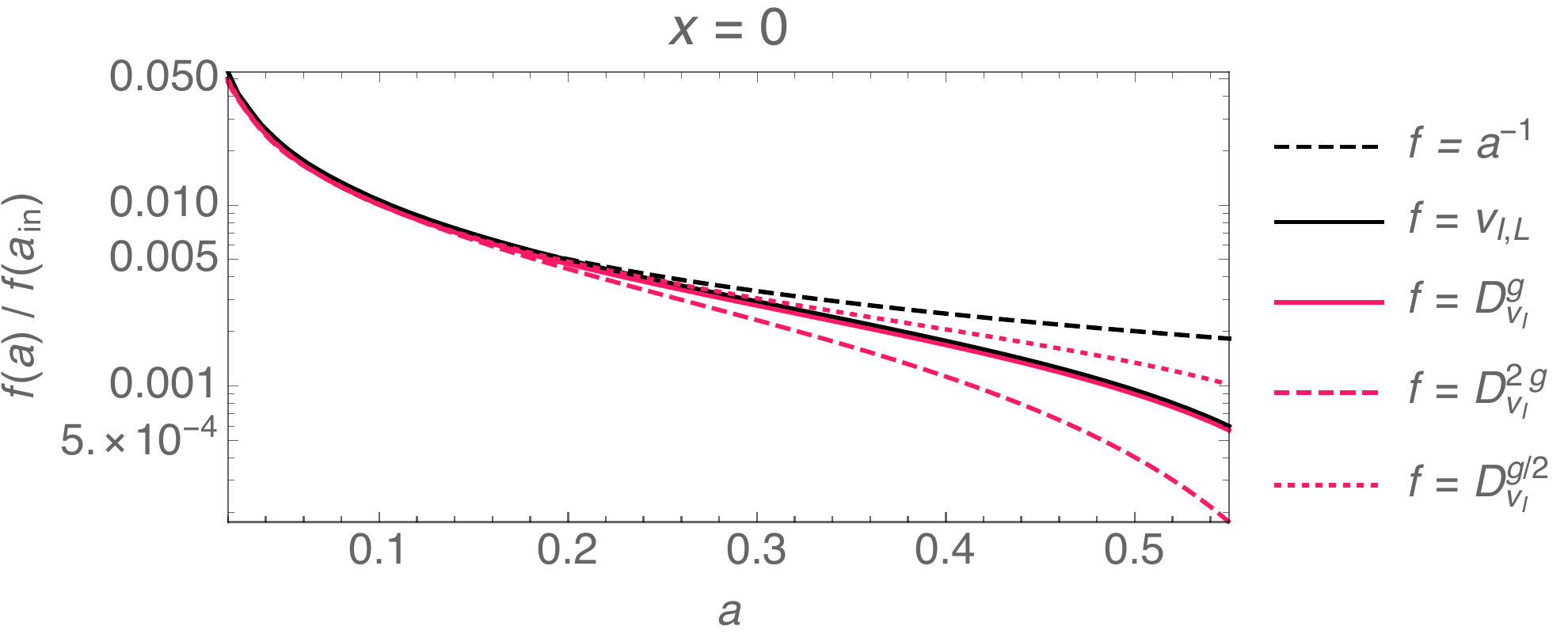} 
\caption{\footnotesize  Various time dependencies for the isocurvature velocity compared to the linear evolution (dashed black).  We see that the smoothed UV solution (solid black) is significantly different from the linear solution, showing that a counterterm is necessary to reproduce the correct time dependence of the long modes.  Using the $g(a)$ that we found from the smoothed solution, we directly reproduce the time dependence of the long-wavelength isocurvature velocity using \eqn{vitimedepsol} (pink solid).  In dashed and dotted pink, we see the effect of using a different $g(a)$ in the solution \eqn{vitimedepsol}.    } \label{velotimdep}
\end{figure}

As a consistency check, consider the linear equation \eqn{gdef} at $x = 0$.  Assuming the $g(a)$ shown in \figref{plotsofg}, we can then solve for the time dependence of $v_{I,L}$ directly.  Defining $v_{I,L}(a , 0) = D^g_{v_I} ( a )$, we have 
\be \label{checkga}
a D_{v_I}^g{}'(a) + (1 - g(a) ) D^g_{v_I}(a) = 0 \ ,
\ee
where $g(a)$ is given in \figref{plotsofg}.   The solution to \eqn{checkga} is 
\be \label{vitimedepsol}
D^g_{v_I} ( a ) = a^{-1} \exp \left\{ \int^a \frac{d a'}{a'} g(a') \right\} \ . 
\ee
In \figref{velotimdep}, we show the time dependence of the isocurvature velocity (for $L = L_2$).  The first thing to notice is that the time dependence of $v_{I,L}$ found by smoothing the UV model is much different, by about $33\%$ at $a = 0.55$, than the linear solution $\propto a^{-1}$.  This should to be compared to the time dependence of $v_{A,L}$ shown in \figref{nlcheckplot}, which is equal to the linear solution to within $10^{-4}$.  This clearly shows that the linear counterterm is only needed in the isocurvature equations, and not the adiabatic ones, as the symmetry arguments of \secref{coupledsec} showed.  Given that the change in the solutions is approximately $33\%$, this means that the effect of counterterm in $g$ can be taken into account perturbatively, as explained in \secref{linearevo}. Alternatively, as we show in \secref{linearevo} and \appref{pertwithlinctsec}, one can construct a formalism that does not rely on treating $g$ perturbatively.  Finally, we see that the solution computed with \eqn{vitimedepsol} matches the smoothed velocity from our UV model, and that changing the $g(a)$ that appears in \eqn{vitimedepsol} has a significant impact on the resulting solution.

{Our two estimates for $g$ are obtained in the case of two pressureless fluids, and they suggest that, even though we detect a non-vanishing $g$ in our numerical run, the effect of the counterterm in $g\, v_I$ is a small or ${\cal{O}}(1)$ effect on the decaying mode. This finding can be verified by inspection of the results obtained using numerical $N$-body codes that simulate two gravitationally-interacting sets of particles with different initial conditions. Several numerical challenges had to be overcome to achieve a satisfactory numerical convergence, see~\cite{Angulo:2013qp,Valkenburg:2016xek,vanDaalen:2019pst,Bird:2020kwe,Hahn:2020lvr}. In these papers, the results of the numerical runs are compared against linear theory at long distances. In this context, clearly, linear theory means our linearized equations setting $g=0$. What is found in these results (see for example Fig. 8~of~\cite{Hahn:2020lvr} or Fig.~2 of~\cite{Bird:2020kwe}) is that the disagreement between linear theory and simulations is much smaller than the constant isocurvature mode, indicating indeed that the decaying isocurvature mode that results from the simulation is still decaying and much smaller than the constant mode. This is consistent with our estimates.

Next, we should consider the fact that baryons are not a set of gravitationally interacting particles but, after star formation begins, they are affected by baryonic effects. Our estimates do not cover this possibility, and to estimate the contribution to the EFT coefficient $g$ from baryonic effects we rely entirely on hydrodynamical simulations. Such simulations with two fluids and accounting for the initial isocurvature mode have been performed in, for example,~\cite{vanDaalen:2019pst,Hahn:2020lvr}. By looking for example at Fig.~2 of~\cite{vanDaalen:2019pst}, we see that the change in the power spectrum at low wavenumbers by the onset of start formation physics is so small that one can bound the size of the decaying isocurvature mode to be at least about an order of magnitude smaller than the constant isocurvature mode. Similar considerations can be obtained by looking at Fig.~9 of~\cite{Hahn:2020lvr}. We therefore conclude that our estimates of $g$ are affected at most of by ${\cal{O}}(1)$ by baryonic effects. This is as expected on theoretical grounds. 
}

%
%
\subsection{Perturbation theory with relative-velocity counterterm} \label{linearevo}

To start the study of perturbation theory  in the presence of the new relative-velocity counterterm $g \, \Theta_I$ that is allowed in the equations of motion, we consider linear evolution.  Contrary to what we did in \secref{sec:realuniverse} for the purposes of estimating in the previous sections, here we take the consistent approach of treating this term not as an interaction. Because it does not enter the adiabatic equations, we focus on the isocurvature equations in this section. 
 In the presence of this new term, allowing for a generic non-local in time counterterm, the linear isocurvature equation becomes
\be \label{newctlineareq}
a^2 \delta^{(1)}_I {}'' ( a, \kvec ) + \left( 2 + \frac{ a \cH'(a)}{\cH(a)} \right) a \delta^{(1)}_I {}' ( a , \kvec)  = \int^a d a_1 g ( a , a_1 ) \,  a_1 \delta^{(1)}_I{}' ( a_1, \kvec) \ .
\ee
{At this point, it is helpful to define the terms in the perturbative expansion of the isocurvature modes proportional to $\epsilon^2$ and $\epsilon^3$ as
\begin{align}
\begin{split}
&\delta^{(n)}_{I} ( a , \kvec)  =  \epsilon^2 \delta^{(n)}_{I_+} ( a , \kvec ) + \epsilon^3 \delta^{(n)}_{I_-} ( a , \kvec ) + \dots \ , \\ 
&  \Theta^{(n)}_{I} ( a , \kvec )  =  \epsilon^2 \Theta^{(n)}_{I_+} ( a , \kvec )  + \epsilon^3 \Theta^{(n)}_{I_-} ( a , \kvec )  + \dots \ ,
\end{split}
\end{align}
for $n\geq1$, where the $\dots$ represent higher orders in $\epsilon$.}  {Since earlier in this section we found that $g \sim \mathcal{O} ( 1 ) - \mathcal{O}(5)$, we can safely take the counting in terms of powers of $\epsilon$ to be the same as the case where $g = 0$, as in \eqn{epsilon3termsedsapprox} for example.}\footnote{{The same argument actually also works if $g \gg 1$, since the powers of $\epsilon$ simply keep track of the different linear solutions' contributions to higher perturbative orders.}}

Now, it is clear that the constant solution for $\delta^{(1)}_{I_+}$ still solves \eqn{newctlineareq}, so we focus on the decaying mode $\delta^{(1)}_{I_-}$.  Since the linear equation is $k$-independent, we can write the solution as $\delta^{(1)}_{I_-} ( a , \kvec ) =  D_{I_-}^g ( a ) \delta^{(1)}_{I_-} ( \kvec ) / D_{I_-}^g ( a_0 ) $, so that the growth factor $ D_{I_-}^g$ (we have included the superscript $g$ to distinguish from the growth factor $D_{I_-}$ when $g=0$, in \eqn{decayinglinsols}) satisfies
\be \label{newctlineareqgrowth}
a^2  D^g_{I_-} {}'' ( a ) + \left( 2 + \frac{ a \cH'(a)}{\cH(a)} \right) a  D^g_{I_-} {}' ( a )  = \int^a d a_1 g ( a , a_1 ) \,  a_1  D^g_{I_-} {}' ( a_1) \ .
\ee
For a general non-local in time interaction $g(a , a_1)$, the above is an infinite order differential equation.

If instead we assume that the interaction is local in time (i.e. $g(a , a_1) = g(a) \delta_D ( a - a_1)$), we can solve \eqn{newctlineareqgrowth} explicitly to find the growth factor 
\be \label{unknowngrowth}
D^{g,{\rm loc.}}_{I_-} ( a ) = \sqrt{ \frac{a_0}{a_{\rm in}}} - \frac{ \cH_0 \Omega_{\rm m,0}^{1/2} }{2} \int^a_{a_{\rm in}} d a_1 \frac{a_0}{a_1^2 \cH ( a_1 ) } \exp \left\{ \int^{a_1} \frac{d a_2}{a_2} g(a_2)  \right\} \ . 
\ee
Because of the unknown function $g(a_2)$ involved in this solution, one cannot explicitly evaluate the growth factor for a specific cosmology, {although since $g \sim \mathcal{O}(1) - \mathcal{O}(5)$, we still expect \eqn{unknowngrowth} to be decaying, and indeed we have checked that this is the case for the numerical solution of $g(a)$ given in \figref{velotimdep}.  Furthermore, if $g(a) < 0$, as is also the case for our solution, \eqn{unknowngrowth} shows that the decaying mode always stays decaying.}  The same is the case for the non-local equation \eqn{newctlineareqgrowth}.  Thus, in perturbation theory, we should treat this growth factor as another unknown, time-dependent coefficient, similar to how we treat the other counterterms in the EFTofLSS.  The $k$-dependence of the linear solution $\delta^{(1)}_{I_-} ( a , \kvec)$, importantly, is still known.

 \begin{figure}[t!]
\centering
\includegraphics[height=5.8cm]{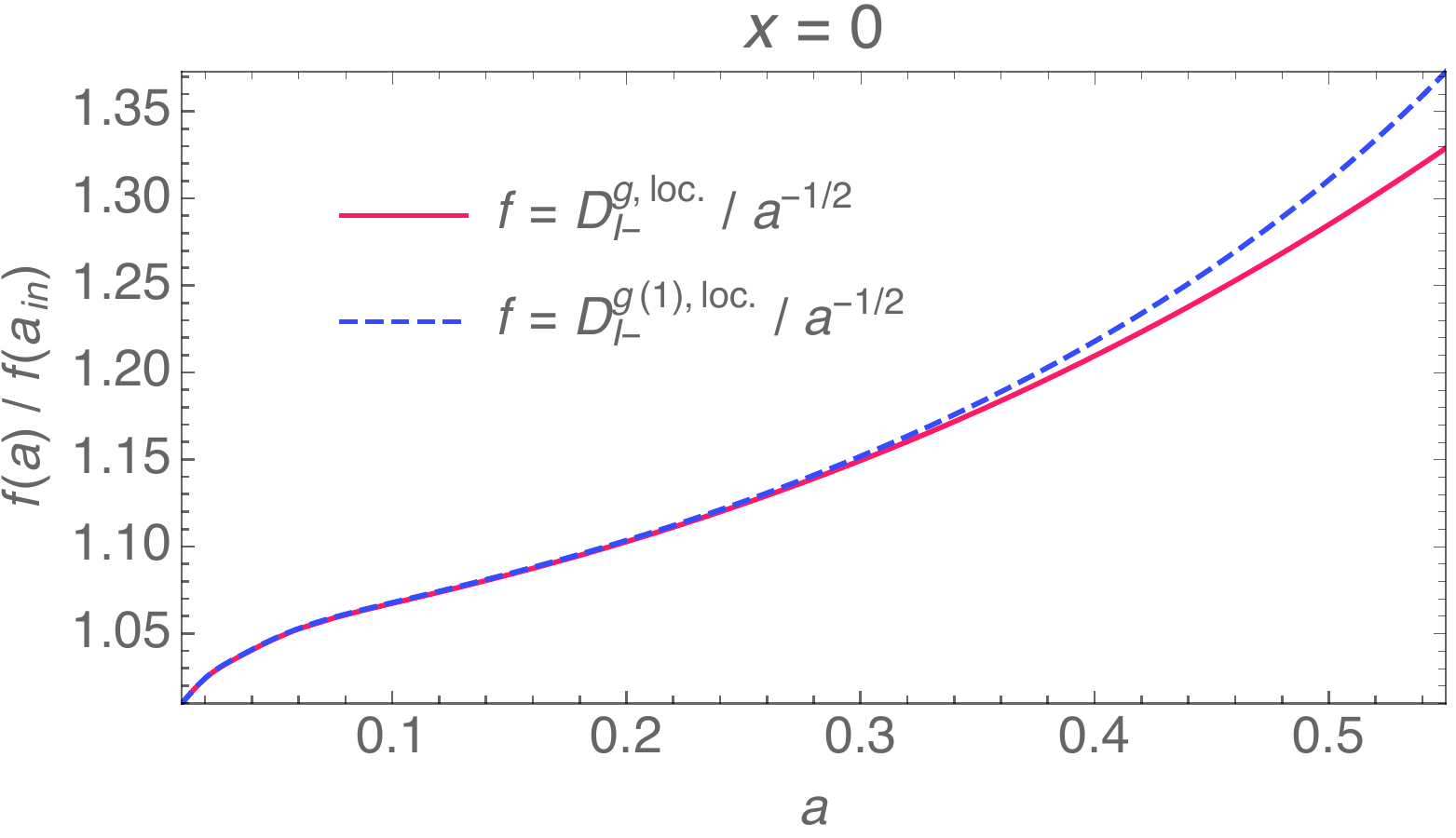} 
\caption{\footnotesize Comparison of the full solution \eqn{unknowngrowth} with the perturbative solution \eqn{unknowngrowthpert}.  We see that overall, the decaying mode is about $30\%$ different from the linear solution, and that the perturbative solution \eqn{unknowngrowthpert} is about $1\%$ different from the full solution \eqn{unknowngrowth}.  } \label{diminusplot}
\end{figure}

{As we found in \secref{sec:sphcoll}, the change in the linear decaying mode is expected to be perturbative.  This means that we can also compute the solution to \eqn{newctlineareqgrowth} perturbatively.  Again assuming that the interaction is local in time, and assuming an EdS universe for simplicity, we can write the solution to \eqn{newctlineareqgrowth} to first order in $g$ as
\be \label{unknowngrowthpert}
D_{I_-}^{g(1),\text{loc.}} ( a ) = \left( \frac{a}{a_{\rm in}} \right)^{-1/2} + \int_{a_{\rm in}}^a d a_1 \, G_{I_-} ( a , a_1) g(a_1) \left( -\frac{1}{2} \left( \frac{a_1}{a_{\rm in}} \right)^{-1/2} \right) \ ,
\ee
where $G_{I_-} ( a , a_1) = - 2 a_1^{-1/2} \left( a^{-1/2} - a_1^{-1/2} \right)$ is the Green's function for the left-hand side of \eqn{newctlineareqgrowth} in an EdS universe.  In \figref{diminusplot}, we compare the above solution to the exact solution \eqn{unknowngrowth} using the $g(a)$ found in \figref{plotsofg} for $L = L_2$.  We see that the perturbative solution is about $1\%$ different from the full solution.  }

{There are  at this points two ways  to proceed in order to formalize a perturbation theory in the presence of the term in $g \, v_I$. The first is by assuming that the contribution of this term is perturbative, and proceed as we did in~(\ref{unknowngrowthpert}), {\it i.e.} consider $g$ as a coupling constant, and treat it as we do for the non-linear term and the other counterterms in the EFTofLSS. This approach does not present conceptual challenges, though one should check that the corrections in $g$ are small, because this is not guaranteed to be true based on first principles. Alternatively, and perhaps more safely, one can set up a perturbation theory where one is not treating the effect of $g$ as a small effect. In fact, we can actually use} $\delta^{(1)}_{I_-} ( a , \kvec) =  D_{I_-}^g ( a ) \delta^{(1)}_{I_-} ( \kvec )  / D_{I_-}^g ( a_0 ) $ in perturbation theory, but now with $D_{I_-}^g ( a )$ as a free function.   Importantly, as shown in \appref{pertwithlinctsec}, the solutions for $\delta^{(n)}_{I_+}$ are not affected by the new counterterm, so they act as normal in perturbation theory {(this is true even if we treat the contribution of $g$ perturbatively)}.
 To solve for the higher order fields $\delta_{I_-}^{(n)}$ for $n \geq 2$, keeping the counterterm as part of the linear equations, we will also need the Green's function related to \eqn{newctlineareq}, which again, is unknown.  However, similar to what was done for the growth factor $D_{I_-}^g ( a )$, we can treat the Green's function formally as an unknown $k$-independent function and proceed with perturbation theory as usual, keeping in mind that all time-dependent functions made from $D_{I_-}^g ( a )$ or the Green's function must be treated as free parameters.  Thus, the solutions for $\delta_{I_-}^{(n)}$ for $n \geq 2$ will be a sum of unknown time-dependent ($k$-independent) functions multiplied by known $k$-dependent kernels.  Since there will be many unknown time-dependent functions {multiplying known functions of wavenumber, the resulting perturbative expansion for $\delta_{I_-}^{(n)}$ will  be reminiscent of the one of a biased tracer.  We discuss this perturbative approach where we do not expand in $g$} in more detail and derive the explicit contribution to $\delta_{I}^{(2)}$ in  \appref{pertwithlinctsec}.

%
%
%
%
  
\section{Comparison to hydro-cosmological simulation}
\label{Comparison}

In this section, we compare our two-loop computation to non-linear data from the hydrodynamical OWLS simulation described in~\cite{vanDaalen:2011xb,Schaye:2009bt}, which includes AGN feedback and is based on WMAP3 cosmological parameters $\{\Omega_m, \Omega_b, \Omega_{\Lambda},\sigma_8,n_s,h\}=\{0.238,0.0418,0.762,0.74,0.951,0.73\}$ \cite{Spergel:2006hy}.  The OWLS project is a collection of different simulations that include various different baryonic effects.  Each simulation has the same cosmological parameters and starts from the same initial conditions at an early time during matter domination.  Then, a fraction $\Omega_b / \Omega_m$ of the particles are labeled as baryons and given specific interactions which mimic star-formation physics, while the rest are kept as CDM particles and interact only through gravity.  Simulations which are referred to as ``dark-matter-only'' do not include any baryonic interactions, and so all of the particles act like CDM (i.e. a standard dark-matter simulation).   In this work, for concreteness, we focus on one OWLS simulation called $AGN$.\footnote{The simulations that we used are available at \href{http://vd11.strw.leidenuniv.nl/}{http://vd11.strw.leidenuniv.nl/}.  {There is a new set of data available at \href{http://powerlib.strw.leidenuniv.nl/}{http://powerlib.strw.leidenuniv.nl/} \cite{vanDaalen:2019pst}.  In this work, we do not explore the range of all the models, but we simply wish to check against a typical model.  We expect, as shown in \cite{Lewandowski:2014rca}, that everything is very similar in terms of accuracy and $k$-reach for various models.}}

We do our analysis in two steps.  First, we find the EFT coefficients for the dark-matter-only power spectrum with the above cosmological parameters.  We do this by comparing the dark-matter-only prediction to the non-linear data from the Coyote interpolator~\cite{Heitmann:2008eq, Heitmann:2009cu, Lawrence:2009uk, Heitmann:2013bra} for 18 different redshifts from $z=0$ to $z=4$. For this analysis, we assume an error of $1\%$ added in quadrature with cosmic variance for a box of size $L = 1 \text{ Gpc}$ on the Coyote data. The IR-resummed two-loop adiabatic power spectrum without baryonic effects $P_{\text{EFT-2-loop}}^{A,\text{DM only}}$  is fit to the Coyote simulation data, which gives a set of coefficients $\{ c^{2}_{s(1)}, c^{2}_{1s}, c^{2}_{4s} \}$ for each of the 18 redshifts (see \appref{tableapp} for parameter values).  Since the Coyote data has more data points  than the OWLS data at low $k$, the former is chosen to determine the dark matter coefficients instead of using the OWLS data directly. {Because the errors are relatively large, and in order to avoid over-fitting, we do the dark-matter-only fits up to the wavenumbers given in \cite{Foreman:2015lca} which used much more precise data.  }

In the second step, we compare our adiabatic and baryonic two-loop predictions to the OWLS-$AGN$ simulation, which includes feedback from active galactic nuclei.   In this simulation, the CDM and baryons are given the same initial conditions, which means that there is no linear isocurvature mode, i.e. $\delta^{(1)}_I = 0$.  Thus, in our computations, we take
\be
P^A_{11} ( a , k) = P^c_{11} ( a , k)  = P^b_{11} ( a , k) = P_{11, \text{CAMB}}^{A} ( a , k ) \ ,  
\ee
where $ P_{11, \text{CAMB}}^{A} $ is the linear total matter power spectrum taken directly from CAMB at each redshift.  All in all, since isocurvature modes are not generated through the standard non-linear interactions,\footnote{Having the same initial conditions for baryons and dark matter implies that $\delta_I^{(1)}=0$ and $\Theta_I^{(1)} = 0$ at linear level.  Looking at the isocurvature equation of motion \eqn{finalisoeqn}, we see that any higher-order isocurvature mode is sourced by a lower order isocurvature mode, so that if $\delta_I^{(1)}=0$ and $\Theta_I^{(1)} = 0$, we have that $\delta_I^{(n)}=0$ and $\Theta_I^{(n)} = 0$ for any higher order $n$.  This  of course comes from the equivalence principle, according to which gravity acts equally on baryons and dark matter, so, once the initial conditions are the same, the evolution is the same.} all of the difference between the CDM and baryon power spectra comes from the different values of the counterterms, i.e. from different UV physics.  Thus, an isocurvature mode is only generated by the different UV physics of the two species in the simulation.

Additionally, we do not  include the linear relative-velocity counterterm $ g \, \Theta_I$ in our fits, since it is expected to be small (see \figref{fig:DESI2loop}), {and since there is no initial isocurvature mode, it would be proportional to the other counterterms, which are suppressed by at least $k^2$, and so would be even smaller than estimated in \secref{sec:realuniverse} and \secref{sec:sphcoll}.}  Indeed we find no indication that it is needed to fit the simulations that we look at in this paper.  The situation may be different, however, for simulations that include an initial isocurvature mode.  We leave exploration of this interesting possibility for future work.\footnote{{The more recent simulations in \cite{vanDaalen:2019pst} include an initial isocurvature mode and it would be interesting to do a similar analysis with that data.  By focusing on simulations where baryons and CDM have identical initial conditions, however,  we are better able to isolate the distinct star-formation effects. } }

\begin{figure}[ht!]
     \begin{center}
     \begin{tabular}{cc}
     \hspace{-.2in}     \includegraphics[width=8cm]{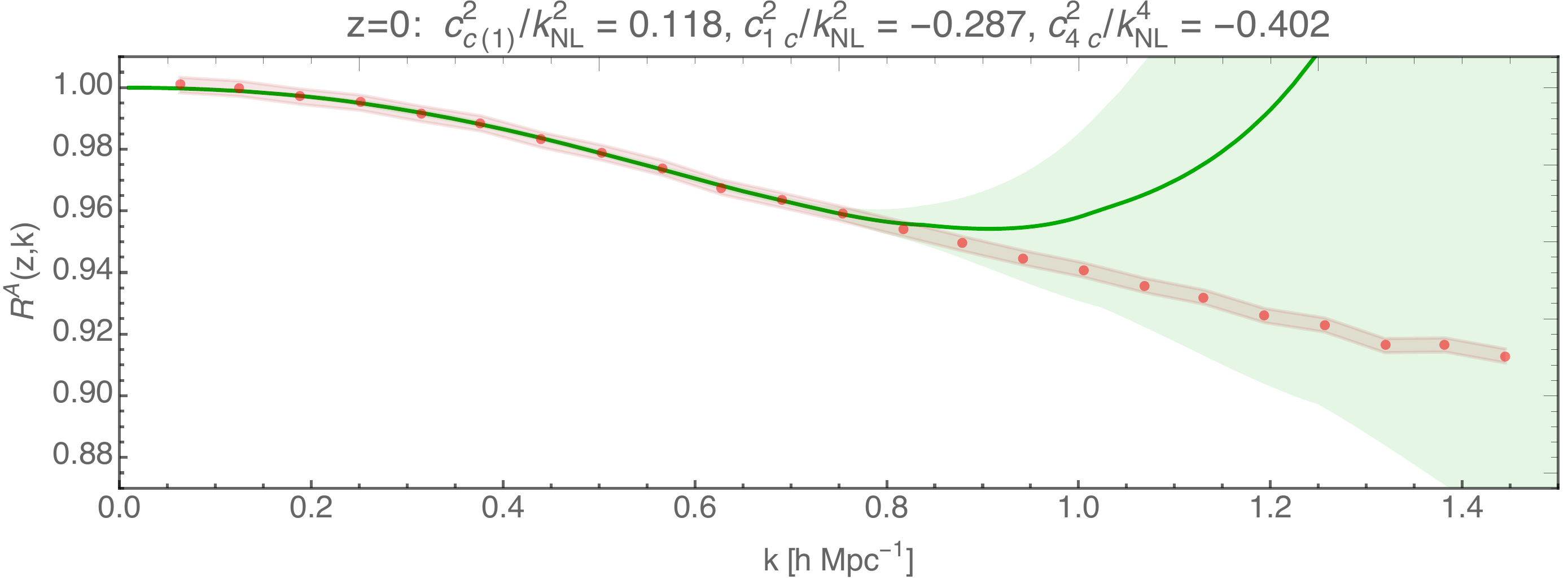} &   \hspace{-.1in} \includegraphics[width=8cm]{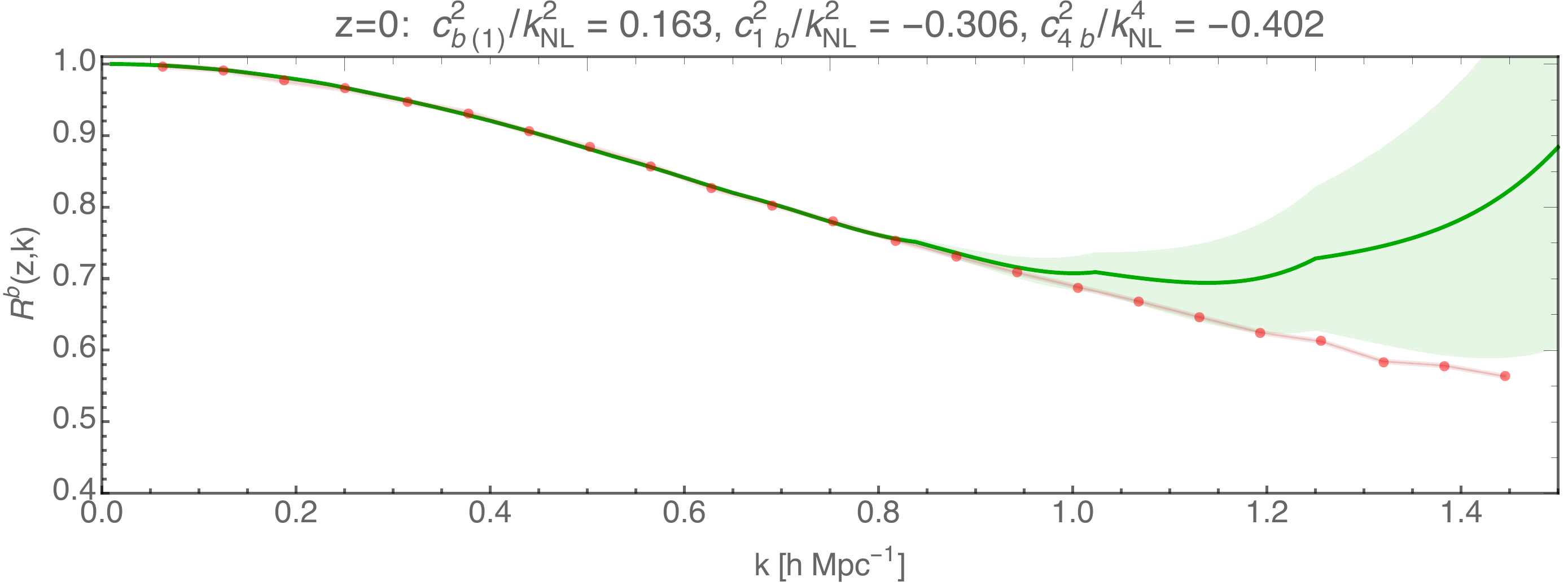}\\
    \hspace{-.2in}         \includegraphics[width=8cm]{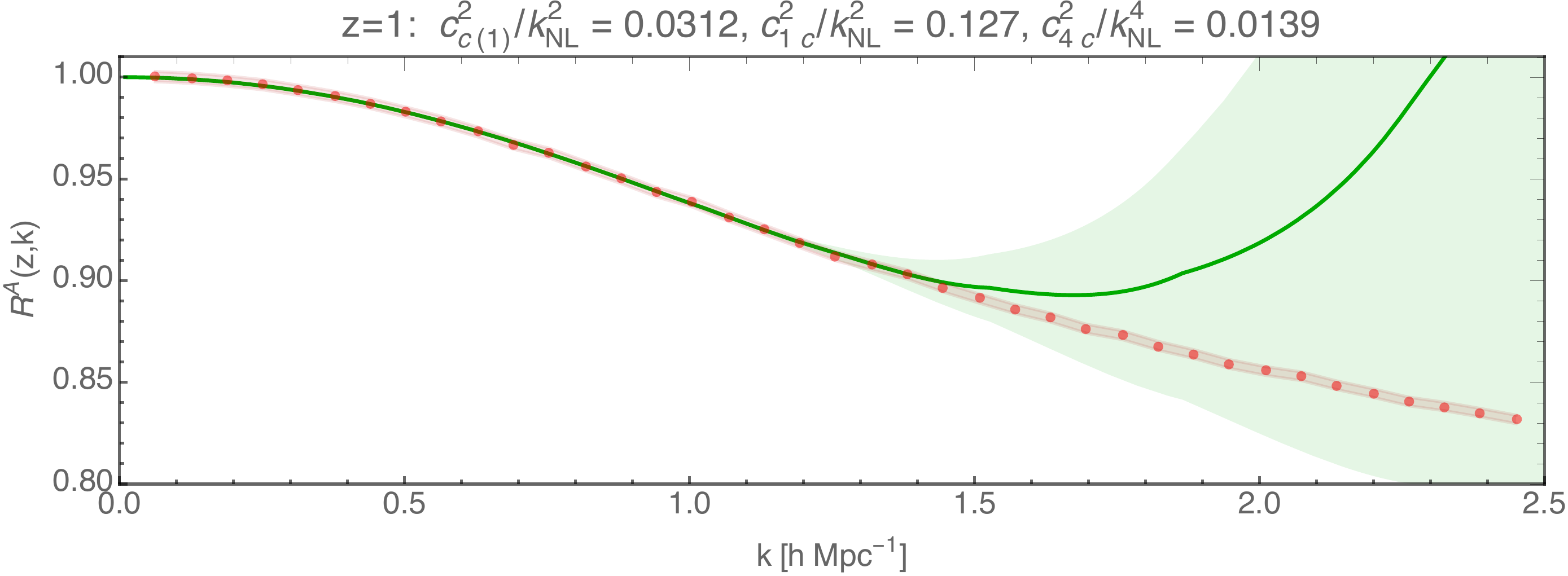} &  \hspace{-.1in} \includegraphics[width=8cm]{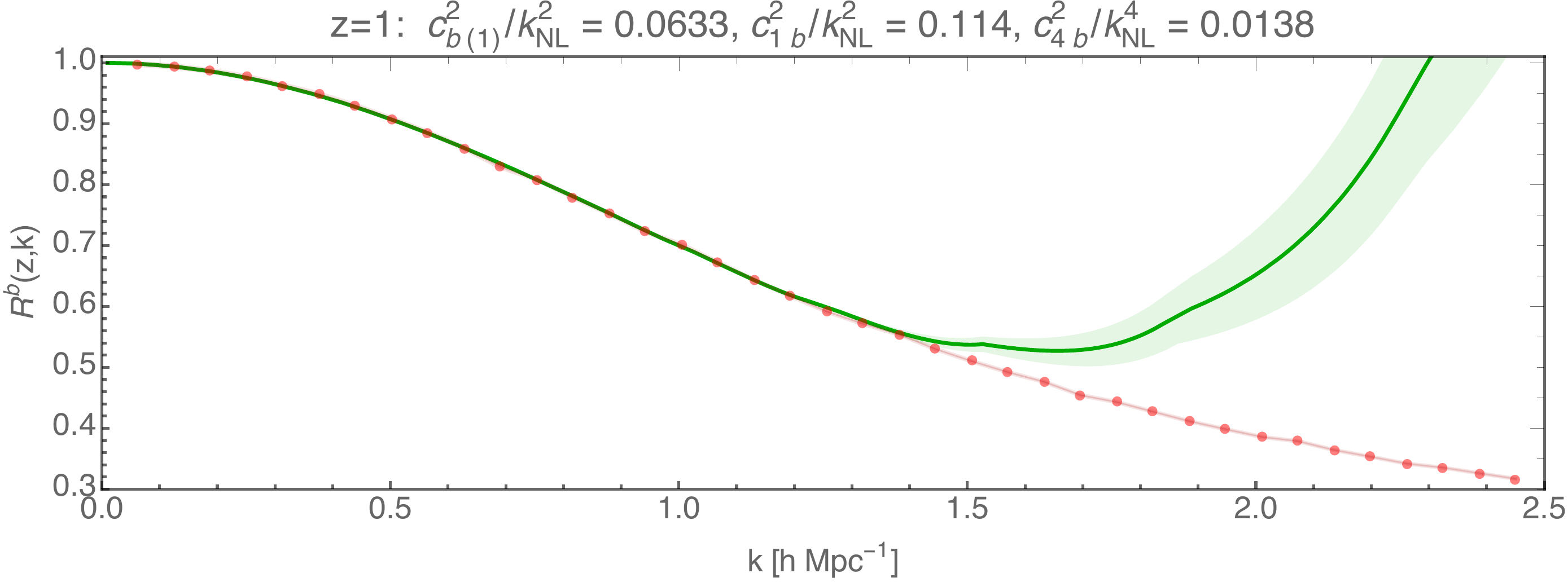}\\
      \hspace{-.2in}        \includegraphics[width=8cm]{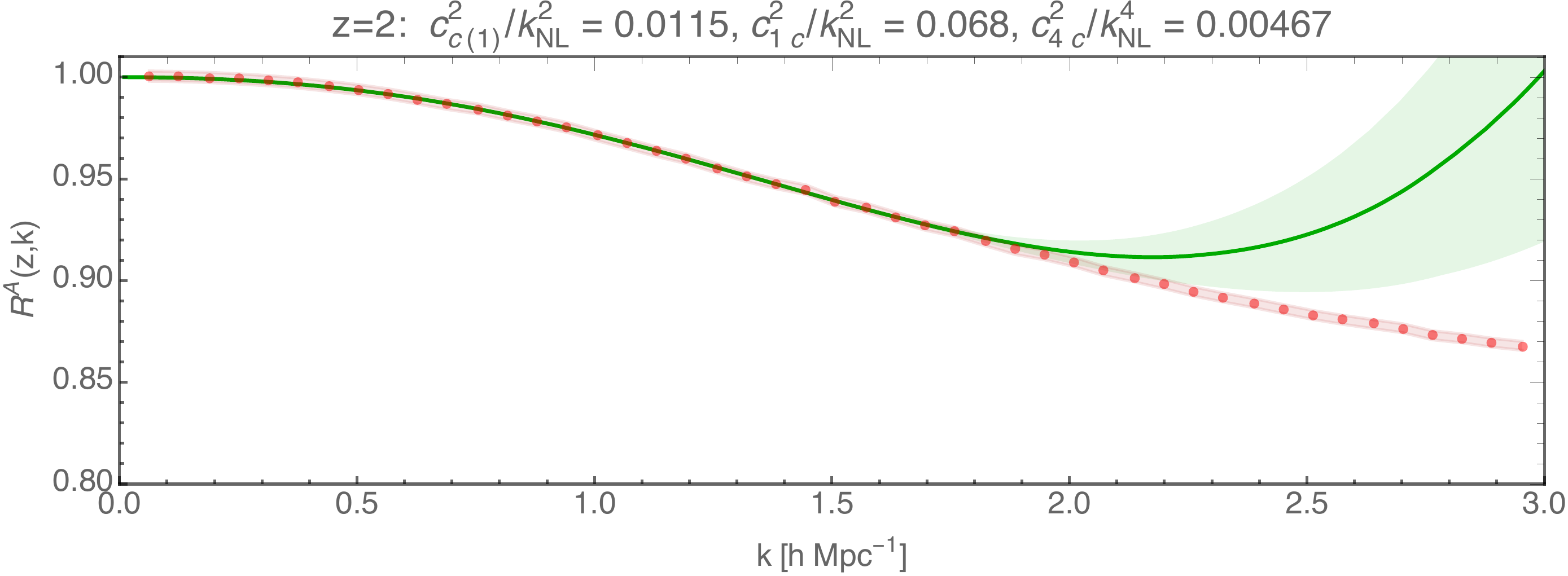} &  \hspace{-.1in} \includegraphics[width=8cm]{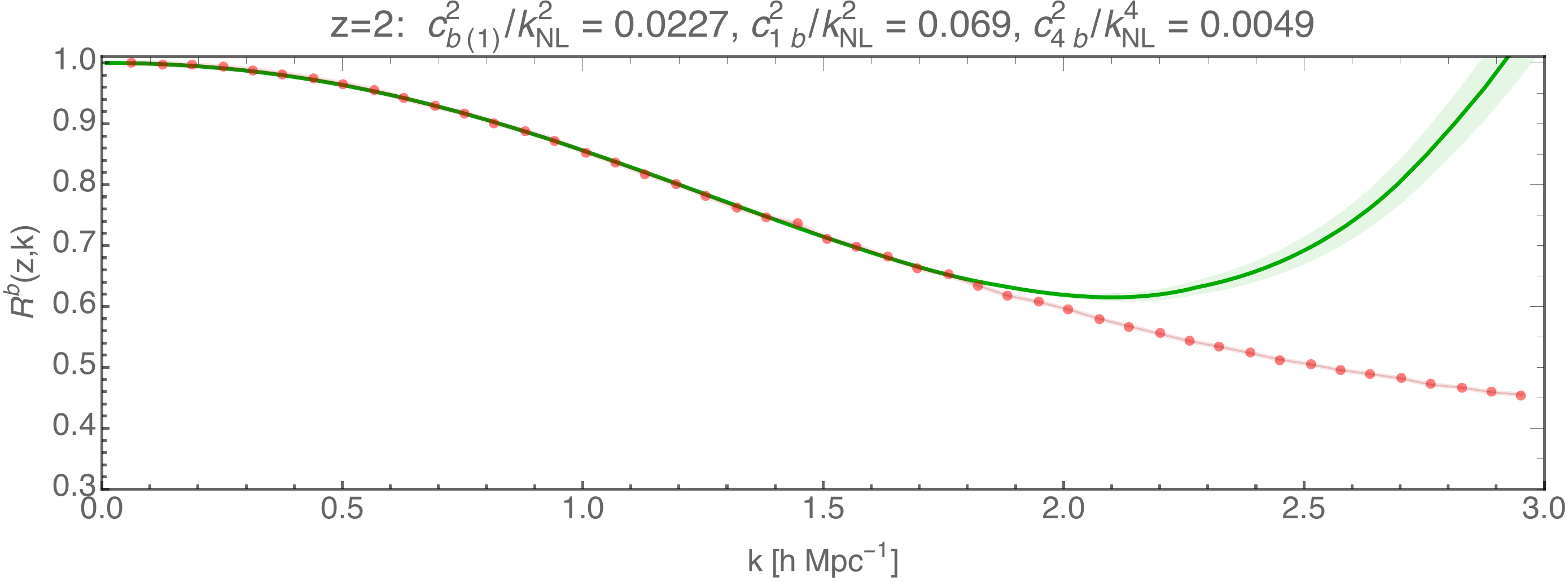}\\
      \hspace{-.2in}        \includegraphics[width=8cm]{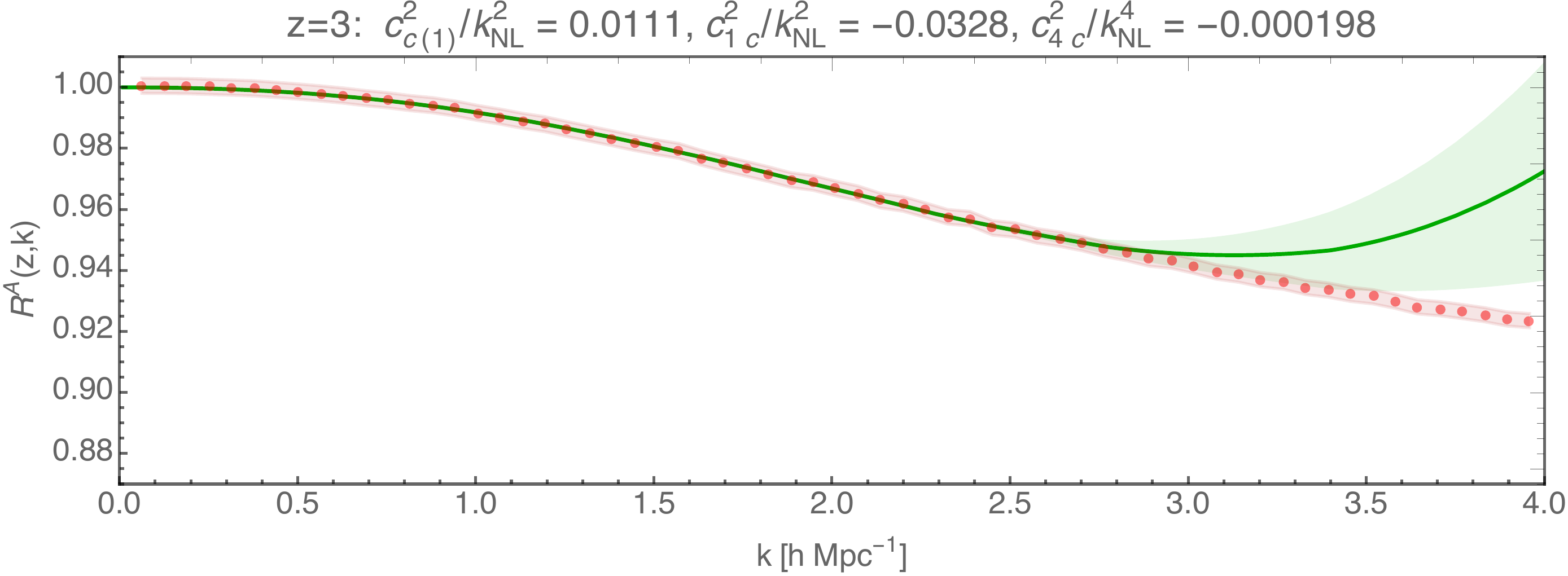} &  \hspace{-.1in} \includegraphics[width=8cm]{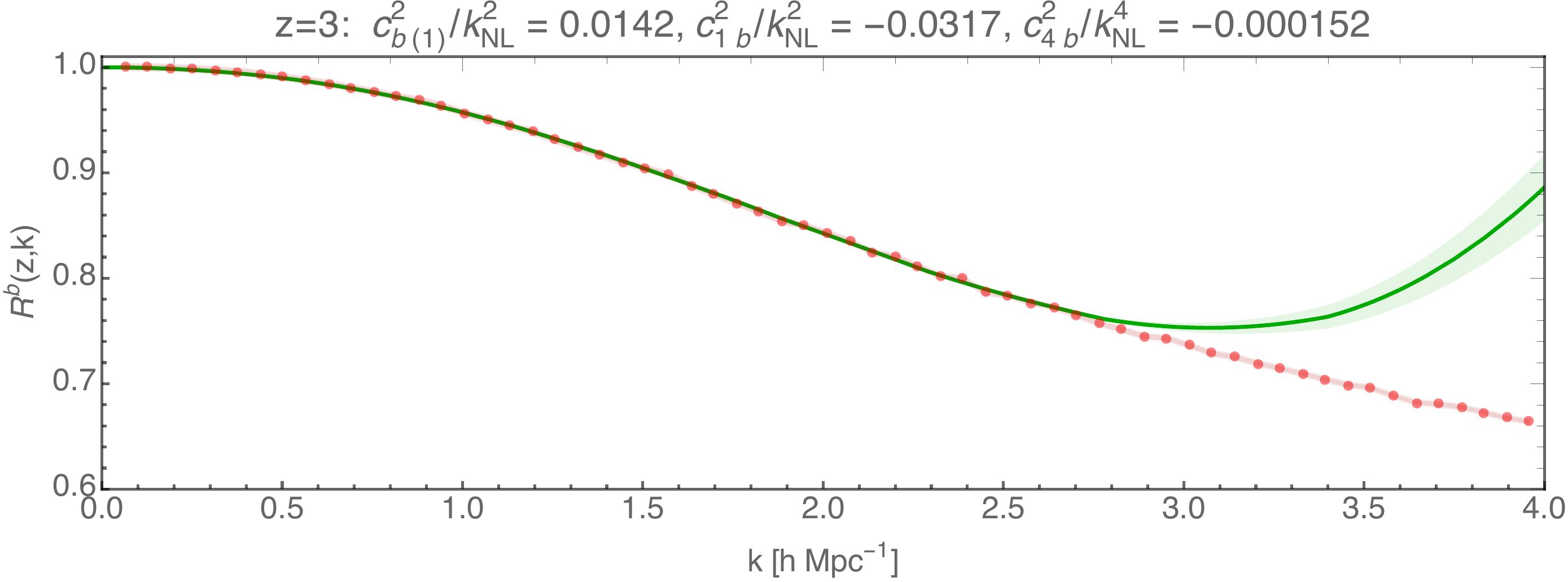}\\
        \hspace{-.2in}   \includegraphics[width=8cm]{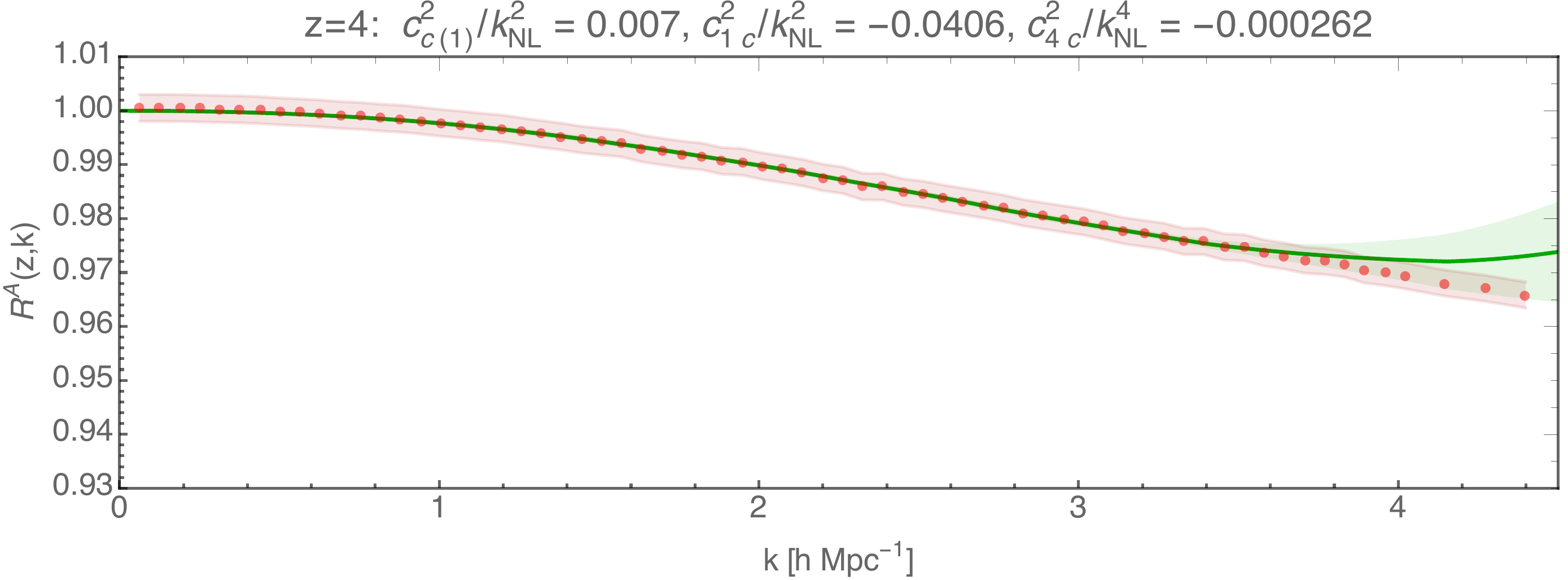} &  \hspace{-.1in} \includegraphics[width=8cm]{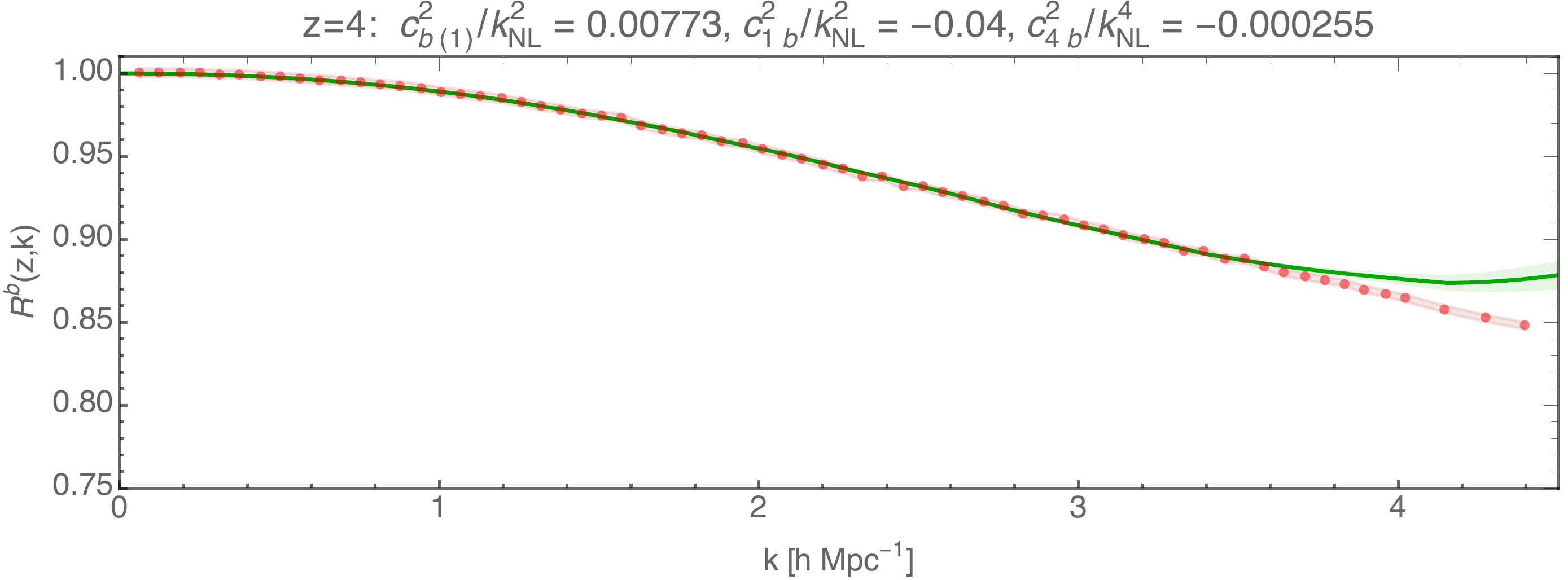}
            \end{tabular} 
    \end{center}
    \caption{ \footnotesize
 Determination of the CDM and baryon EFT parameters $\{ c_{c (1)}^2, c_{1c}^2, c_{4c}^2 \}$ and $\{ c_{b (1)}^2, c_{1b}^2, c_{4b}^2 \}$ for $z=\{ 0, 1, 2, 3, 4 \}$ by fitting $R^{A}_{\text{EFT}}$ and $R^{b}_{\text{EFT}}$ (green solid) to $R^{A}_{\text{OWLS}}$ and $R^{b}_{\text{OWLS}}$ (red dots) up to $k_{\rm fit}$ (described in the main text). The red region is the estimated error of $0.25 \%$ on the data.  The green band is the theory error coming from the $1\sigma$ errors on the fit parameters at $ k_{\rm fit}$.  All parameters presented are dimensionless, so the numerical values given above are in the appropriate units of $\unitsk$ coming from the explicit factors of $\knl$.}
   \label{cAfit}
\end{figure}

{Given the dark-matter-only coefficients $\{ c^{2}_{s(1)}, c^{2}_{1s}, c^{2}_{4s} \}$  found earlier, we then define the differences
\be \label{defdiff}
\Delta c^{2}_{\sigma(1)} \equiv c^{2}_{\sigma(1)} - c^{2}_{s(1)}  \ , \quad \Delta c^{2}_{1\sigma} \equiv c^{2}_{1\sigma} - c^{2}_{1s } \andd   \Delta c^{2}_{4\sigma} \equiv c^{2}_{4\sigma} - c^{2}_{4s } \  ,
\ee
for $\sigma \in \{ A,c,b\}$.  The fitting procedure then consists of determining the sets of CDM and baryon coefficients $\{ \Delta c^{2}_{c(1)},  \Delta c^{2}_{1c}, \Delta c^{2}_{4c} \}$ and $\{ \Delta c^{2}_{b(1)}, \Delta c^{2}_{1b}, \Delta c^{2}_{4b} \}$ for each redshift by performing the fits
\begin{align}
\begin{split}  \label{ratiofittingfns}
    R^{A}_{\text{NL}}\equiv\frac{P_{\text{OWLS}}^{{A}}}{P_{\text{OWLS}}^{\text{DM only} }}   \quad &\leftrightarrow \quad   R^{A}_{\text{EFT}}\equiv \frac{P_{\text{EFT-2-loop}}^{A}[ c^{2}_{s(1)}+ \Delta c^{2}_{A(1)}, c^{2}_{1s}+ \Delta c^{2}_{1A}, c^{2}_{4s}+ \Delta c^{2}_{4A}  ]  }{P_{\text{EFT-2-loop}}^{A,\text{DM only}}  [  c^{2}_{s(1)}, c^{2}_{1s}, c^{2}_{4s}  ]  }      \, ,  \\
        R^{b}_{\text{NL}}\equiv\frac{P_{\text{OWLS}}^{ b }}{P_{\text{OWLS}}^{{\text{DM only}}}}     \quad &\leftrightarrow \quad  R^{b}_{\text{EFT}}\equiv \frac{P_{\text{EFT-2-loop}}^{b}[ c^{2}_{s(1)}+ \Delta c^{2}_{b(1)}, c^{2}_{1s}+ \Delta c^{2}_{1b}, c^{2}_{4s}+ \Delta c^{2}_{4b}  ]  }{P_{\text{EFT-2-loop}}^{A,\text{DM only}}  [  c^{2}_{s(1)}, c^{2}_{1s}, c^{2}_{4s}  ]  }   \, , 
\end{split}
\end{align}
where we Taylor expand the ratios $R^A_{\rm EFT}$ and $R^b_{\rm EFT}$ up to two-loop terms, and $\{ \Delta c^{2}_{A(1)},  \Delta c^{2}_{1A}, \Delta c^{2}_{4A} \}$ are defined in terms of $\{ \Delta c^{2}_{c(1)},  \Delta c^{2}_{1c}, \Delta c^{2}_{4c} \}$ and $\{ \Delta c^{2}_{b(1)},  \Delta c^{2}_{1b}, \Delta c^{2}_{4b} \}$ using \eqn{cavaluedefs} and \eqn{defdiff}.  Explicitly, the Taylor expanded ratio up to two loops, $R^\sigma_{\rm EFT} |_2$ at $z = 0$ is
\begin{align}
\begin{split} \label{taylorexpandratio}
R^\sigma_{\rm EFT} |_2  = 1 - 4 \pi \Delta c_{\sigma ( 1 ) }^2 \frac{k^2}{\knl^2} & + \frac{2\pi }{P_{11}^A} \left( \Delta c_{\sigma (1)}^2 \left( 2 \frac{k^2}{\knl^2} P_{1\text{-loop}}^A +P_{1\text{-loop}}^{A,(c_s)} \right) + \Delta c_{1\sigma}^2 P_{1\text{-loop}}^{A,(\text{quad,1})}  \right)  \\
&  + \frac{8 \pi^2}{17} \frac{k^4}{\knl^4} \left(  14 [ \Delta c_{\sigma(1)}^2 ]^2 - 6 c_{s(1)}^2 \Delta c_{\sigma(1)}^2 + 17 \Delta c_{4\sigma}^2   \right)   \ ,
\end{split}
\end{align}
where we have used $\xi_\sigma = 3$ in \eqn{candbtwoloop}, and $\sigma \in \{c,b,A\}$.  We see that all terms which do not contain any counterterms cancel in the ratio as expected because the numerator and denominator only differ in the counterterms.  Additionally, most parameters appear directly as the difference from the dark-matter-only parameters, except for $c_{\sigma(1)}^2$, which is why we have chosen the parametrization \eqn{defdiff}.}  We directly fit these ratios because the cosmic variance is greatly reduced in the non-linear data after taking the ratio.  In order to avoid over-fitting, we use the safe-fitting routine described in \cite{Foreman:2015lca}.  In each fit, we call the maximum $k$ included in the fit $k_{\rm max}$, then we vary $k_{\rm max}$, finding the best fit and errors for the parameters for each $k_{\rm max}$.  Finally, we choose $k_{\rm fit}$ to be the maximum value of $k_{\rm max}$ where the parameter best fit values are still consistent with the lower $k_{\rm max}$ fits.  In \appref{tableapp}, we give an example of how we perform this procedure at $z = 2$. {In \appref{changekfitapp}, we consider our fits using smaller values of $k_{\rm fit}$. }

As mentioned, by using the ratio of power spectra, the cosmic variance of the simulation data is greatly reduced. As an estimate of all of the residual errors, we assume an error of $0.25\%$ on the ratio data, and the results of our fits are given in \figref{cAfit}.      We find that at two loops, the ratios fit the data up to $ k \approx 0.8 \unitsk$ at $z = 0$, for example, and to higher $k$ at higher redshifts.   {As discussed in \cite{Lewandowski:2014rca} for one loop, since we are dealing with the ratio of EFT predictions, the reach of $R_{\text{EFT}}$ is increased with respect to the dark-matter-only fits.  {In general, this is essentially because $\Delta c^2 / c^2$ is small for the various quantities in \eqn{defdiff}, as one can see in the tables in \appref{tableapp}.  Sometimes, though, we see that $\Delta c^2 / c^2 \sim \mathcal{O}(1)$, in which case we have to look a bit closer at the specific terms in the expansions.}  

As an estimate of the increase in $k$-reach that we can expect, we look at the highest order terms included in the expansions, which at two loops, {for example, are proportional to $k^4 P_{11} $}.  For dark-matter-only, we have
\be
\frac{P^{\text{DM only}}_{\rm EFT-2-loop}}{P^\sigma_{11}} \supset \frac{8 \pi^2}{17} \frac{k^4}{\knl^4} ( 14 c_{s(1)}^4 +17 c_{4s}^2 )   \ .
\ee
Using the values of the parameters obtained from our fits (see \appref{tableapp}, we focus on the baryon fit because it is the larger deviation at $z=0$), this term is numerically $-31 (k/\knl)^4$.  On the other hand, the $k^4$ term in \eqn{taylorexpandratio} is $0.1 (k/\knl)^4$, so it is clear that the scales controlling the expansions are quite different.   {We then rescale $k$ in each expansion so that the coefficient of $k^4$ is unity, and we find that the coefficients of all of the lower powers of $k$ are of order unity, confirming that all the counterterms are suppressed in a similar way.}  Calling $k_{\rm fit}^{R}$ the reach of the ratio \eqn{taylorexpandratio}, and calling $k_{\rm fit}^P$ the reach of the two-loop power spectrum fit, we thus expect $k_{\rm fit}^R \sim (31/0.1)^{1/4} k_{\rm fit}^P$.  Using $k_{\rm fit}^P \approx 0.26 \unitsk$, this gives $k_{\rm fit}^R \approx 1.07 \unitsk$, which is in qualitative agreement with what we found (see \appref{tableapp} for parameter values).}\footnote{{As more extreme examples of when $\Delta c^2 / c^2 \sim \mathcal{O}(1)$, we can look at the baryon parameters at $z =0.5$ and $z=1$ (see \tabref{cs1table} for example).  The ratio of the $k^4$ coefficients of the baryon ratio expansion to the dark-matter only expansion at $z = 0.5$ is $124.7$, and at $z = 1$ is $21.8$.  This means that we expect $k^R_{\rm fit} (z = 0.5) / k^P_{\rm fit} ( z = 0.5) \sim 3.34  $, and $k^R_{\rm fit} (z = 1) /  k^P_{\rm fit} ( z = 1 ) \sim 2.16$.  Looking at \tabref{kreachtab}, we have $k^R_{\rm fit} (z = 0.5) / k^P_{\rm fit} ( z = 0.5) \approx 3.19  $  and  $k^R_{\rm fit} (z = 1) / k^P_{\rm fit} ( z = 1) \approx 2.4  $, which again, is in agreement with our estimates.}}   These fits are quite a remarkable improvement over other analytic descriptions of baryons in LSS.

{As an application of our above results, we see that we have an analytic way of adding baryonic effects to a dark-matter-only simulation.  If $P_{\rm NL}^{ \text{DM only}} ( a , k)$ is the non-linear output of some dark-matter-only simulation, then we can parametrize the effects of baryons by using the ratio \eqn{taylorexpandratio} to get
\be \label{dataratio}
P^\sigma_{\rm NL} ( a ,  k ) = R^\sigma_{\rm EFT} |_2  ( a , k ) \, P_{\rm NL}^{\text{DM only}} ( a , k ) \ , 
\ee 
which is valid up to the high scale $k^R_{\rm fit} ( a )$ shown in \figref{cAfit} for two loops, for example.   }

\begin{figure}[t!]
\centering
\includegraphics[width=15cm]{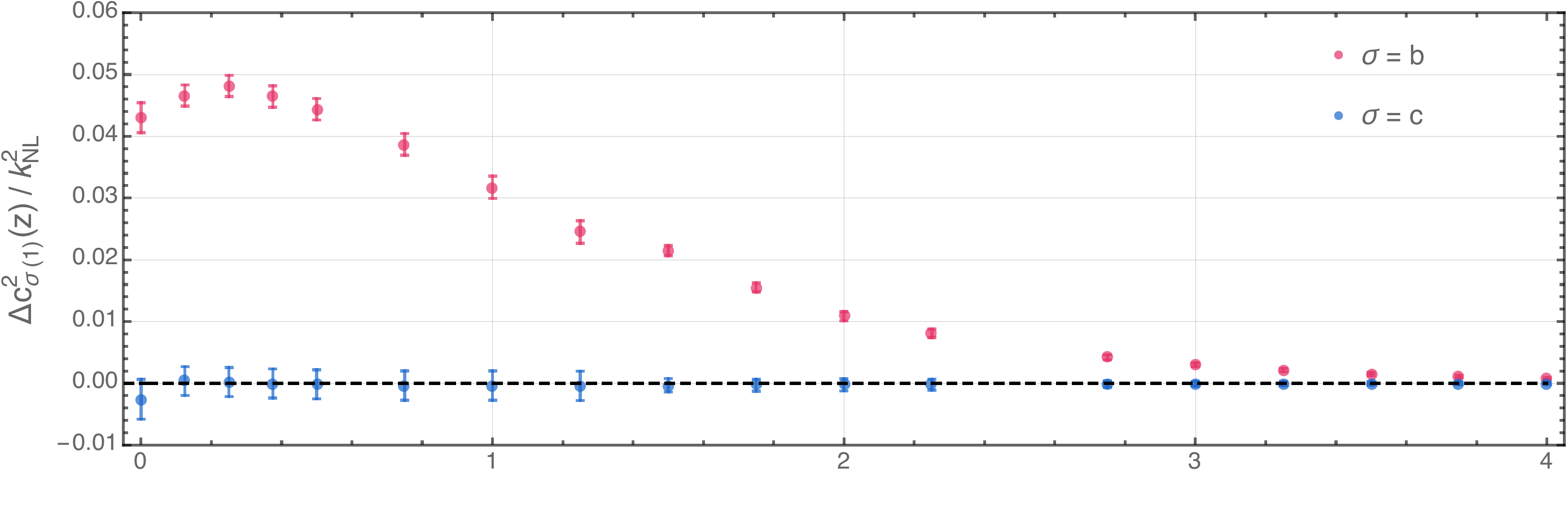}\\
\includegraphics[width=15cm]{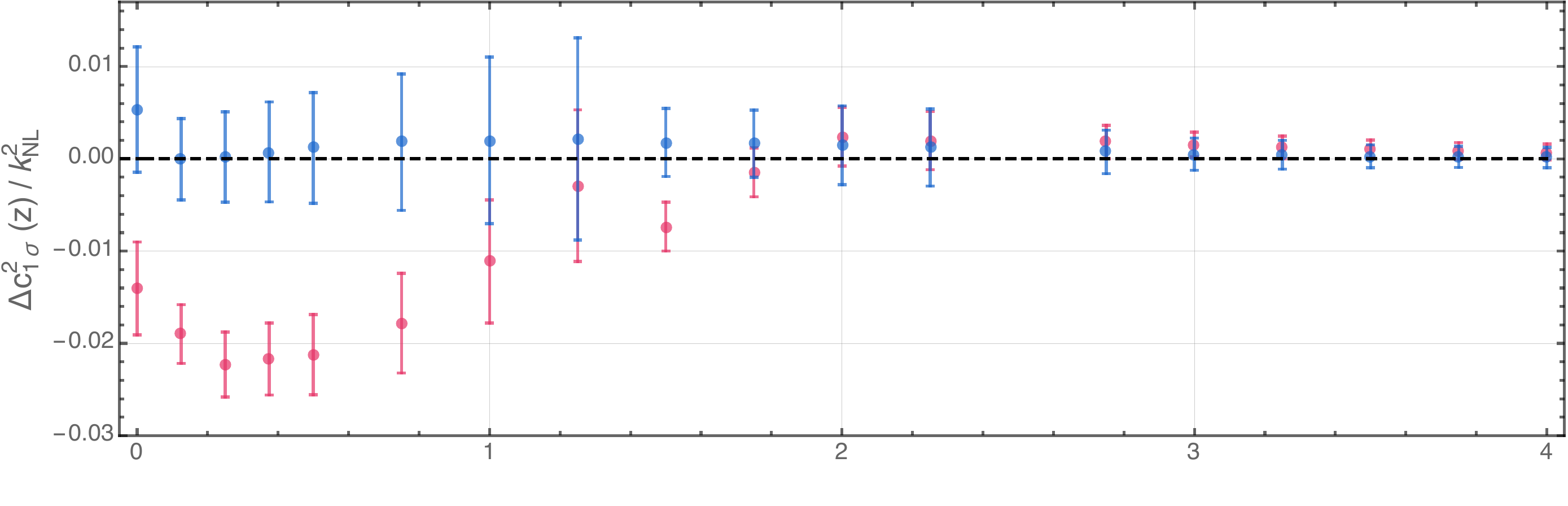}\\
\includegraphics[width=15cm]{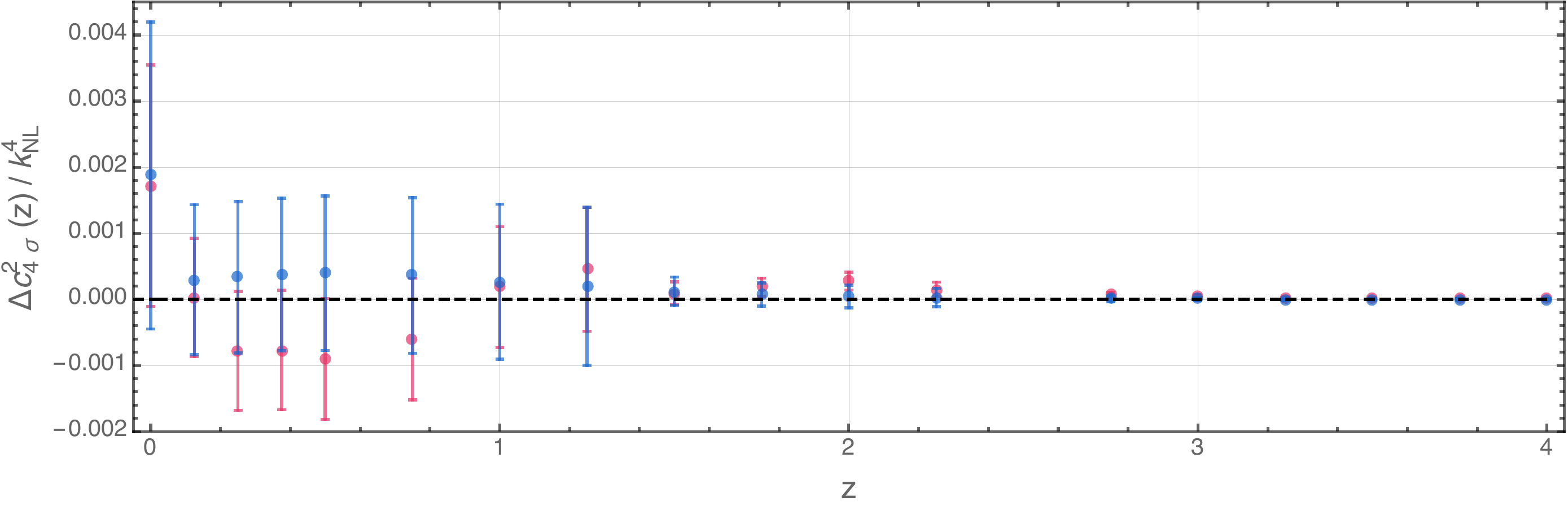}
\caption{ \footnotesize The dark matter and baryonic EFT parameters $\Delta c_{c(1)}^2$,  $\Delta c_{b(1)}^2$,  $\Delta c_{1c}^2$, $\Delta c_{1b}^2$ and $\Delta c_{4c}^2$, $\Delta c_{4b}^2$ as a function of redshift (baryons are in pink, and CDM is in blue). The error-bars are the $1\sigma$ errors at $0.75 k_{\rm fit}$ for each parameter.   We plot the differences \eqn{defdiff} here because these are the parameters best measured by fitting to the ratios.  We see explicitly from the first plot that baryonic physics starts to kick in between $z \approx 3$ and $z \approx 2$.   All parameters presented are dimensionless, so the numerical values given above are in the appropriate units of $\unitsk$ coming from the explicit factors of $\knl$. } \label{cbcc}
\end{figure}

We present the time-dependence of the parameters $\{ c_{c (1)}^2, c_{1c}^2, c_{4c}^2 \}$ and $\{ c_{b (1)}^2, c_{1b}^2, c_{4b}^2 \}$  in \figref{cbcc}. One can clearly see the onset of star-formation physics coming from baryonic processes in the simulation. At early times, the coefficients $c_{c(1)}^2$ and $c_{b(1)}^2$ for dark matter and baryons are about the same and start to differ between $z \approx 3$ and $z \approx 2$.  {We also see that the dark-matter EFT coefficients do not change much in the presence of baryons with respect to the dark-matter-only simulation.  The baryons, on the other hand, have a sizable difference from the dark-matter-only coefficients, as expected since they have additional star-formation interactions in the UV.  In general, {for most $c^2$ coefficients, we find that $\Delta c^2 / c^2$ is usually small}, which implies that a large fraction of EFT parameters are determined by gravitational effects, which are the same for dark matter and baryons.}

 \begin{figure}[t!]
\centering
\includegraphics[width=15cm]{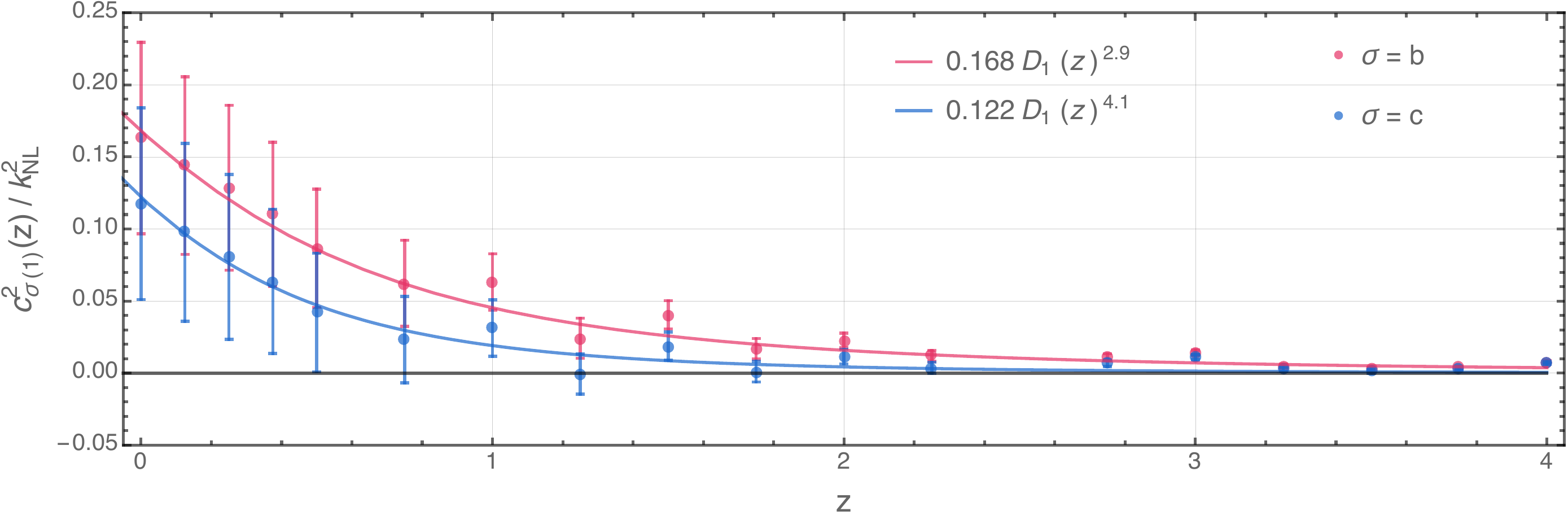}
\caption{\footnotesize {Fit for the time dependence of $c_{c(1)}^2$ and $c_{b(1)}^2$ using \eqn{ansatzcb}.  Because the errors on the dark-matter-only data are larger than the errors on the ratio data, the uncertainty in the dark-matter-only parameters $\{ c^{2}_{s(1)}, c^{2}_{1s}, c^{2}_{4s} \}$ is larger than that in the difference parameters shown in \figref{cbcc}.  
Since this figure shows the total parameters, the errors here are larger than in \figref{cbcc}. } } \label{cbccfit1}
\end{figure}

Next, we find an approximate parametrization for the time dependence of the coefficients $c_{c(1)}^2$ and $c_{b(1)}^2$.  The values of  $c_{c(1)}^2$ and $c_{b(1)}^2$ start by being approximately zero at early times and then increase with time. This behavior is characteristic of the EFTofLSS: at late times non-linearities start to grow proportional to the non-linear scale $k_{\text{NL}}^{-1}$. In the effective field theory, these effects are under control and accounted for in the mildly non-linear regime through the counterterms.  In general, the time dependence of the counterterms is free, but has the structure
\be
c^2\left(z, \Lambda \right)=c^{2}_{\text{finite}}\left(z,k_{\text{ren}}\right)+c^{2}_{\text{ct}}\left(z,\Lambda \right) \ , 
\ee
where $\Lambda$ is the UV cutoff of the theory, and $k_{\rm ren}$ is the renormalization scale where the coefficient is fit to the data.  The term $c_{\rm ct}^2 ( z , \Lambda)$ is needed to cancel the cutoff dependence of physical observables, and so must have the same time dependence as the UV parts of the loop integrals.  In our case, because we use the UV-improved loop integrals, we expect this contribution to be negligible.  The time dependence of the finite part of the counterterm, $c_{\rm finite}^2$, however, is in general unconstrained theoretically, but one expects them to behave with the same time scale as the system, which is $H$.  Because of this, we choose to parametrize the time dependence of the finite parts of the counterterms as the power laws \cite{Foreman:2015lca}
\be
c_{c(1)}^2\left(z\right)=A_c D_1\left(z\right)^{\alpha_c}  \andd c_{b(1)}^2\left(z\right)=A_b D_1\left(z\right)^{\alpha_b}\, .
\label{ansatzcb}
\ee
The fits with the corresponding coefficients are plotted in \figref{cbccfit1}.  See \appref{tableapp} for tables of all of the parameter values. The EFTofLSS thus provides an analytic understanding of the evolution of star-formation physics.

%
%
 %
 %

\section{The effect of baryons on the lensing potential}
\label{lensing}

The purpose of this section is twofold.  First, we discuss the inclusion of baryonic physics in the computation of the lensing potential in the EFTofLSS and describe a strategy which significantly lowers theoretical error bars up to $\ell \approx 2000$.   Second, we explicitly compute the effects on the lensing potential due to baryonic physics for the WMAP3, OWLS-$AGN$ simulation that we studied above.  

Overall, our results suggest that a proper understanding of baryonic effects will be important for interpreting data from upcoming lensing surveys, including lensing of the CMB in the CMB-S4 effort \cite{Abazajian:2016yjj,Abitbol:2017nao,Abazajian:2019eic}.  An accurate understanding of baryonic effects will be crucial in, for example, neutrino mass constraints \cite{Natarajan:2014xba,Chung:2019bsk}.  Previous studies (see for example \cite{Chung:2019bsk} for the CMB, and \cite{Huang:2018wpy} for weak lensing) have reached similar conclusions by studying the impact of baryonic physics on lensing using the outputs of hydrodynamical simulations, like the OWLS simulation that we used in this paper.  The advantage of our EFT approach, though, is that we have analytic control over our predictions on large scales.  The functional form of baryonic effects on large scales, i.e. as a function of $k$ (or $\ell$), is completely fixed by symmetries and is organized in a controlled derivative expansion.  In this way, one can continue to improve the computation, up to non-perturbative effects, by including higher order terms.  The details about the small-scale physics, both baryonic and gravitational, are contained in a set of free parameters which must be fit to data.  For predictions on large scales, we view this as a significant advantage over having to run hydrodynamical simulations, which themselves must make motivated, but ultimately ad hoc, assumptions about the unknown short-scale physics.  On smaller scales, of course, our analytic approach breaks down, and one is forced to use other methods such as simulations.  However, there is still plenty of information in the large-scale modes.  For example, much of the constraining power for a neutrino mass sum of less than $120 \text{ MeV}$ comes from $\ell \lesssim 2100$ \cite{Abazajian:2016yjj}.

Now we turn to our computation, and refer to \cite{Lewis:2006fu} for a detailed discussion of weak gravitational lensing.  Using the small-angle-approximation allows us to write the lensing potential as 
\begin{eqnarray}
\psi (\hat{n})=-2\int_{0}^{\chi_{\ast}}d\chi \frac{f_{K}(\chi_{\ast}-\chi)}{f_{K}(\chi_{\ast})f_{K}(\chi )}\Phi ( \eta_0-\chi , \chi {\hat{n}} )\ ,
\end{eqnarray}
where $\chi$ is the comoving distance,\footnote{This is explicitly defined by 
\be
\chi ( a ) = \int_a^1 \frac{d a'}{(a')^2 H(a')} \ . 
\ee}  $\chi_*$ is the distance to the emitting source, $\hat n$ is the line-of-sight unit vector, and $\eta_0 - \chi$ is the conformal time at which the photon was at position $\chi \hat n$.

Transforming the lensing potential to harmonic space by using $\psi(\hat{n})=\sum_{\ell m}\psi_{\ell m}Y_{\ell m}(\hat{n})$ gives
\begin{eqnarray}
\left<\psi(\hat{n})\psi(\hat{n}')\right>=\sum_{\ell \ell'mm'}  \left<\psi_{\ell m}\psi^{\ast}_{\ell'm'}\right>  Y_{\ell m}(\hat{n})Y^{\ast}_{\ell m}(\hat{n}')\  .
\label{aps2}
\end{eqnarray}
We then define the lensing potential power spectrum $C_\ell^\psi$ from 
\be
 \left<\psi_{\ell m}\psi^{\ast}_{\ell'm'}\right> =\delta_{\ell \ell '}\delta_{mm'}C_{\ell}^{\psi} \ .
\ee
Assuming flatness of the universe ($f_{K}\left(\chi\right)=\chi$), and using the \textit{Limber approximation} one can write the lensing potential power spectrum as
\begin{eqnarray} \label{clpsi}
C_{\ell}^{\psi} = \frac{8\pi^2}{\ell^3}\int_{0}^{\chi_{\ast}} d\chi  \, \chi  \, P_{\Phi}\left( a\left(\chi\right)   , k=\frac{\ell}{\chi}   \right)\left(\frac{\chi_{\ast}-\chi}{\chi_{\ast} \chi}\right)^2\,,
\label{C}
\end{eqnarray}
where $P_{\Phi}$  is the power spectrum of the gravitational potential,\footnote{The power spectrum of the gravitational potential is normalized in the following way
\be
\langle \Phi ( a , \kvec ) \Phi ( a , \kvec') \rangle = \delta_D ( \kvec+ \kvec' ) \frac{2 \pi^2}{k^3} P_{\Phi} ( a , k) \ . 
\ee} which is related to the adiabatic power spectrum by
\be \label{potentialps}
P_\Phi ( a , k ) = \frac{9 \, \Om ( a )^2 \cH(a)^4 }{8 \pi^2 } \frac{P^A ( a , k)}{k} \ . 
\ee

 \begin{figure}[t] 
\centering
\begin{tabular}{cc}
\includegraphics[width=10cm]{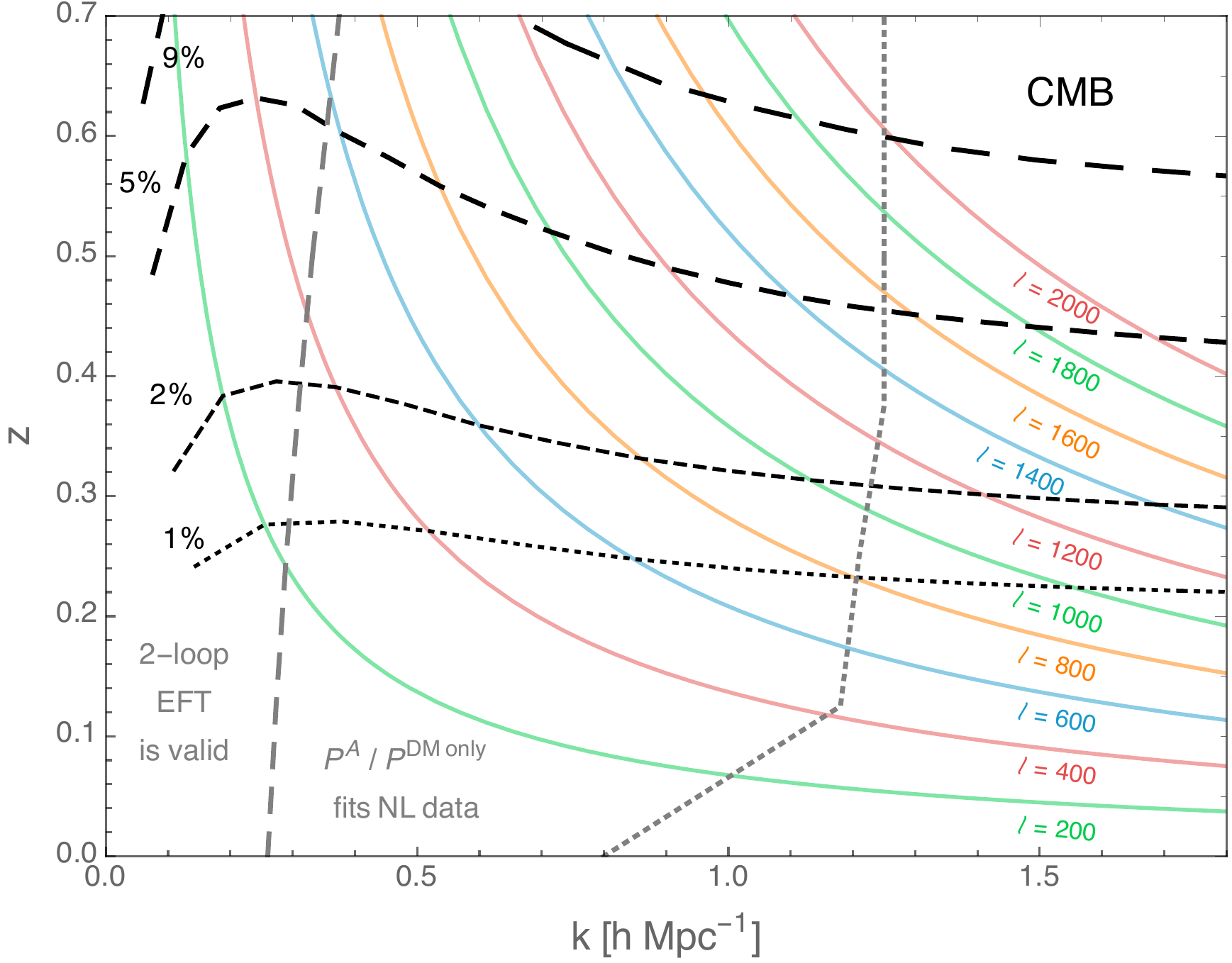} 
\end{tabular} 
\caption{\footnotesize Size of the contribution to the lensing potential \eqn{clpsi} for lensing of the CMB.  The solid lines are the curves in the $(k,z)$-plane where the integral $C_{\ell}^{\psi}$ is evaluated for different $\ell$.   In the region to the left of the gray dashed line, the two-loop EFT prediction for $P^{A}_{\text{EFT-2-loop}}$ is valid, based on a precise comparison to non-linear data \cite{Foreman:2015lca}.  In the region to the left of the gray dotted line, the ratio $P^{A}/P^{A, \text{DM only}}$ fits the non-linear data with $0.25\%$ error on the data (see \figref{cAfit}).  The dashed curves marked $p\%$ delineate the regions of the plane which contribute $(100-p) \%$ (above) and $p \%$ (below) to the lensing integral \eqn{clpsi} for each $\ell$.   }
\label{lensing1}
\end{figure}

\begin{figure}[t]
\centering
\begin{tabular}{cc}
\includegraphics[width=10cm]{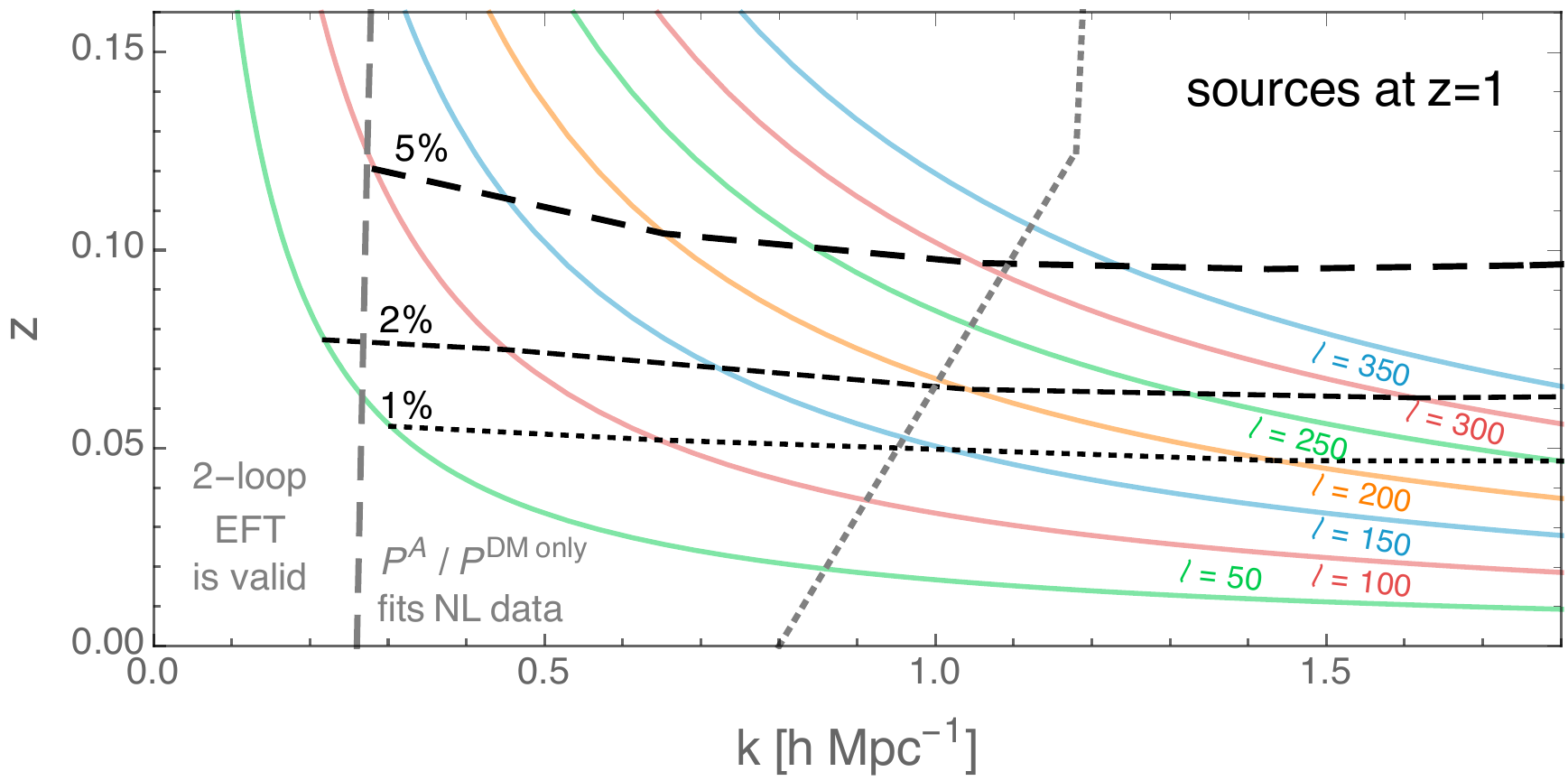} \\ 
\includegraphics[width=10cm]{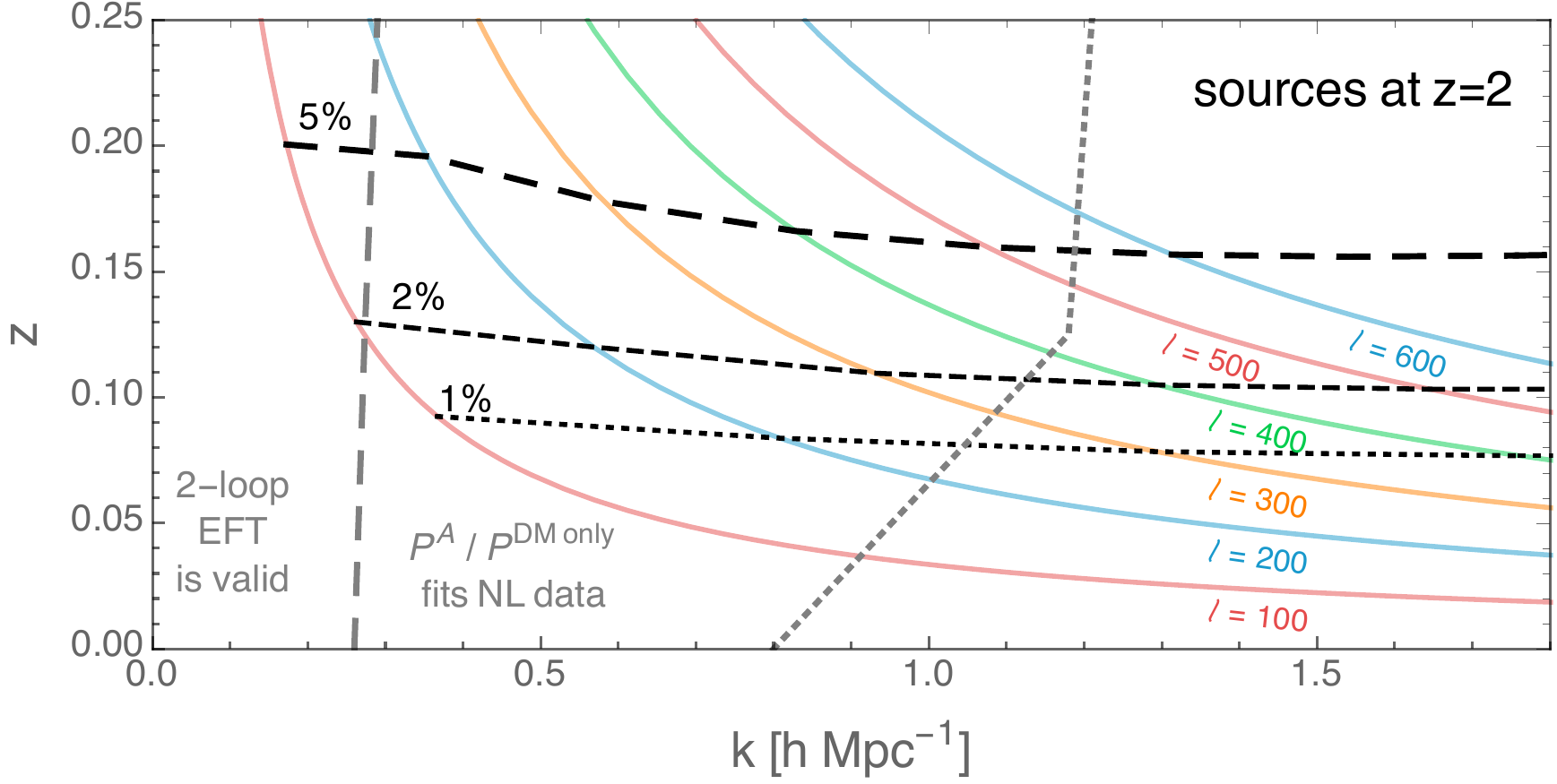} 
\end{tabular} 
\caption{\footnotesize Size of the contribution to the lensing potential \eqn{clpsi} for lensing of sources at $z=1$ and $z=2$.  The solid lines are the curves in the $(k,z)$-plane where the integral $C_{\ell}^{\psi}$ is evaluated for different $\ell$.   In the region to the left of the gray dashed line, the two-loop EFT prediction for $P^{A}_{\text{EFT-2-loop}}$ is valid, based on a precise comparison to non-linear data \cite{Foreman:2015lca}.  In the region to the left of the gray dotted line, the ratio $P^{A}/P^{A, \text{DM only}}$ fits the non-linear data with $0.25\%$ error on the data (see \figref{cAfit}).  The curves marked $p\%$ delineate the regions of the plane which contribute $(100-p) \%$ (above) and $p \%$ (below) to the lensing integral \eqn{clpsi} for each $\ell$. }
\label{lensing1p}
\end{figure}

Notice that in the expression for $C_\ell^\psi$ in \eqn{clpsi} for a fixed $\ell$, the power spectrum must be integrated over a range of times $a ( \chi )$ and a range of wavenumbers $k = \ell / \chi$.  As we have discussed, at a given time, the EFTofLSS can only be trusted up to a certain wavenumber.  Thus, we would like to examine, for a given $\ell$, how much of the integrand in \eqn{clpsi} can be trusted within the two-loop computation that we have presented above.  To do that, we follow \cite{Foreman:2015uva}.    For the moment, consider a universe (or simulation) without baryons.  The basic procedure is to first compute the $C_\ell^\psi$ with the non-linear power spectrum from Halofit \cite{Takahashi:2012em} for our fiducial WMAP3 cosmology to get an estimate of which region of modes and redshifts contributes $95\%$ and $5\%$ of the integral.  Then, knowing where the EFTofLSS fails at each redshift, we can deduce for which multipoles $\ell$ we can reliably compute $C_\ell^\psi$ with less than a $5\%$ error (and $2\%$ and $1\%$ as shown in \figref{lensing1} and \figref{lensing1p}).

The situation is even better than this, however.  We know that we can obtain a $5\%$ error by simply setting $P_{\Phi} = 0$ in the integral \eqn{clpsi} after the EFTofLSS fails (call this scale $k_{\rm fit}^P(a)$ for concreteness).  However, if we instead use the Halofit model,\footnote{Or any other emulator, for example the recent one developed for Euclid \cite{Knabenhans:2018cng}.} which is about $10\%$ different from the true power spectrum that includes baryonic effects, in the integral \eqn{clpsi} for the modes \emph{after} the EFTofLSS fails, then we will effectively have a $0.5\%$ error on the computation.  This small theoretical error is much smaller than the errors on the Planck lensing data \cite{Aghanim:2018oex}, and so would not be expected to improve the Planck analysis.  In fact, the errors on the Planck data are such that using Halofit for the entire power spectrum in \eqn{clpsi} is sufficient for the Planck analysis~\cite{Aghanim:2018oex}.  {However, the situation is different for CMB-S4, which has much smaller errors (see~\cite{Chung:2019bsk}, for example, for a comparison of the errors)}.

In \figref{lensing1} and \figref{lensing1p}, we show the regions of validity of the EFT relevant for the computation of the lensing power spectrum for the CMB (assuming the last scattering surface at $z\sim 1100$ as a single lens source plane) and for photons originating from sources at $z=1$ and $z=2$.  The solid, colored curves are the paths in the $(k,z)$ plane that must be integrated over for a given $\ell$ in \eqn{clpsi}.    In the region to the left of the gray dashed line, the two-loop EFT prediction for $P^{A}_{\text{EFT-2-loop}}$ is valid, based on a precise comparison to non-linear data \cite{Foreman:2015lca}.  In the region to the left of the gray dotted line, the ratio $P^{A}/P^{A, \text{DM only}}$ fits the non-linear data with $0.25\%$ error on the data (see \figref{cAfit}).  The curves marked $p\%$ delineate the regions of the plane which contribute $(100-p) \%$ (above) and $p \%$ (below) to the lensing integral \eqn{clpsi} for each $\ell$.

Looking at \figref{lensing1} and \figref{lensing1p}, we see that the two-loop EFT contributes with less than $5\%$ error to $C_{\ell}^{\psi}$ for $\ell \lesssim 600$ for CMB photons, $\ell \lesssim 100$ for $z=1$ sources, and $\ell \lesssim 160$ for $z = 2$ sources. The ratio fits the non-linear data much better.  If the dark-matter power spectrum is known, then one can use the ratio to compute $C_{\ell}^{\psi}$ with less than $5\%$ error for $\ell \lesssim 1600$ for CMB photons, $\ell \lesssim 300$ for $z=1$ sources, and $\ell \lesssim 550$ for $z=2$ sources for $0.25 \%$ error on the ratio of the OWLS simulation data.  As just discussed, however, these errors can be made much smaller by using an approximation of the non-linear power spectrum (such as Halofit) for the modes after the EFT fits fail.

This leads us to the following prescription for computing the baryon correction to the lensing power spectrum with a reduced theoretical error using the two-loop EFT prediction starting from a simulation power spectrum.  Since we are focused on baryonic effects in this paper, we assume that the non-linear dark-matter power spectrum $P_{\rm NL}^{\text{DM only}} ( a , k)$ is known.  This can be taken from the EFTofLSS, from a simulation, or from an emulator {under the conditions that the prediction is accurate enough}.  Then using this, we fit the ratio as described near \eqn{ratiofittingfns} to obtain $R^A_{\rm EFT} ( a , k) $, which is valid up to a high scale $k_{\rm fit}^{ R} ( a )$.  To compute the lensing power spectrum without the effects of baryons, $C_{\ell}^{\psi, \text{DM only}}$, we take $P^A ( a , k) \rightarrow P_{\rm NL}^{\text{DM only}} ( a , k) $ in \eqn{potentialps}, and to compute the lensing power spectrum with the effects of baryons, $C_{\ell}^{\psi,A}$, we take $P^A ( a , k) \rightarrow P_{\rm NL}^{\text{DM only}} ( a , k) R^A_{\rm EFT} ( a , k)$ in \eqn{potentialps}.  {Since this last replacement is only valid for $k \lesssim k_{\rm fit}^{R} ( a )$, we use a linear extrapolation of $R^A_{\rm EFT} ( a , k)$ from $k_{\rm fit}^{R} ( a )$ to $2 k_{\rm fit}^{R} ( a )$, and then the constant value $R^A_{\rm EFT} ( a , 2 k_{\rm fit}^{R} ( a ))$ above $2 k_{\rm fit}^{R} ( a )$, as an estimate of the full power spectrum at higher wave numbers.}   {Had we not performed the extrapolation, the error would have scaled as the integral in (\ref{clpsi}) with $P^A$ replaced by $\left(R^A_{\rm EFT} ( a , k)-1\right)P_{\rm NL}^{\text{DM only}} ( a , k)$, and integrated in the range of $k$'s above $k_{\rm fit}^{R} ( a )$. We can then bound $\left(R^A_{\rm EFT} ( a , k)-1\right)\lesssim 0.1$, as can be seen from hydrodynamical simulations (see e.g.~\cite{vanDaalen:2019pst}). Finally, we assume that using the extrapolation for $R^A_{\rm EFT} ( a , k)$ rather than its maximum} gives us another factor of two smaller error bars.  {Thus, to find the error for each $\ell$, we look at the contribution plots \figref{lensing1} and \figref{lensing1p}, find what the percentage contribution is where the solid curve crosses the ratio-fit curve (gray dotted), and multiply by  $0.1\times 0.5=0.05 $.  We call this error the `high-$k$ approximation' error, since it comes from using an approximate form of the power spectrum for wavenumbers larger than $k_{\rm fit}^R(a)$.  This error, along with the estimated error coming from the three-loop EFT terms in the ratio fit, are plotted as the gray and teal bands, respectively, in \figref{lensing3}.} {In \appref{changekfitapp}, we consider the effect of using smaller values of $k_{\rm fit}^R(a)$.}

The resulting ratio of the adiabatic and the dark-matter-only lensing-potential power spectra is shown in \figref{lensing3}.\footnote{{We thank S. Foreman for communication comparing our results with \cite{Chung:2019bsk}.}}  Including baryonic effects in the power spectrum clearly has more than a percent level effect on the lensing power spectrum, and for the specific simulation that we studied in this paper, the effect of baryons is larger than the estimated CMB-S4 error bars for $\ell \gtrsim 1000$.  For reference, we have also included the direct numerical integration of the simulation data in the plot, which gives us an indication of the systematic theoretical error in our calculation.  

We would like to stress, however, that although we fit to a specific baryonic simulation in this work, we believe that our strategy is much more flexible than using baryonic simulations to match observed data.  Regardless of the true nature of baryonic physics in our universe, we have shown that the two-loop EFT is able to capture these effects in a set of time-dependent parameters, and that this can be reliably used to compute observables like the lensing power spectrum.  {Explicitly, {if one wants to use dark-matter simulations to analyze real data}, one can use \eqn{dataratio} with unspecified counterterms in the expression for $ R^\sigma_{\rm EFT} |_2$ to get the total power spectrum in the presence of baryons, and from there the lensing potential.  Then one can extract the counterterms as well as the cosmological parameters directly by fitting to data (as recently done for galaxy clustering in  \cite{DAmico:2019fhj,Ivanov:2019pdj,Colas:2019ret}).}

%
%
 %
 %

\section{Conclusion}
	\label{sec:Conclusion}

In this work, we have discussed many important effects of baryonic physics on large-scale clustering.  First, in \secref{theory}, we showed how a new counterterm, proportional to the relative velocity $v^i_I$ and not derivatively suppressed, is generically allowed in the EFT of two fluid-like species, showing up specifically in the equation of motion for the isocurvature mode.  This term is consistent with the separate conservation of mass of each species, the conservation of total momentum, and diffeomorphism invariance.  We presented our arguments in two ways, first in a more bottom-up EFT construction based on symmetries in \secref{theory}, and second in a more top-down approach based directly on the Einstein equations and the conservation of the pseudo stress tensor in \appref{GRsec}.   We then explicitly constructed the effective force and effective stress tensors needed to compute the various power spectra of this system up to two-loop order.  

In \secref{linrelvelsec}, we examined in much more detail the effects of the new linear counterterm proportional to $v^i_I$.  We first pointed out that the new counterterm is in fact \emph{necessary} to have a well defined mathematical framework.  This is because the one-loop term shown in \eqn{eq:UVdiv} is actually UV divergent for typical small scale CDM behavior, explicitly going like $(\log \Lambda_{\rm UV})^3$ where $\Lambda_{\rm UV}$ is a UV cutoff, as we discussed near \eqn{uvdivergencereal}, {and this counterterm is needed to absorb this divergence.}  Apart from being necessary to cancel this UV divergence, the new counterterm also has a finite contribution that changes the linear equation of motion for the isocurvature mode.  In \secref{sec:realuniverse} and \secref{sec:sphcoll}, we estimated the size of the finite contribution using perturbation theory and {the non-linear solution of a one-dimensional UV model}, respectively, and showed that the effect of this counterterm on the power spectrum is subdominant to other higher order effects that we neglected in this paper, showing that it can safely be ignored at the level at which we work. We also estimated that its effect can be probably neglected for future observations. Finally, in \secref{linearevo} and \appref{pertwithlinctsec}, we discussed how, even when present, the new counterterm can be included consistently in perturbation theory.  The end result is that the linear equation for the decaying isocurvature mode is modified by an unknown time-dependent function, such that in the perturbative expansion of this mode, all of the time-dependent coefficients are unknown (although the $k$-dependent kernels are known).    

In \secref{Comparison}, we compared our two-loop EFT prediction to the OWLS-$AGN$ simulation, specifically to the ratio of power spectra in a simulation including baryonic effects to a simulation with only dark matter, at $18$ different redshifts between $z =0$ and $z=4$.   We found quite a remarkable fit to the data (see \figref{cAfit}), and we discussed why the fit of the ratio at two-loops has a larger $k$-reach than fitting the power spectrum directly (for example the ratio fits {well up} to $k \approx 0.8 \unitsk$ at $z = 0$, and $k \approx 3.6 \unitsk$ at $z = 4$).  Between $z \approx 2$ and $z \approx 3$, we saw that the baryon EFT parameter associated with the $k^2 P_{11}(k)$ counterterm starts to deviate significantly from the analogous CDM parameter, signaling the onset of star-formation physics.  This shows that the EFTofLSS provides a powerful analytic description of baryonic effects in LSS.

Finally, in \secref{lensing}, we used our two-loop EFT prediction to compute the power spectrum of CMB lensing for the simulation studied earlier (see \figref{lensing3}).  {We did this simply by predicting the correction due to baryonic physics to a dark-matter power spectrum (which could be obtained by an $N$-body simulation for example).} Given that baryonic effects are expected to be important for analyzing CMB lensing with CMB-S4 data, we provided an analytic recipe for parametrizing baryonic effects on CMB lensing.  We showed that the two-loop EFT correctly predicts the lensing power spectrum, including baryonic effects, up to $\ell \approx 2000$ with a theoretical error that is about $1/4$ of the size of the CMB-S4 errors at $\ell = 2000$.  Since our analytic approach ultimately does not rely on baryonic simulations, we view this as an important and actionable step forward for our understanding of {the effect of baryonic physics on the lensing observable.}

Now that we have shown that the two-loop EFT correctly describes baryonic physics, both in the matter power spectrum and in lensing, we mention a few interesting directions for future study.  There are many potential improvements to the actual computation of the lensing power spectrum.  {First, as discussed near \eqn{taylorexpandratio}, terms which do not involve counterterms cancel in the ratio $R^\sigma_{\rm EFT}$.  This means that, using only two-loop integrals, one can compute $R^\sigma_{\rm EFT}|_3$, since the three-loop integrals will cancel in the ratio.  This could provide a way to extend the EFT prediction for the ratio to higher wavenumbers with relatively little extra computational effort than was used in this paper.}
Second, one could use an updated template for the non-linear dark-matter-only power spectrum, such as the one developed in \cite{Knabenhans:2018cng} for Euclid.  Third, one could study the time dependence of the EFT parameters (as in \figref{cbcc}) for various different baryonic simulations and come up with a reliable prescription for the time dependence of the parameters.  

%
%
 %
 %

\subsubsection*{Acknowledgments}

We are greatly indebted to J. Schaye and M. P. van Daalen for providing us with the data of their simulations. We thank S. Foreman for very useful conversations and for sharing some of his Mathematica codes, and J.~J.~Carrasco, A.~Refregier, and M. Zaldarriaga for interesting conversations. M.L. thanks M. Millea for interesting conversations during the completion of this work.  L.S. is partially supported by the Simons Foundation Origins of the Universe program (Modern Inflationary Cosmology collaboration) and by NSF award 1720397.  M.L. acknowledges the Northwestern University Amplitudes and Insight group, Department of Physics and Astronomy, and Weinberg College for support. 

\appendix

%
%
%
\section{Equations of motion from pseudo stress tensor} \label{GRsec}

%
%
\subsection{Einstein tensor}
 Here, we collect some useful expressions for the Einstein tensor $G^\mu{}_\nu \equiv R^\mu{}_\nu - \half R \delta^\mu{}_\nu$, where $R_{\mu \nu}$ is the Ricci tensor, and $R$ is its trace,  for the metric \eqn{metric} at background, first, and second orders.  As we will see, this is all we will need since there are at most two derivatives in the Einstein tensor, so terms with more than two fields are suppressed by relativistic corrections. 
 We will typically keep track of $\Phi$ and $\Psi$ in the expressions for the Einstein tensor, connection,  etc., but will take $\Phi = \Psi$ in our final equations, as is justified in our setting.  
For the background, we have (here and elsewhere, an overbar denotes a background value),
 \be
 \bar G_{ij} = -a^2 ( 3 H^2 + 2 \dot H) \delta_{ij} \ , \quad \bar G_{00} = 3 H^2 \ , \quad \bar G^\mu{}_\mu = -6 (2 H^2 + \dot H) \  ,
 \ee
 at linear order, we have
 \begin{align}
 \begin{split} \label{linearG}
&   G_{\rm L}{}^0{}_0 = 6 H (H \Phi + \dot \Psi ) - 2 a^{-2} \partial^2 \Psi  \ , \\
&   G_{\rm L}{}^0{}_i = - 2 \partial_i (H \Phi + \dot \Psi )  \ , \\
& G_{\rm L}{}^i{}_j = 2  \delta_{ij}   \left\{ (3 H^2 + 2 \dot H) \Phi  + H ( \dot \Phi + 3 \dot \Psi )  + \ddot \Psi      \right\}  + a^{-2} (\delta_{ij} \partial^2 - \partial_i \partial_j ) ( \Phi - \Psi ) \ ,
 \end{split}
 \end{align}
 and at second order, we have 
  \begin{align}
 \begin{split} \label{quadG}
&   G_{(2)}{}^0{}_0 = - 3(4 H^2 \Phi^2 + 4 H \dot \Psi ( \Phi - \Psi ) + \dot \Psi^2   ) - 8 a^{-2} \Psi \partial^2 \Psi -3a^{-2} (\partial \Psi)^2  \ , \\
& G_{(2)}{}^0{}_i  = 8 H \Phi \partial_i \Phi + 2 \dot \Psi \partial_i (\Phi - 2 \Psi ) + 4 (\Phi - \Psi ) \partial_i \dot \Psi  \ , \\
& G_{(2)}{}^i{}_j  = \delta_{ij} \left\{  - 4( 3 H^2 + 2 \dot H )  \Phi^2  - 2 \dot \Phi \dot \Psi  + \dot \Psi^2 - 4 H ( -3 \Psi \dot \Psi + 2 \Phi \dot \Phi + 3 \Phi \dot \Psi     ) - 4 (\Phi - \Psi )  \ddot \Psi  \right\} \\
& \hspace{1in} + a^{-2} \Big\{ \delta_{ij} ( \partial_k \Phi \partial_k \Phi + 2  \Psi \partial^2 ( \Phi - \Psi) )   \\
& \hspace{1in} - 2 \Psi \partial_i \partial_j ( \Phi - \Psi ) + \partial_i \Psi \partial_j \Psi - \partial_i \Phi \partial_j \Phi - \partial_i \Phi \partial_j \Psi - \partial_i \Psi \partial_j \Phi  \\
& \hspace{1in}  + \left(  \partial_i \partial_j -  \delta_{ij} \partial^2  \right) ( \Phi^2 +  \Psi^2  )   \Big\}   \ . 
  \end{split}
 \end{align}

%
%
\subsection{Pseudo stress tensor in FRW}

Here, we show that in the non-relativistic, Newtonian limit ($v/c \ll 1$ and $c \, \partial_i / H \gg 1$, where $c$ is the speed of light, which we take to be equal to unity in this work), the Einstein equations\footnote{Here, $\mpl^2 = 1/(8 \pi G_N)$ is the Planck mass squared, and $G_N$ is the Newton constant.}
 \be \label{einsteinequations}
 \mpl^2 G^\mu{}_\nu = T^\mu{}_\nu \ ,
 \ee
for the FRW metric in the Newtonian gauge \eqn{metric} can be written as the conservation of a pseudo stress tensor $t^\mu{}_\nu$ in the sense that
  \be \label{pseudocons2}
 a^{-3} \partial_\mu ( a^3 t^{\mu}{}_\nu ) \approx 0 \ , 
 \ee
where here and elsewhere, we use the symbol $\approx$ to mean equal up to relativistic corrections in the Newtonian limit.  In general, the $\nu = 0$ equation is of order $H \partial^2 \Phi$, while the $\nu = i$ equation is of order $H^2 \partial_i \Phi$.  This is similar to what is done in \cite{weinberggravandcosmo}, for example, in flat space.

To proceed, we write the various terms in \eqn{einsteinequations} in powers of perturbations like 
\begin{align}
\begin{split}
& G^\mu{}_\nu = \bar G^\mu{}_\nu + G_{\rm L}{}^\mu{}_\nu + G_{\rm NL}{}^\mu{}_\nu  \ , \\
& T^{\mu}{}_\nu = \bar T^\mu{}_\nu + \delta T^\mu{}_\nu \ ,
\end{split}
\end{align}
where $G_{\rm L}{}^\mu{}_\nu$ is linear in perturbations, $G_{\rm NL}{}^\mu{}_\nu$ is non-linear, and $\delta T^\mu{}_\nu$ contains all the perturbations of the stress-energy tensor. Writing the spatial part of the background stress-energy  tensor as $\bar T_{ij} = \delta_{ij} a^2 \bar p$, the zeroth order Einstein equations give
 \be \label{backgroundee}
 3 H^2 \mpl^2 = \bar T_{00} \andd  -2 \dot H \mpl^2 = \bar T_{00} +\bar  p \ ,
 \ee
 while the perturbed Einstein equations are 
 \be \label{perturbedeinstein}
 \mpl^2 ( G_{\rm L}{}^\mu{}_\nu + G_{\rm NL}{}^\mu{}_\nu ) = \delta T^{\mu}{}_\nu \ . 
 \ee

 Next, since the Bianchi identity $\nabla_\mu G^\mu{}_\nu  = 0$ is a strict identity, regardless of any equations of motion, it is true at each order in perturbations.\footnote{The covariant derivative is $\nabla_\mu G^\nu{}_\rho = \partial_\mu G^\nu{}_\rho + \Gamma^\nu_{\mu \lambda} G^\lambda{}_\rho  - \Gamma^\lambda_{\mu \rho} G^\nu{}_\lambda. $ }  In particular, it is true at first order
 \be \label{linearbianchi} 
 \bar \nabla_\mu G_{\rm L}{}^\mu{}_\nu + \nabla^{\rm L}_\mu \bar G^\mu{}_\nu = 0 \ , 
 \ee
 where we have expanded the covariant derivative $\nabla_\mu = \bar \nabla_\mu + \nabla^{\rm L}_\mu + \dots$ (the actual derivative $\partial_\mu$ only appears in $\bar \nabla_\mu$).  Now, we evaluate the two terms in \eqn{linearbianchi}.  First, we look at
 \be \label{linearnabla}
 \bar \nabla_\mu  G_{\rm L}{}^\mu{}_\nu  = a^{-3} \partial_\mu ( a^3 G_{\rm L}{}^\mu{}_\nu )  - \bar \Gamma^\rho_{\mu \nu} G_{\rm L}{}^\mu{}_\rho  \ .
 \ee
 For the metric \eqn{metric}, we have
 \begin{align}
 \begin{split}
& \bar \Gamma^\rho_{\mu i} G_{\rm L}{}^\mu{}_\rho = 0   \\
&  \bar \Gamma^\rho_{\mu 0} G_{\rm L}{}^\mu{}_\rho =  6 H ( (3 H^2  + 2 \dot H) \Phi + H ( \dot \Phi + 3 \dot \Psi)  + \ddot \Psi  ) + 2 H  a^{-2} \partial^2 ( \Phi - \Psi ) \ .
\end{split}
 \end{align}
 Since the second line above will be in the equation involving $\partial_0 G_{\rm L}{}^0{}_0 \sim H \partial^2 \Psi$, only the last term is relevant in the Newtonian limit.  However, since we will eventually set $\Psi = \Phi$, we drop that term now to get
 \be \label{linearnabla2}
  \bar \nabla_\mu  G_{\rm L}{}^\mu{}_\nu  \approx a^{-3} \partial_\mu ( a^3 G_{\rm L}{}^\mu{}_\nu )  \ .
 \ee
 
 The next term we need in \eqn{linearbianchi} is $\nabla^{\rm L}_\mu \bar G^\mu{}_\nu$.  We have
 \be
 \nabla^{\rm L}_\mu \bar G^\mu{}_0 = - 6 \dot H \dot \Psi \ , \quad \text{and} \quad   \nabla^{\rm L}_\mu \bar G^\mu{}_i = - 2 \dot H \partial_i \Phi \ ,
 \ee
 which to leading order is (no sum on $i$),
 \be
 \nabla^{\rm L}_\mu \bar G^\mu{}_\nu \approx - 2 \dot H \delta_{\nu i} \partial_i \Phi \ .  
 \ee
 Then, combining this with \eqn{linearnabla2} we have
 \begin{align}
 \begin{split}
 a^{-3} \partial_\mu ( a^3 G_{\rm L}{}^\mu{}_\nu )  - 2 \dot H \delta_{\nu i } \partial_i \Phi  \approx 0  \ ,
\end{split}
 \end{align}
 and using the Einstein equations \eqn{perturbedeinstein} to replace $G_{\rm L}{}^\mu{}_\nu$ gives,
 \be \label{intermediate9}
  a^{-3} \partial_\mu ( a^3 ( \mpl^{-2} \delta T^\mu{}_\nu - G_{\rm NL}{}^\mu{}_\nu) )  - 2 \dot H \delta_{\nu i } \partial_i \Phi \approx 0 \ .
 \ee
 
 Finally, we determine which parts of $G_{\rm NL}{}^\mu{}_\nu$ contribute at leading order in the Newtonian limit.  First of all, $G^\mu{}_\nu$ only contains two derivatives, and since $G_{\rm NL}{}^\mu{}_\nu$ starts at second order, it has at least two powers of the gravitational potentials.  Thus, since the $\nu = 0$ equation contains terms going like $H \partial^2 \Phi$, $G_{\rm NL}{}^\mu{}_0$ is always a higher relativistic order term in that equation.  In the $\nu = i$ equation, however, the term that appears is $\partial_j G_{\rm NL}{}^j{}_i$, and since we have $G_{(2)}{}^j{}_i \sim  \partial_i \Phi \partial_j \Phi$, this term goes like $\partial_i \Phi  \partial^2 \Phi$, which is a leading term in the relativistic counting ($\partial_0 G_{\rm NL}{}^0{}_i$ has only one spatial derivative, and so it is negligible in the leading non-relativistic order).  All higher orders in $G_{\rm NL}{}^j{}_i$ are subleading, though.  Thus, we have shown that \eqn{intermediate9} is exactly of the form  \eqn{pseudocons2}, with\footnote{ The indices on actual tensors, like the stress-energy tensor $T_{\mu \nu}$ and the Einstein tensor $G_{\mu \nu}$, are raised and lowered by the metric $g_{\mu \nu}$, but indices on non-tensor quantities, like the Kronecker delta function $\delta_{ij}$, spatial derivatives $\partial_i \equiv \partial / \partial x^i$, the fluid velocity $v^i$, the momentum density $\pi^i$, and the pseudo-tensor $t^i{}_j$, will not be raised or lowered with the metric.  This is why the placement of indices is not always the same on both sides of an equation.}  
  \begin{align}
  \begin{split}
& t^0{}_0 \approx \delta T^0{}_0 \ , \quad t^0{}_i \approx \delta T^0{}_i \ , \quad t^i{}_0 \approx \delta T^i{}_0 \ , \\
&t^i{}_j \approx \delta T^i{}_j - \mpl^2 G_{\rm (2)}{}^i{}_j - 2 \mpl^2 \dot H \delta^i{}_j \Phi  \ ,
\end{split}
 \end{align}
 and 
\be \label{gnl}
a^2 G_{\rm (2)}{}^i{}_j \approx \delta_{ij}  \partial_k \Phi \partial_k \Phi  - 2 \partial_i \Phi \partial_j \Phi   +  2 \left(  \partial_i \partial_j -  \delta_{ij} \partial^2  \right)  \Phi^2   \ ,
\ee
 for $\Phi = \Psi$.

%
%
\subsection{Equations of motion}

Next, we use the conservation of the pseudo stress tensor derived above to derive the equations of motion for the dark-matter and baryon fluids.  We start by writing the total stress-energy tensor as the sum of the CDM and baryon stress-energy tensors
\be
T^\mu{}_\nu = T_c^\mu{}_\nu + T_b^\mu{}_\nu \ . 
\ee
{We first consider the $\nu = 0$ case of \eqn{pseudocons2}.  In the Newtonian limit, which is what we use for the long-wavelength modes, one can take energy conservation and mass conservation to be the same \cite{Baumann:2010tm}.  For the case of dark matter and baryons, there is a separate mass conservation for each species.  Taking into account this additional conservation law, we can split the $\nu = 0$ equation of \eqn{pseudocons2} into two separate equations    }
\be \label{conteqs1}
a^{-3} \partial_0 ( a^3 \delta T_c^0{}_0 ) + \partial_i \delta T_c^i{}_0 = 0   \ , \quad \text{and} \quad a^{-3} \partial_0 ( a^3 \delta T_b^0{}_0 ) + \partial_i \delta T_b^i{}_0 = 0  \ . 
\ee
These equations express the conservation of the total number of particles of each species separately, and we will write them in a more familiar form below.  

The $\nu = i$ equation in \eqn{pseudocons2}, on the other hand, gives
\be \label{totalmomcons1}
a^{-3} \partial_0 \left( a^3 \left( \delta T_c^0{}_i + \delta T_b^0{}_i \right) \right) + \partial_j \left( \delta T_c^j{}_i + \delta T_b^j{}_i \right) + T^{00} \partial_i \Phi = 0  \ , 
\ee
where we have used that 
\be
- \partial_j \left(    \mpl^2 G_{\rm NL}{}^j{}_i + 2 \mpl^2 \dot H \delta_{ji}\Phi  \right) = T^{00} \partial_i \Phi  \ ,
\ee
which can be shown using the expression for $G_{\rm NL}{}^j{}_i$ in \eqn{gnl}, the background equation for a pressureless field $ -2 \dot H \mpl^2 = \bar T^{00} $, and the $(00)$ Einstein equation (Poisson equation), which is, as usual, at leading relativistic order,
\be
\delta T^{00} = 2 a^{-2} \mpl^2 \partial^2 \Phi \ . 
\ee
{Next, we observe that we should have two separate equations of motion for baryons and dark matter. While this is quite intuitive, one can formally establish this by noting that in the  early universe the two fluids are weakly coupled, and so they give rise to two degrees of freedom. Clearly, each equation of motion should be invariant with respect to diffeomorphisms, as they come, at least formally, from the variation of a diffeomorphism-invariant action.}
Therefore, we can write the separate evolution equations for $\delta T_c^0{}_i$ and $\delta T_b^0{}_i$ in a general way that automatically satisfies the total conservation of momentum \eqn{totalmomcons1}  as 
\begin{align}
\begin{split} \label{momconseq1}
& a^{-3} \partial_0 \left( a^3  \delta T_c^0{}_i  \right) + \partial_j  \delta T_c^j{}_i  + \lambda T^{00} \partial_i \Phi  + \varphi_i = 0 \ , \\
& a^{-3} \partial_0 \left( a^3  \delta T_b^0{}_i  \right) + \partial_j  \delta T_b^j{}_i  + (1- \lambda) T^{00} \partial_i \Phi  - \varphi_i = 0  \ , 
\end{split}
\end{align}
{for any time-dependent $\lambda$ and functional of the fields $\varphi_i$, and then we can constrain the forms of $\lambda$ and $\varphi_i$ by demanding diffeomorphism invariance for each of the two equations.  Notice that at this point, nothing specific has been assumed about each species' equation of motion, since $\varphi_i$ is generic.   Then, by demanding that each equation is separately diffeomorphism invariant, we are assuming that {the two species are independent degrees of freedom, as it is evident for dark matter and baryons by thinking about their early universe dynamics.}}

{The subset of diffeomorphisms that preserve the Newtonian gauge and that is relevant for the Newtonian limit are the so-called Galilean transformations}
\be
t \rightarrow t+a^2 n^i(t) x^i \  , \quad \text{and} \quad x^i \rightarrow x^i + n^i ( t ) \ , 
\ee
for generic time dependent $n^i ( t )$, which act on the terms present in the equations of motion as\footnote{These transformations are simply inherited from the relevant diffeomorphism transformation rules.} 
\begin{align} 
\begin{split} \label{fluidreps}
& \partial_i  \rightarrow \partial_i \ , \quad  \partial_0   \rightarrow \partial_0 - \dot n^i    \partial_i  \ ,  \quad  \Phi    \rightarrow \Phi  - a^2 ( \ddot n^i  + 2 H \dot n^i  ) x^i   \ , \\ 
 & \delta T_\sigma^{00} \rightarrow \delta T_\sigma^{00} \ , \quad \delta T_\sigma^{0i} \rightarrow \delta T_\sigma^{0i} + \dot n^i  T_\sigma^{00}  \ , \quad  \delta T_\sigma^{ij} \rightarrow \delta T_\sigma^{ij} + \dot n^j \delta T_\sigma^{i0} + \dot n^i \delta T_\sigma^{0j} + \dot n^i \dot n^j  T_\sigma^{00}  \ ,
\end{split}
\end{align}
where here and elsewhere, $\sigma \in \{ c, b\}$, {and where we neglected terms that contribute only at relativistic level}.  It is straightforward to see that the continuity equations \eqn{conteqs1} are invariant, so we move to the momentum density conservation equations \eqn{momconseq1}.  Imposing that each equation in \eqn{momconseq1} is Galilean invariant and defining the transformation $\varphi_i \rightarrow \varphi_i + \Delta \varphi_i$, we find {the single constraint}
\be
a^2 ( \ddot n^i + 2 H \dot n^i ) \left( T_c^{00} - \lambda T^{00} \right) + \Delta \varphi_i = 0 \ .
\ee
From this, we learn that we can set $\varphi_i \equiv \varphi \partial_i \Phi + \gamma^i$, where $\varphi$ and $\gamma^i$ are Galilean scalars, which then imposes the constraint
\be
\lambda T^{00} + \varphi = T_c^{00} \ . 
\ee
This means that the equations of motion \eqn{momconseq1} are
\begin{align}
\begin{split} \label{momconseq2}
& a^{-3} \partial_0 \left( a^3  \delta T_c^0{}_i  \right) + \partial_j  \delta T_c^j{}_i  + T^{00}_c \partial_i \Phi  + \gamma^i = 0 \ ,  \\
& a^{-3} \partial_0 \left( a^3  \delta T_b^0{}_i  \right) + \partial_j  \delta T_b^j{}_i  + T^{00}_b \partial_i \Phi  - \gamma^i = 0  \ . 
\end{split}
\end{align}
In order to write the equations \eqn{conteqs1} and \eqn{momconseq2} in a more familiar form, we parametrize the stress tensors as 
\be \label{stressdef2}
T_\sigma^0{}_0 = -\rho_\sigma \ , \quad T_\sigma^0{}_i =  a \pi^i_\sigma \ , \quad \text{and} \quad  T_\sigma^i{}_j =  \frac{ \pi_\sigma^i \pi_\sigma^j }{\rho_\sigma}  + \tau_\sigma^{ij} \ ,
\ee
which finally gives the standard \eqns{finalcont1}{momeqb1}.  {The form of this parametrization is of course general, but the usefulness comes from two realizations.  The first is that the effect of $\tau^{ij}_\sigma$ on $\rho_\sigma$ and $\pi^i_\sigma$ is perturbative at long distances, and the second is that dark matter and baryons do not move too much in the history of the universe.  This means that $\tau^{ij}_\sigma$ can be written as local-in-space powers and derivatives of $\rho_\sigma$ and $\pi^i_\sigma$ \cite{Baumann:2010tm}.}  In turn, this means that the counterterms for the EFTofLSS come in through the effective stress tensors $\tau^{ij}_c$ and $\tau^{ij}_b$, and the effective force $\gamma^i$, which are all Galilean scalars.  The important new possibility is a term $\gamma^i \sim H  \pi^i_I$, which is allowed by the symmetries, and is in fact generically needed to cancel UV divergences in the one-loop power spectrum, as we explicitly show in \secref{relvel}.  

Finally, we stress that we did not include counterterms in the continuity equations \eqn{finalcont1}.  This is because our system is coupled to gravity, and the stress-energy pseudo-tensor $t^{\mu}{}_\nu$ in \eqn{pseudocons2} is symmetric.  Operationally, what this means is that whatever field appears in the gradient $\partial_i$ in the continuity equations \eqn{finalcont1} should appear in the time derivative in the momentum equations \eqns{momeqc1}{momeqb1}, because $t^0{}_i = -a^2 t^i{}_0$.  For example, if one wanted to add counterterms of the form $- a^{-1} \partial_i F_\sigma^i$ to the right-hand side of \eqn{finalcont1}, then we should have $\partial_0 (\pi^i_\sigma + F^i_\sigma ) $ appearing in the momentum equations \eqns{momeqc1}{momeqb1}.  Ultimately, this means that the equations take the same form as those that we have presented.

%
%
%
\section{Perturbative solutions for two fluids} \label{solsec}

%
\subsection{General equations} \label{geneqsapp}

Here we study the structure of the perturbative solutions for two fluids. 
We start by neglecting the role of the counterterms, to which we return in \secref{pertwithlinctsec}. Therefore, neglecting counterterms, we start with the equations of motion in the adiabatic-isocurvature basis.  The continuity equations are 
\begin{align}
\begin{split}   \label{conteq2}
& a \delta'_A - \Theta_A = \partial_i \left(  \delta_A \frac{\partial_i \Theta_A}{\partial^2} + w_b w_c \delta_I \frac{\partial_i \Theta_I}{\partial^2} \right)  \ , \\
&  a \delta'_I  - \Theta_I = \partial_i \left(  \delta_A \frac{\partial_i \Theta_I}{\partial^2} + \delta_I \frac{\partial_i \Theta_A}{\partial^2}  + (w_b-w_c) \delta_I \frac{\partial_i \Theta_I}{\partial^2} \right)  \ .
\end{split}
\end{align}
To derive the two-derivative equations in the most convenient form, it is easiest to start with the momentum equations \eqn{momeqc1} and \eqn{momeqb1}, then use the continuity equation in the form \eqn{finalcont1}, from which we straightforwardly get
\begin{align}
\label{finaladiabaticeqn}
& a^2 \delta_A'' + \left( 2 + \frac{a \cH'}{\cH} \right) a \delta_A' - \frac{3}{2} \Om  \delta_A =     \partial_i \partial_j \Bigg\{ \frac{3}{2} \Om  \left(  \frac{\partial_i \delta_A}{\partial^2} \frac{\partial_j \delta_A}{\partial^2}  - \frac{\delta_{ij} }{2}  \frac{\partial_k \delta_A}{\partial^2} \frac{\partial_k \delta_A }{\partial^2}   \right) \\
& \hspace{.4in} +    (1 + \delta_A) \frac{\partial_i \Theta_A}{\partial^2} \frac{\partial_j \Theta_A}{\partial^2}    + 2 w_b w_c \delta_I  \frac{\partial_{i} \Theta_A}{\partial^2} \frac{\partial_{j} \Theta_I}{\partial^2}   + w_b w_c \left[1 + \delta_A + ( w_b-w_c) \delta_I  \right] \frac{\partial_i \Theta_I}{\partial^2} \frac{\partial_j \Theta_I}{\partial^2}   \Bigg\}  \ ,  \nonumber
\end{align}
and
\begin{align}
\label{finalisoeqn}
& a^2 \delta_I'' + \left( 2 + \frac{a \cH'}{\cH} \right) a \delta_I' =  \frac{3}{2}\Om   \partial_i \left( \delta_I \frac{\partial_i \delta_A}{\partial^2} \right) + \partial_i \partial_j \Bigg\{  (1 + \delta_A) \left( 2 \frac{\partial_{i} \Theta_A}{\partial^2} \frac{\partial_{j} \Theta_I}{\partial^2}   + ( w_b - w_c)  \frac{\partial_i \Theta_I}{\partial^2} \frac{\partial_j \Theta_I}{\partial^2}    \right) \nonumber   \\
& \hspace{.6in} + \delta_I \Bigg(  \frac{\partial_i \Theta_A}{\partial^2} \frac{\partial_j \Theta_A}{\partial^2}  +2 (w_b - w_c)   \frac{\partial_{i} \Theta_A}{\partial^2} \frac{\partial_{j} \Theta_I}{\partial^2}   + (w_b^2 - w_b w_c -w_c^2)  \frac{\partial_i \Theta_I}{\partial^2} \frac{\partial_j \Theta_I}{\partial^2}     \Bigg)    \Bigg\}  \ ,
\end{align}
where $\Om \equiv \bar \rho_A / ( 3 \mpl^2 H^2)$ is the time-dependent matter fraction, $\mpl$ is the Planck mass, and we have used the dimensionless velocity divergence 
\be
\Theta_\Upsilon \equiv - \partial_i v^i_\Upsilon  / \cH \ ,
\ee
where $\Upsilon \in \{ A , I\}$.  We have also used 
\be
\partial_i \Phi = \frac{3}{2} \Om \cH^2 \frac{\partial_i \delta_A}{\partial^2} \ , 
\ee
and the identity
\be
\partial_i \left( \delta_A \frac{\partial_i \delta_A}{\partial^2} \right) =  \partial_i \partial_j \left(  \frac{\partial_i \delta_A}{\partial^2} \frac{\partial_j \delta_A}{\partial^2}  - \frac{\delta_{ij}}{2}  \frac{\partial_k \delta_A}{\partial^2} \frac{\partial_k \delta_A }{\partial^2}   \right)   \ , 
\ee
to write the right-hand side of \eqn{finaladiabaticeqn} as a double total spatial derivative.  As we will see, the derivative structure on the right-hand sides of the above equations determines the UV structure of the solutions.

%
\subsection{Linear solutions and EdS expansion} \label{linearandedsapp}
In $\Lambda$CDM, we expand the linear solutions as    
\begin{align}
\begin{split} \label{adiabaticisolinearsols}
&\delta_A^{(1)} ( a , \kvec ) = \frac{D_{A_+} ( a )}{D_{A_+} ( a_0 )} \delta_{A_+}^{(1)} ( \kvec ) +  \epsilon^5  \frac{ D_{A_-} ( a ) }{  D_{A_-} ( a_0 )   } \delta_{A_-}^{(1)} ( \kvec ) \ , \\  
&\delta_I^{(1)} ( a , \kvec ) = \epsilon^2 \frac{ D_{I_+} ( a )}{D_{I_+} ( a_0 )} \delta_{I_+}^{(1)} ( \kvec ) + \epsilon^3 \frac{D_{I_-} ( a ) }{D_{I_-} ( a_0 )} \delta_{I_-}^{(1)} ( \kvec )  \ ,
\end{split}
\end{align}
where $\epsilon \equiv \sqrt{a_{ \rm in} / a_0}$.  Note that for standard adiabatic initial conditions from inflation, we have $\delta_{A_-}^{(1)}(\kvec) = 0$, but we keep that term for completeness here.  We have normalized the linear solutions so that at early times $a_{\rm in}$, where there is a sizable isocurvature mode, the various free functions contribute at comparable levels.  We typically have in mind $a_{\rm in} \sim 5 \times 10^{-3}$, a bit after recombination  when $\delta_A \approx \delta_I$, so we can think of $\epsilon \approx 0.07$.
 Neglecting radiation, the growth factors satisfy the second order equations
\begin{align}
 \label{reallineareqs1}
& a^2 D_{A_\pm}'' + \left( 2 + \frac{a \cH'}{\cH} \right) a D_{A_\pm}' - \frac{3}{2} \Om ( a ) D_{A_\pm} = 0 \ , \\
&   a^2 D_{I_\pm}'' + \left( 2 + \frac{a \cH'}{\cH} \right)  a D_{I_\pm}' = 0 \ .  \label{reallineareqs2}
\end{align}
The growing and decaying solutions to \eqn{reallineareqs1} are given by \cite{dodelson}
\be
D_{A_+}  ( a ) = \frac{5}{2} \Omega_{{\rm m},0} \cH_0^2 \frac{\cH(a)}{a} \int_0^a \frac{d  a_1}{\cH( a_1 )^3} \andd D_{A_-} ( a ) = \frac{H(a)}{H_0 \Omega_{{\rm m},0}^{1/2} } \ ,
\ee
while the growing and decaying solutions to \eqn{reallineareqs2} are given by 
\be \label{decayinglinsols}
D_{I+}(a) = \text{const.} \andd D_{I_-} ( a ) = { \sqrt{\frac{a_0}{a_{\rm in}}}  } - \frac{\cH_0 \, \Omega_{\rm m, 0}^{1/2} }{2} \int_{a_{\rm in}}^a da_1 \frac{a_0}{a_1^2 \cH(a_1) }   \ . 
\ee
{We have chosen all of the normalizations above so that the solutions approach their standard EdS forms in the early universe (see \eqn{linearsols}).}

To investigate the loop structure, we work in the EdS approximation for simplicity, where $\Om = 1$ and $a \cH' / \cH = -1/2$.  In that case, the linear equations are
\begin{align} \label{lineareqseds}
& a^2 \delta_A^{(1)}{}'' + \frac{3}{2}a \delta_A^{(1)}{}' - \frac{3}{2} \delta_A^{(1)} = 0 \andd  a^2 \delta_I^{(1)}{}'' + \frac{3}{2} a \delta_I^{(1)}{}' = 0 \ . 
\end{align}
The general solutions to the above equations are 
\begin{align}
\begin{split} \label{linearsols}
& \delta_A^{(1)}( a , \kvec ) =  \frac{a}{a_0} \delta_{A_+}^{(1)} ( \kvec )  +  \epsilon^5 \left( \frac{a}{a_0} \right)^{-3/2}   \delta_{A_-}^{(1)} ( \kvec)  \ , \\
& \delta_I^{(1)} ( a , \kvec ) =   \epsilon^2  \delta_{I_+}^{(1)} ( \kvec)  +   \epsilon^3 \left( \frac{a}{a_0} \right)^{-1/2}  \delta_{I_-}^{(1)} ( \kvec)  \ .
\end{split}
\end{align}
  Finally, with \eqn{linearsols}, we can also find the solutions for the linear velocity divergences, 
\begin{align}
\begin{split} \label{linearthetasols}
& \Theta_A^{(1)}( a , \kvec ) =    \frac{a}{a_0}   \delta_{A_+}^{(1)} ( \kvec ) - \frac{3}{2} \epsilon^5 \left( \frac{a}{a_0} \right)^{-3/2}   \delta_{A_-}^{(1)} ( \kvec) \ , \\
&  \Theta_I^{(1)} ( a , \kvec ) = - \frac{1}{2} \epsilon^3 \left( \frac{a}{a_0} \right)^{-1/2}  \delta_{I_-}^{(1)} ( \kvec)\ .
\end{split}
\end{align}
Notice that the constant isocurvature solution $\delta_{I_+}^{(1)}$ does not contribute to the isocurvature velocity divergence, so the leading term for the isocurvature velocity is decaying.

Next, {we would like to find the perturbative solutions in the EdS expansion to \eqn{finaladiabaticeqn} and \eqn{finalisoeqn}.  Let us start with the second order equations, and look at the right-hand-sides of \eqn{finaladiabaticeqn} and \eqn{finalisoeqn}.  At each perturbative order, each expression can be organized in an expansion in powers of $\epsilon$, and all of the terms with a fixed power of $\epsilon$ will have the same time dependence.\footnote{One can realize this by noticing that the dependence in $a$ is through the combination $a/a_0$, and there is a symmetry of rescaling $a_0$ and $\epsilon$. Therefore, for any dependence in $a$, there is an associated dependence in $a_0$, and therefore in $\epsilon$.}  This allows us to write the solutions for $\delta_A^{(2)}$ and $\delta_{I}^{(2)}$ as sums of terms with the appropriate $\epsilon$ and $a$ dependence.  Then, the solutions at higher and higher orders will contain more and more distinct powers of $\epsilon$.   } 
  For example, looking at \eqn{finaladiabaticeqn} and \eqn{finalisoeqn}, we see that
\begin{align}
\begin{split}
& \delta_A^{(2)}  \sim a^2 + \epsilon^5 a^{-1/2} + \epsilon^{10} a^{-3} \ , \\
& \delta^{(2)}_I \sim \epsilon^2 a + \epsilon^3 a^{1/2} + \epsilon^6 a^{-1} + \epsilon^7 a^{-3/2} + \epsilon^8 a^{-2} \ .
\end{split}
\end{align}
This leads us to make the following ansatz for the growing and first-decaying modes of the solutions
\begin{align}
\begin{split} \label{deltaansatz}
& \delta_A^{(n)} ( a  , \kvec ) = \left( \frac{a}{a_0} \right)^{n-1} \left( \frac{a}{a_0} \delta^{(n)}_{A_+} ( \kvec) +\epsilon^5 \left( \frac{a}{a_0} \right)^{-3/2} \delta^{(n)}_{A_-} ( \kvec) + \dots  \right)  \ , \\
& \delta_I^{(n)} ( a , \kvec ) = \left( \frac{a}{a_0} \right)^{n-1}  \left( \epsilon^2 \delta_{I_+}^{(n)} (\kvec) + \epsilon^3 \left( \frac{a}{a_0} \right)^{-1/2} \delta_{I_-}^{(n)} ( \kvec)   + \dots  \right) \ . 
\end{split}
\end{align}
While the general expansion contains many higher powers of $\epsilon$, which can easily be taken into account in a code for example, we will focus on the terms shown above.  The ansatz for the velocity divergences, then, is
\begin{align}
\begin{split} \label{thetaansatz}
& \Theta_A^{(n)} ( a  , \kvec ) = \left( \frac{a}{a_0} \right)^{n-1} \left( \frac{a}{a_0} \Theta^{(n)}_{A_+} ( \kvec) +\epsilon^5 \left( \frac{a}{a_0} \right)^{-3/2} \Theta^{(n)}_{A_-} ( \kvec) + \dots  \right)  \ , \\
& \Theta_I^{(n)} ( a , \kvec ) = \left( \frac{a}{a_0} \right)^{n-1}  \left( \epsilon^2 \Theta_{I_+}^{(n)} (\kvec) + \epsilon^3 \left( \frac{a}{a_0} \right)^{-1/2} \Theta_{I_-}^{(n)} ( \kvec)   + \dots  \right) \ . 
\end{split}
\end{align}

%
\subsection{UV limit of one-loop terms} \label{uvlimapp}

With the form of the equations in \secref{geneqsapp}, we can easily compute the leading UV behavior of the one-loop power spectra.  Looking at the adiabatic equation \eqn{finaladiabaticeqn}, we see that the adiabatic power spectrum will always have the standard form.  This is guaranteed because the interaction terms on the right-hand side are all under $\partial_i \partial_j$, and because of wave-vector conservation in Fourier space, these derivatives will always turn into the total external wavenumber.

This means that the first place for us to look for the new terms in the UV is in  $\langle \delta_A ( \kvec ) \delta_I ( \kvec' ) \rangle $.
We first look at the $(22)$ diagram.  Using \eqn{finaladiabaticeqn} and \eqn{finalisoeqn},  we have the schematic contributions 
\be
\delta_A^{(2)} \sim \partial_i \partial_j \left(  \frac{\partial_i \delta_A^{(1)}}{\partial^2}  \frac{\partial_j \delta_A^{(1)}}{\partial^2 } \right) \  , \quad \text{and} \quad \delta^{(2)}_I \sim \partial_i \left( \delta_I^{(1)} \frac{\partial_i \delta_A^{(1)}}{\partial^2 } \right) \ ,
\ee
which in the UV limit $k_1 \gg k$ gives
\begin{align}
\begin{split}
P^{AI}_{22} ( a , k ) \rightarrow  - \epsilon^2 \left( \frac{a}{a_0} \right)^3 \int_{\kvec_1}   \left( \frac{\kvec \cdot \kvec_1 }{k_1^2}  \right)^2 \left( \frac{\kvec' \cdot \kvec_1}{k_1^2} \right) P_{++}^{AA} ( k_1) P_{++}^{A I}(k_1)  \ ,
\end{split}
\end{align}
which is zero because it is odd in $\kvec_1$.  This means that this contribution starts at $k^4$, as anticipated.  

To find a $k^2$ contribution, we look at $\langle \delta^{(2)}_I ( a , \kvec ) \delta^{(2)}_I ( a , \kvec') \rangle$.  For concreteness, we compute the $\epsilon^4$ contribution.  Calling $\delta^{(2)}_{I,\rm{st.}}$ the part of $\delta^{(2)}_{I}$ that is relevant to compute the $\epsilon^4 k^2$ stochastic contribution, we have
\be
\label{delta2st}
a^2 \delta_{I,\rm{st.}}^{(2)}{}'' + \frac{3}{2} a \delta_{I,\rm{st.}}^{(2)}{}'   = \frac{3}{2} \partial_i \left( \delta_I^{(1)} \frac{\partial_i \delta^{(1)}_A }{\partial^2 } \right) \ , 
\ee
and taking the $\epsilon^2$ part, we have\footnote{To save space, we introduce the notation
\be
\int_{\kvec_1 , \dots ,  \kvec_n}^{\kvec} = \int_{\kvec_1} \dots \int_{\kvec_n} (2 \pi)^3 \delta ( \kvec - \sum_{i=1}^n \kvec_i ) \ . 
\ee}
\be
\delta_{I,\rm{st.}}^{(2)} ( a , \kvec ) = \epsilon^2 \frac{a}{a_0} \int_{\kvec_1 , \kvec_2}^{\kvec} \frac{ \kvec \cdot \kvec_2}{k_2^2} \delta^{(1)}_{I_+} ( \kvec_1 ) \delta^{(1)}_{A_+} ( \kvec_2 ) \ ,
\ee
which gives
\be
P^{II}_{22} ( a , k ) \rightarrow \epsilon^4 \left( \frac{a}{a_0} \right)^2 \int_{\kvec_1} \frac{(\kvec \cdot \kvec_1)^2}{k_1^4} \left( P_{++}^{A I} ( k_1)^2 - P_{++}^{A A } ( k_1 ) P_{++}^{I I} ( k_1) \right) \ ,
\ee
in the UV limit $k_1 \gg k$.  This goes as $k^2$, as expected, and vanishes if there is no isocurvature mode.

Now we turn to the $(13)$ diagram.  First of all, the term $\langle \delta_A^{(3)} (a ,  \kvec ) \delta_I^{(1)} ( a ,  \kvec ') \rangle$ will go like $k^2 P_{++}^{AI} ( k )$ because $\delta_A^{(3)}$ has the standard form, as discussed above.  This means that we should look at 
\be \label{lookatthis}
\langle \delta_A^{(1)} ( a , \kvec ) \delta_I^{(3)} ( a , \kvec ') \rangle \ . 
\ee
Using some simple observations, we can compute this contribution in the UV limit without fully solving the equations of motion.  
First of all, terms going like $k^0 P_{\pm \pm}^{AA} ( k )$ must absent in \eqn{lookatthis} because of  {Galilean invariance}, which our general construction in \secref{coupledsec} shows (and we have explicitly verified with the solution in \secref{recursionsec}).\footnote{{This is because the relevant counterterm that enters the $\delta^{(3)}_I$ equation is $\partial_i \gamma^i$, and $\gamma^i$ is Galilean invariant.  This means that $\gamma^i \sim \pi^i_I$, as opposed to $\pi^i_A$, and so the only available counterterm is $\propto \partial_i \pi^i_I \sim \Theta_I$.  In practice, the reason that the $k^0 P^{AA}_{\pm \pm} (k)$ term does not show up in the full perturbative calculation is non-trivial and relies on the cancellation of many terms, whose coefficients are related because of Galilean invariance.  }}   So we look for contributions going like $k^0 P_{+ +}^{AI} ( k )$ or $k^0 P_{+ -}^{AI} ( k )$ (in reality, it is only the latter that is present, as we show below, and indeed is expected on symmetry arguments, as explained in \secref{coupledsec}).  This means that one of the factors of $\delta_I^{(1)} $ is an external leg, and the terms with the most factors of $\partial_i \delta_I^{(1)} / \partial^2$ will be the ones with the $k^0$ contribution.  

We start by writing the solution for $\delta^{(3)}_{I,\rm{UV}}$, where we use the subscript UV to denote the solution relevant for the UV limit.  Notice that all of the vertices with $\partial_i \partial_j$ in \eqn{finalisoeqn} will contribute $k^2$, and so will not produce the desired effect.  
Thus, we immediately see that 
\be \label{delta3uv}
a^2 \delta_{I,\rm{UV}}^{(3)}{}'' + \frac{3}{2} a \delta_{I,\rm{UV}}^{(3)}{}'   = \frac{3}{2}  \partial_i \left(  
\delta^{(2)}_I \frac{\partial_i \delta_{A}^{(1)} }{\partial^2}  \right)  \ .
\ee
The term in $\delta^{(2)}_I$ with the most derivatives in the denominator of the isocurvature mode comes from 
\be
a^2 \delta_{I,\rm{UV}}^{(2)}{}'' + \frac{3}{2} a \delta_{I,\rm{UV}}^{(2)}{}' = \partial_i \partial_j  \left( 2  \frac{\partial_i \Theta_A^{(1)}}{\partial^2} \frac{\partial_j \Theta_I^{(1)}}{\partial^2}   + ( w_b - w_c)  \frac{\partial_i \Theta_I^{(1)}}{\partial^2} \frac{\partial_j \Theta_I^{(1)}}{\partial^2}    \right)  \ .
\ee
Notice that this UV contribution is proportional to the isocurvature velocity $\Theta_I^{(1)} \sim \delta_{I_-}^{(1)}$, confirming what we found in \secref{coupledsec} based on general symmetry arguments.

Plugging in the linear solutions from \eqn{linearsols} and \eqn{linearthetasols}, we have terms going like $\epsilon^3$, $\epsilon^6$, and $\epsilon^8$ in $\delta^{(2)}_{I,\rm{UV}}$, all of which are easy to keep track of, but we will compute only the term leading in $\epsilon$ here.  We have
\be
a^2 \delta_{I,\rm{UV}}^{(2)}{}'' + \frac{3}{2} a \delta_{I,\rm{UV}}^{(2)}{}'  =  2 \partial_i \partial_j  \left(  \frac{\partial_i \Theta_A^{(1)}}{\partial^2} \frac{\partial_j \Theta_I^{(1)}}{\partial^2}     \right) \ ,
\ee
which has the solution
\be
\delta^{(2)}_{I,\rm{UV}} ( a , \kvec ) = - 2  \epsilon^3 \left( \frac{a}{a_0} \right)^{1/2}  \int_{\kvec_1 , \kvec_2}^{\kvec} \frac{\kvec \cdot \kvec_1 \kvec \cdot \kvec_2}{k_1^2 \, k_2^2} \delta_{A+}^{(1)} ( \kvec_1 ) \delta_{I-}^{(1)} ( \kvec_2 ) \ . 
\ee
Then, plugging this into \eqn{delta3uv}, we find
\be
\delta^{(3)}_{I,\rm{UV}} ( a  , \kvec) = - \epsilon^3 \left( \frac{a}{a_0} \right)^{3/2} \int_{\kvec_1 , \kvec_2}^{\kvec} \int_{\vec p_1  , \vec p_2}^{\kvec_1}  \frac{\kvec \cdot \kvec_2 \, \kvec_1 \cdot \pvec_1 \, \kvec_1 \cdot \pvec_2}{k_2^2 \,  p_1^2 \, p_2^2} \delta^{(1)}_{A+} ( \pvec_1 ) \delta^{(1)}_{I-} ( \pvec_2 ) \delta^{(1)}_{A+} ( \kvec_2 )  \ . 
\ee
Finally, this gives, for $k_1 \gg k$, 
\be \label{k0countertermcomp}
P_{13}^{AI} (a ,  k )  \rightarrow \epsilon^3 \left( \frac{a}{a_0} \right)^{5/2} P^{AI}_{+-}( k )    \int \frac{d^3 k_1}{(2 \pi)^3} (\hat k \cdot \hat k_1 )^2  P^{AA}_{++} ( k_1 )  \ ,
\ee
which we have confirmed with the full computation (i.e. using the recursion relations in \secref{recursionsec}).  One can find similar UV divergences in, for example, $\langle \delta_I ( a,  \kvec) \delta_I( a , \kvec') \rangle$, all of which, at one loop, come from the solution for $\delta^{(3)}_{I,\rm{UV}}$ that we found above and so are adjusted by the same counterterm.  Notice, as mentioned in the main text, that this divergence is suppressed by a higher power of $\epsilon$ than if it were proportional to $\delta_{I+}^{(1)}$.

%
\subsection{Recursion relations} \label{recursionsec}
In this subsection, we derive recursion relations for the perturbative solutions at leading orders in $\epsilon$.  For the standard recursion relations for dark matter, see \cite{Goroff:1986ep, Jain:1993jh}.  Specifically, we will work up to order $\epsilon^3$, which is necessary to find the counterterm needed to correct \eqn{k0countertermcomp}, for example.  From \eqn{deltaansatz}, we see that terms with two factors of $\delta^{(1)}_I$ starts at $\epsilon^4$, so we can ignore those terms for our purposes.  Thus, the relevant equations for us are, in the EdS approximation, 
\begin{align}
\begin{split} \label{eqnsforrecursion}
& a \delta_A' ( a ,   \kvec ) - \Theta_A ( a ,   \kvec)  = \int_{\kvec_1 , \kvec_2}^{\kvec}  \alpha ( \kvec_1 , \kvec_2 ) \delta_A ( a ,  \kvec_1) \Theta_A (a ,  \kvec_2 )    \ , \\
& a \Theta_A ' (a ,  \kvec) + \half \Theta_A ( a , \kvec) - \frac{3}{2} \delta_A ( a , \kvec) =  \int_{\kvec_1 , \kvec_2}^{\kvec}  \beta ( \kvec_1 , \kvec_2 )  \Theta_A (a ,  \kvec_1 ) \Theta_A (a ,  \kvec_2 )  \ ,   \\
& a \delta_I' (  a , \kvec ) - \Theta_I (  a , \kvec)  =  \int_{\kvec_1 , \kvec_2}^{\kvec}   \alpha ( \kvec_1 , \kvec_2 )  \left( \delta_I(a , \kvec_1) \Theta_A(a , \kvec_2) + \delta_A (a , \kvec_1 ) \Theta_I (a , \kvec_2)   \right) \ ,    \\
& a \Theta_I ' (a ,  \kvec) + \half \Theta_I (a ,  \kvec)  = \int_{\kvec_1 , \kvec_2}^{\kvec} 2  \beta ( \kvec_1 , \kvec_2 )   \Theta_A (a ,  \kvec_1) \Theta_I (a ,  \kvec_2)      \ . 
\end{split}
\end{align}

Next, we write the perturbative solutions, up to order $\epsilon^3$, in the form
\begin{align}
\begin{split} \label{epsilon3terms}
& \delta_A^{(n)} ( a , \kvec) = \left( \frac{a}{a_0} \right)^n \delta^{(n)}_{A_+} ( \kvec)  \ , \quad \Theta_A^{(n)} ( a , \kvec) =  \left( \frac{a}{a_0} \right)^n \Theta^{(n)}_{A_+} ( \kvec) \\
& \delta_I^{(n)}(a , \kvec) = \epsilon^2 \left( \frac{a}{a_0} \right)^{n-1} \delta_{I_+}^{(n)} ( \kvec ) + \epsilon^3 \left( \frac{a}{a_0} \right)^{n-3/2} \delta_{I_-}^{(n)} ( \kvec) \ , \\ 
& \Theta_I^{(n)}(a , \kvec) =  \epsilon^3 \left( \frac{a}{a_0} \right)^{n-3/2} \Theta_{I_-}^{(n)} ( \kvec) \ ,
\end{split}
\end{align} 
and expand the momentum dependent parts as
\begin{align}
\begin{split} \label{ordernansatz}
& \delta^{(n)}_{\Gamma} ( \kvec) = \int_{\kvec_1, \dots , \kvec_n}^{\kvec} F^{\Gamma}_n ( \kvec_1 , \dots , \kvec_n) \delta_{\Gamma}^{(1)} ( \kvec_1) \delta_{A_+}^{(1)} ( \kvec_2 ) \cdots \delta_{A_+}^{(1)} ( \kvec_n )  \ ,  \\ 
& \Theta^{(n)}_{\Gamma} ( \kvec) = \int_{\kvec_1, \dots , \kvec_n}^{\kvec} G^{\Gamma}_n ( \kvec_1 , \dots , \kvec_n) \delta_{\Gamma}^{(1)} ( \kvec_1) \delta_{A_+}^{(1)} ( \kvec_2 ) \cdots \delta_{A_+}^{(1)} ( \kvec_n ) \ , 
\end{split}
\end{align}
where $\Gamma \in \{ A_+ , I_+ , I_- \}$.  The first-order kernels satisfy $F^\Gamma_1 = 1$ for $\Gamma \in  \{ A_+ , I_+ , I_- \}$,  $G^{A_+}_1 = 1$, $G^{I_+}_1 = 0$, and $G^{I_-}_1 = -1/2$.   At the order that we work, the equations for the adiabatic mode in \eqn{eqnsforrecursion} are exactly the same as the standard dark-matter case, so the kernels $F^{A_+}_n$ and $G^{A_+}_n$ are the same as in, for example \cite{Goroff:1986ep, Jain:1993jh}.  Thus, we concentrate on the isocurvature solution below.

Plugging the ansatz \eqn{ordernansatz} into the equations of motion \eqn{eqnsforrecursion} and matching orders in $\epsilon$, we find 
\begin{align}
&G_n^{I_+} ( \kvec_1 , \dots , \kvec_n ) =  0 \ , \\
&F_n^{I_+} ( \kvec_1 , \dots , \kvec_n) = \frac{1}{n-1} \sum_{m=1}^{n-1}  \alpha \left( \kvec_{1;m} , \kvec_{m+1;n } \right) F^{I_+}_m ( \kvec_1 , \dots , \kvec_m ) G^{A_+}_{n-m} ( \kvec_{m+1} , \dots , \kvec_n)  \nonumber \ , 
\end{align}
and
\begin{align}
&G_n^{I_-} ( \kvec_1 , \dots , \kvec_n ) =  \frac{2}{n-1} \sum_{m=1}^{n-1} \beta \left( \kvec_{1;m} , \kvec_{m+1;n}     \right) G^{I_-}_{m} ( \kvec_1 , \dots , \kvec_m ) G^{A_+}_{n-m} ( \kvec_{m+1}  , \dots , \kvec_n ) \ , \\
&F_n^{I_-} ( \kvec_1 , \dots , \kvec_n) = \frac{1}{n-3/2} \sum_{m=1}^{n-1} \Bigg[ \alpha \left( \kvec_{1;m} , \kvec_{m+1;n } \right) F^{I_-}_m ( \kvec_1 , \dots , \kvec_m ) G^{A_+}_{n-m} ( \kvec_{m+1} , \dots , \kvec_n) \nonumber \\
& \hspace{.7in} +  \alpha \left(\kvec_{m+1;n } , \kvec_{1;m}  \right) G^{I_-}_m ( \kvec_1 , \dots , \kvec_m ) F^{A_+}_{n-m} ( \kvec_{m+1} , \dots , \kvec_n) \Bigg] + \frac{1}{n-3/2} G^{I_-}_n(\kvec_1 , \dots , \kvec_n) \nonumber \ , 
\end{align}
where we have defined $\kvec_{i;j} \equiv \kvec_i + \dots + \kvec_j$.\footnote{{During the completion of this work, \cite{Rampf:2020ety} appeared which derived the recursion relation for $F^{I_+}_n$ above.}}  The fact that $G_n^{I_+} = 0$ was to be expected, since from \eqn{eqnsforrecursion}, the isocurvature velocity divergence is sourced by a term proportional to the isocurvature velocity divergence itself, for which the $\delta_{I_+}$ contribution is initially zero.  Of course, these kernels are applicable to the case when $g=0$, i.e. the linear counterterm is absent.  We discuss the case for $g\neq 0$ in the next subsection.

As a final comment, in a general cosmology, one normally uses the $k$-dependence of the EdS solution and replaces $a \rightarrow D_{A_+}(a)$.  This is percent-level accurate \cite{Takahashi:2008yk}, and relies on 
\be
\frac{a^2 D_{A_+}'(a)^2}{\Om(a) D_{A_+}(a)^2 } \approx 1 \ . 
\ee
The analogous procedure for implementing this for the solutions in \eqn{epsilon3terms} is
\begin{align}
\begin{split} \label{epsilon3termsedsapprox}
& \delta_A^{(n)} ( a , \kvec) = \left( \frac{D_{A_+}(a)}{ D_{A_+}(a_0)} \right)^n \delta^{(n)}_{A_+} ( \kvec)  \ , \quad \Theta_A^{(n)} ( a , \kvec) =    \frac{a D'_{A_+}(a)}{ D_{A_+}(a_0)}  \left( \frac{D_{A_+}(a)}{ D_{A_+}(a_0)} \right)^{n-1} \Theta^{(n)}_{A_+} ( \kvec) \\
& \delta_I^{(n)}(a , \kvec) = \epsilon^2 \left(\frac{D_{A_+}(a)}{ D_{A_+}(a_0)}  \right)^{n-1} \delta_{I_+}^{(n)} ( \kvec ) + \epsilon^3   \frac{D_{I_-}(a)}{ D_{I_-}(a_0)}  \left( \frac{D_{A_+}(a)}{ D_{A_+}(a_0)}  \right)^{n-1} \delta_{I_-}^{(n)} ( \kvec) \ , \\ 
& \Theta_I^{(n)}(a , \kvec) = - 2  \epsilon^3  \frac{a D'_{I_-}(a)}{ D_{I_-}(a_0)}  \left( \frac{D_{A_+}(a)}{ D_{A_+}(a_0)}  \right)^{n-1} \Theta_{I_-}^{(n)} ( \kvec) \ ,
\end{split}
\end{align} 
which additionally relies on 
\be
\frac{ D'_{I_-} ( a ) / D_{I_-}(a)}{D'_{A_+} ( a ) / D_{A_+}(a)} \approx - \half \ . 
\ee
This is approximately true for the cosmology used in this paper, where the ratio goes from $-0.5$ at early times, to $-0.46$ at the current time.

%
%
\subsection{Expansion with linear counterterm} \label{pertwithlinctsec}

When one includes the linear relative-velocity counterterm, as in \eqn{newctlineareq}, the perturbative expansion is somewhat different from what we have described previously in this section.  This is because there is not generally an EdS-like expansion where the time dependence of all of the terms can be made the same (or approximately the same, as is done in the EdS approximation).  As remarked in \secref{linearevo}, although the growth factor $D^g_{I_-}$ for the decaying isocurvature mode must be treated as a free function, one still knows the $k$-dependence of the solutions.  To see how this plays out in perturbation theory, let us show an explicit example, namely the computation of {$\delta^{(2)}_{I}( a , \kvec) $.   

First, we show that even in the presence of the linear relative-velocity counterterm, the $\epsilon^2$ contribution to $\Theta_I$ is absent at every order in perturbation theory.  The equations of motion for the isocurvature mode, including the relative-velocity counterterm, up to order $\epsilon^3$, are
\begin{align}
\begin{split}   \label{deltaiexpand}
&  a \delta'_I (a) - \Theta_I (a )  = \partial_i \left(  \delta_A(a)  \frac{\partial_i \Theta_I ( a ) }{\partial^2} + \delta_I (a ) \frac{\partial_i \Theta_A ( a ) }{\partial^2}   \right)  \ , \\
& a \Theta_I ' ( a )  + \left( 1 + \frac{a \cH'}{\cH} \right) \Theta_I ( a )  - \int^a d a_1 \, g(a , a_1) \Theta_I ( a_1 ) =  \partial_i \partial_j \left( \frac{\partial_i \Theta_A ( a ) }{\partial^2 } \frac{\partial_j \Theta_I ( a ) }{\partial^2 } \right)  \ ,
\end{split}
\end{align}
where we have suppressed the obvious spatial coordinates for notational convenience.  As discussed in \secref{linearevo}, the linear solutions are $\delta^{(1)}_{I} ( a , \kvec)  = \epsilon^2 \delta_{I+}^{(1)} ( \kvec) +  \epsilon^3 D_{I_-}^g ( a ) \delta_{I_-}^{(1)} ( \kvec)  / D_{I_-}^g ( a_0 ) $ which implies that $\Theta_I^{(1)} ( a , \kvec ) = \epsilon^3 a D^g_{I_-}{}' ( a ) \delta^{(1)}_{I_-} ( \kvec)  / D_{I_-}^g ( a_0 ) $.  Now, looking at the second equation in \eqn{deltaiexpand}, it is clear that since the $\epsilon^2$ piece is absent in $\Theta_I^{(1)}$, it will be absent in all higher orders in perturbation theory, since the right-hand side is proportional to $\Theta_I$.  Thus, we have $\Theta^{(n)}_{I_+} = 0$ for $n \geq 1$.  

There \emph{is} an $\epsilon^2$ contribution to $\delta^{(n)}_I$, for $n \geq 2$, from \eqn{deltaiexpand}, though.  Taking the $\epsilon^2$ piece of the first equation in \eqn{deltaiexpand}, and using that $\Theta^{(n)}_{I_+} = 0$, we have
\be
a \delta^{(n)}_{I_+} {}'  = \sum_{m=1}^{n-1} \partial_i \left( \delta_{I_+}^{(m)} \frac{\partial_i \Theta_{A_+}^{(n-m)}}{\partial^2} \right) \ , 
\ee
which can be solved using the expansion in \eqn{epsilon3termsedsapprox}, in the same way as if there were no relative-velocity counterterm.  

Now we focus on the decaying isocurvature mode, proportional to $\epsilon^3$, which is affected by the new counterterm.  At linear order, we have $\delta^{(1)}_{I_-} ( a , \kvec ) = D^g_{I_-} ( a ) \delta^{(1)}_{I_-} ( \kvec )  / D_{I_-}^g ( a_0 )$. }
 The equation for the second-order field is given by (ignoring terms with two isocurvature modes) 
\be \label{newctlineareq2}
a^2 \delta_{I_-}^{(2)} {}'' ( a, \kvec ) + \left( 2 + \frac{ a \cH'(a)}{\cH(a)} \right) a \delta^{(2)}_{I_-} {}' ( a , \kvec)  - \int^a d a_1 g ( a , a_1 ) \,  a_1 \delta^{(2)}_{I_-} {}' ( a_1, \kvec)  =   S^{(2)}_{I_-} ( a , \kvec)   \ ,
\ee
where the second-order source term is given by (this is easiest seen by working directly with the momentum equations \eqn{momeqc1} and \eqn{momeqb1})
\begin{align}
& S^{(2)}_{I_-} ( a , \kvec ) = \int^{\kvec}_{\kvec_1 , \kvec_2}    \Bigg( \frac{3}{2} \Om(a) \alpha ( \kvec_1 , \kvec_2 ) \delta^{(1)}_{I_-}(a , \kvec_1 ) \delta_{A_+}^{(1)} ( a , \kvec_2 ) \\
& \hspace{1.2in} +  2 \alpha( \kvec_1 , \kvec_2 ) \alpha( \kvec_2 , \kvec_1 ) \Theta^{(1)}_{I_-} ( a , \kvec_1 ) \Theta^{(1)}_{A_+} ( a , \kvec_2 )  \nonumber\\
& \hspace{1.2in} - \int^a d a_1 g ( a , a_1 ) \alpha ( \kvec_1 , \kvec_2) \left( \delta^{(1)}_{A_+} ( a_1 , \kvec_1) \Theta^{(1)}_{I_-} ( a_1  , \kvec_2 ) + \delta^{(1)}_{I_-} ( a_1 , \kvec_1) \Theta^{(1)}_{A_+} ( a_1 , \kvec_2)    \right)  \Bigg) \nonumber \ ,
\end{align}
where $\delta^{(1)}_{A_+} ( a , \kvec ) \equiv D_{A_+} ( a ) \delta^{(1)}_{A_+} ( \kvec) / D_{A_+} ( a_0 ) $,  $\Theta_{A_+}^{(1)} ( a , \kvec ) \equiv a \delta^{(1)}_{A_+} {}' ( a , \kvec ) $, and $\Theta^{(1)}_{I_-} ( a , \kvec ) \equiv a \delta^{(1)}_{I_-} {}' ( a , \kvec )$.

Next, we must assume that there is a Green's function, which we call $G_g ( a , a_1)$, to the non-local linear equations.  Then, after plugging in the liner solutions, we find
\begin{align}
& \delta^{(2)}_{I_-} ( a , \kvec )  =  \\
& \quad \int_{\kvec_1 , \kvec_2}^{\kvec} \left[ \left( T_1 ( a ) + T_4 ( a ) \right) \alpha ( \kvec_1 , \kvec_2 ) + T_2 ( a ) \alpha(\kvec_1 , \kvec_2 ) \alpha( \kvec_2 , \kvec_1 ) + T_3 ( a ) \alpha( \kvec_2 , \kvec_1 )  \right]   \delta^{(1)}_{I_-}( \kvec_1 ) \delta^{(1)}_{A_+}( \kvec_2) \nonumber
\end{align}
where
\begin{align}
\begin{split}
T_1 ( a ) & = \int^a d a_1 G_g ( a , a_1 ) \frac{3}{2} \Om ( a_1 ) \frac{ D_{I_-}^g ( a_1 ) D_{A_+}(a_1)}{D^g_{I_-}(a_0) D_{A_+}(a_0)}  \ , \\
T_2 ( a )  & =  \int^a d a_1G_g ( a , a_1) 2 a_1^2  \frac{D_{I_-}^g{}' ( a_1 ) D_{A_+}'(a_1)}{D^g_{I_-}(a_0) D_{A_+}(a_0)} \ , \\
T_3 ( a ) & = - \int^a d a_1 \int^{a_1} d a_2 G_g ( a , a_1 ) g(a , a_2) a_2 \frac{ D_{I_-}^g{}'(a_2 ) D_{A_+} ( a_2)  }{D^g_{I_-}(a_0) D_{A_+}(a_0)}\ , \\
T_4 ( a ) & = - \int^a d a_1 \int^{a_1} d a_2 G_g ( a , a_1 ) g(a , a_2) a_2 \frac{D_{I_-}^g(a_2 )  D_{A_+}' ( a_2) }{D^g_{I_-}(a_0) D_{A_+}(a_0)}\ .
\end{split}
\end{align}
However, since both $G_g ( a , a_1)$ and $g(a,a_1)$ are unknown, we can just treat $T_{1,2,3,4} ( a )$  as free functions.  Although the time dependence is unknown, the structure of the equations still fixes some of the momentum dependence.   Higher orders in perturbation theory can be found in an analogous manner.  Realistically, though, the decaying isocurvature mode  is expected to be very small compared to other contributions in the power spectrum, so we leave exploration of this topic to future work.  We simply stress here that perturbation theory is not spoiled, at least as a matter of principle, by the presence of the linear relative-velocity counterterm.

%
\section{Coefficient relationships} \label{coefficientsapp}

In this appendix, we relate the coefficients in the stress tensors \eqn{stress1} and \eqn{stress2} to the coefficients appearing in the power spectra \eqn{candbtwoloop}.  Here, we work in an EdS universe, where the counterterms are assumed to have the time dependences needed to cancel UV divergences.  Thus, these relationships are not strictly true in the real universe, but we include them here to give an indication of the various contributions.  First, we define
\begin{align}
\begin{split}
& c_{cc}^2 \equiv a^{-2} \left( w_c c_{c,g}^2 + c_{c,v}^2 \right) \ , \quad c_{cb}^2 \equiv a^{-2} w_b c_{c,g}^2  \ ,  \\
& c_{bc}^2 \equiv a^{-2} w_c c_{b,g}^2 \ , \quad   c_{bb}^2 \equiv a^{-2} \left( w_b c_{b,g}^2 + c_{b,v}^2 + c_{\star(1)}^2 \right)   \ , \\
& \bar c_c^2 \equiv c_{cc}^2 + c_{cb}^2 \andd \bar c_b^2 \equiv c_{bc}^2 + c_{bb}^2 \ . 
\end{split}
\end{align}
We also define the adiabatic and isocurvature combinations
\be
\bar c_A^2 \equiv w_c \bar c_c^2 + w_b \bar c_b^2 \andd \bar c_I^2 \equiv \bar c_c^2 - \bar c_b^2 \ . 
\ee
Then we have
\be
c_{\sigma (1)}^2 = \bar c_\sigma^2 \ , 
\ee
for $\sigma \in \{A,c,b\}$, and the coefficient related to non-locality in time is given by 
\be
\frac{\xi_\sigma + \frac{5}{2} }{2 \left( \xi_\sigma + \frac{5}{4} \right)} = \frac{9}{13} \ . 
\ee
Next, we have
\begin{align}
\begin{split}
& c^2_{4c} = \frac{9}{26} \left( \frac{1}{2 \pi  } \left(  c_{4c,g}^2 + c_{4c,v}^2 \right) - c_{cb}^2 \bar c_I^2 \right) \ , \\  
& c^2_{4b} = \frac{9}{26} \left(\frac{1}{2 \pi } \left( c_{4b,g}^2 + c_{4b,v}^2 + c_{4\star}^2 \right) + c_{bc}^2 \bar c_I^2 \right)\ ,  \\ 
& c^2_{4A} = w_c c^2_{4c} + w_b c^2_{4b} + \frac{9}{26} w_b w_c \bar c_I^4 \  ,
\end{split}
\end{align}
and finally
  \begin{align}
  \begin{split}
  c_{1c}^2 = \frac{36}{33}a^{-2} \left( c_{1c}^{cc}  +  c_{ 1c}^{cb} + c_{1c}^{bb} \right)  \andd    c_{1b}^2 = \frac{36}{33}a^{-2} \left( c_{1b}^{cc}  +  c_{ 1b}^{cb} + c_{1b}^{bb} \right) \ . 
  \end{split} 
  \end{align}
  
  %
  %
  \section{Fitting procedure and tables of fit values} \label{tableapp}
{First, we give an example of how our fitting procedure for the ratio works at $z=2$ to determine  $\{ \Delta c^{2}_{c(1)},  \Delta c^{2}_{1c}, \Delta c^{2}_{4c} \}$ and $\{ \Delta c^{2}_{b(1)}, \Delta c^{2}_{1b}, \Delta c^{2}_{4b} \}$.  As mentioned in the text, $k_{\rm max}$ is the maximum $k$ that is included in the fit.  We then choose $k_{\rm fit}$ to be the maximum value of $k_{\rm max}$ where the best fit values are consistent with the lower $k_{\rm max}$ fits.  In the following plots, the connected dotted points are the best fit values, the dashed lines are the $1\sigma$ error on the parameter, and the solid lines are the $2\sigma$ error.   We say that the fit at a given $k_{\rm max}$ is consistent with the lower $k_{\rm max}$ fits if the best fit value at $k_{\rm max}$ lies within the $2\sigma$ error bands of all lower $k_{\rm max}$ fits.  As can be seen in the following plots, it is the baryon fit that determines $k_{\rm fit}$, because the baryon ratio varies more over the range of $k$'s considered (see \figref{cAfit}, for example).  If the parameters drift outside of the $2\sigma$ regions at different values of $k_{\rm max}$, we use the lowest value for $k_{\rm fit}$.  We see that the baryon parameters start to drift outside of the $2\sigma$ bands of the lower $k_{\rm max}$ fits at approximately $k_{\rm max} \approx 1.8 \unitsk$, which is what we use for $k_{\rm fit}^R$ (the $k_{\rm fit}$ of the ratio, see \tabref{kreachtab}).    Finally, we also present the values of the EFT parameters and errors measured in the fits in \secref{Comparison}.}

%

  \begin{figure}[ht!]
     \begin{center}
     \begin{tabular}{cc}
         \includegraphics[width=12cm]{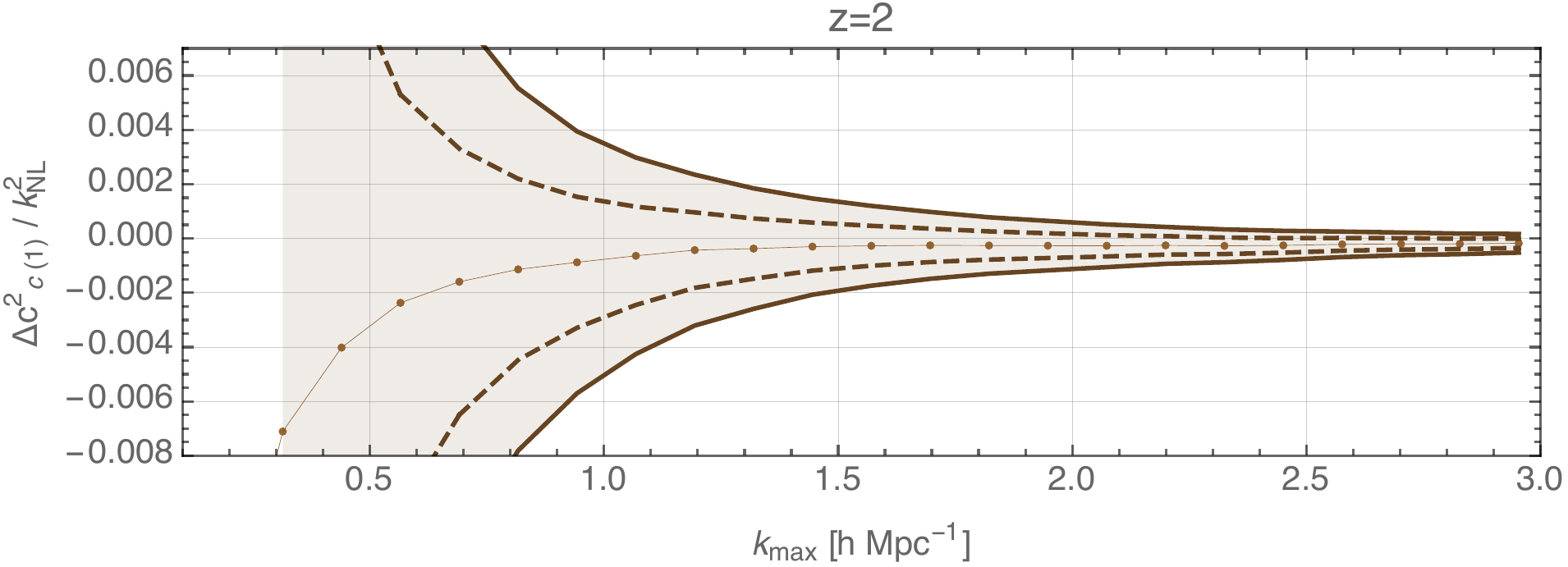}  \\ \includegraphics[width=12cm]{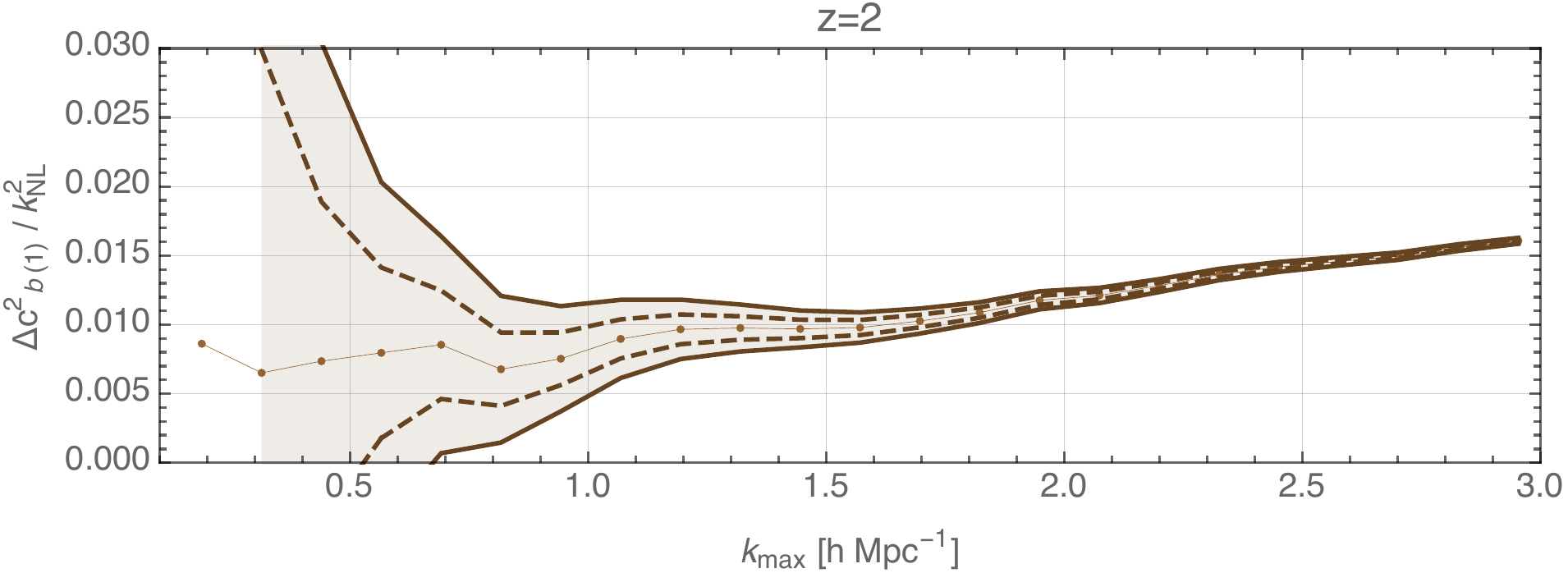}
            \end{tabular} 
    \end{center}
    \caption{ \footnotesize  Determination of $\Delta c^2_{\sigma (1)}$ and $k_{\rm fit}^R$. }
   \label{triangleplots1}
\end{figure}

  \begin{figure}[ht!]
     \begin{center}
     \begin{tabular}{cc}
       \includegraphics[width=12cm]{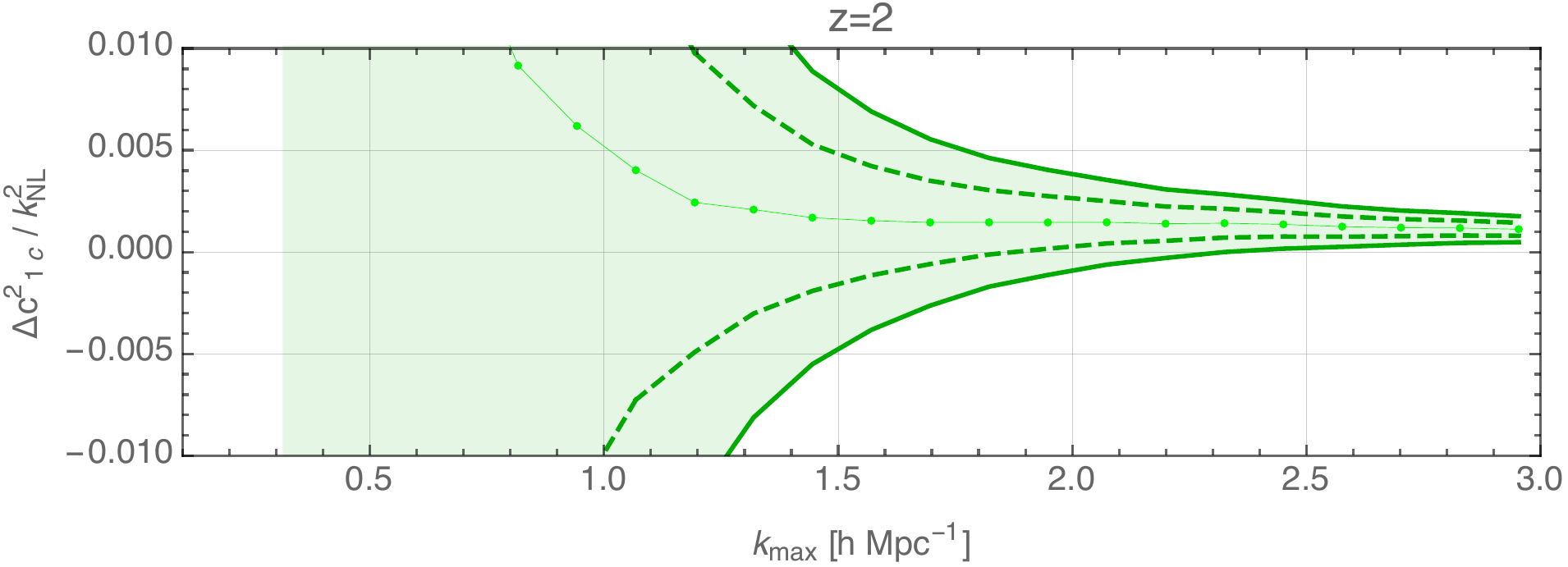} \\  \includegraphics[width=12cm]{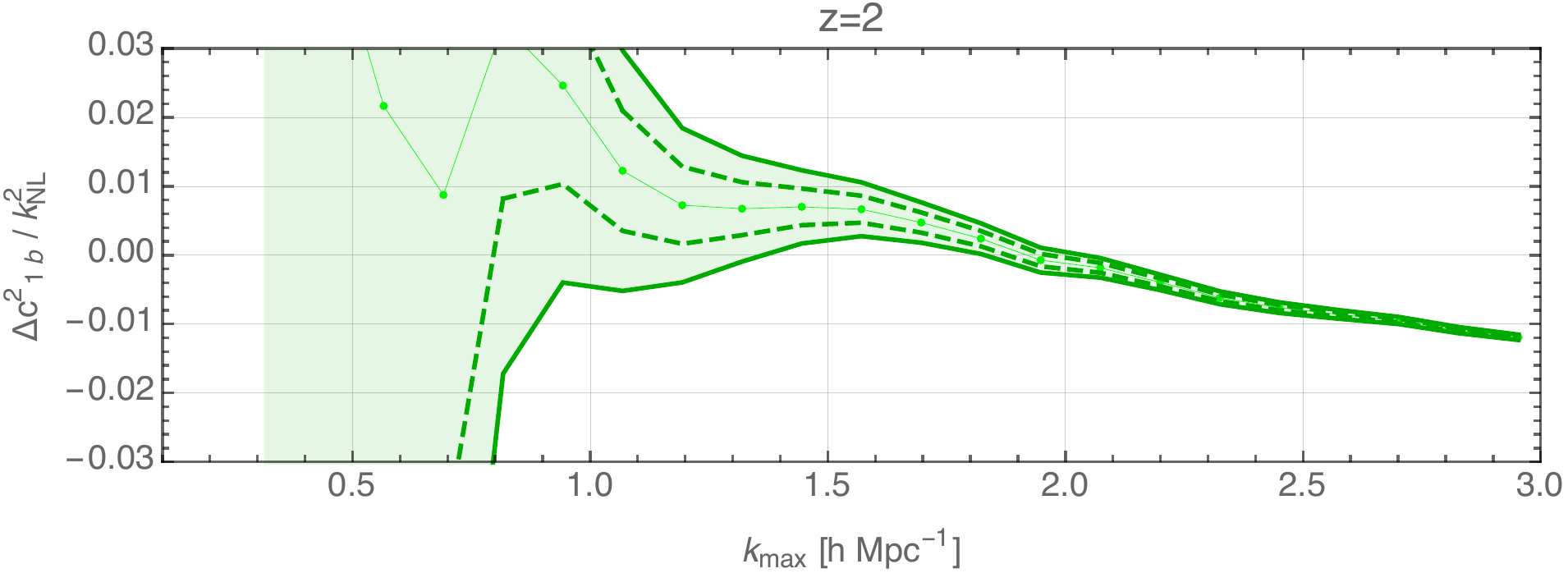}
            \end{tabular} 
    \end{center}
    \caption{ \footnotesize  Determination of $\Delta c^2_{1 \sigma }$ and $k_{\rm fit}^R$.  }
   \label{triangleplots2}
\end{figure}

  \begin{figure}[ht!]
     \begin{center}
     \begin{tabular}{cc}
       \includegraphics[width=12cm]{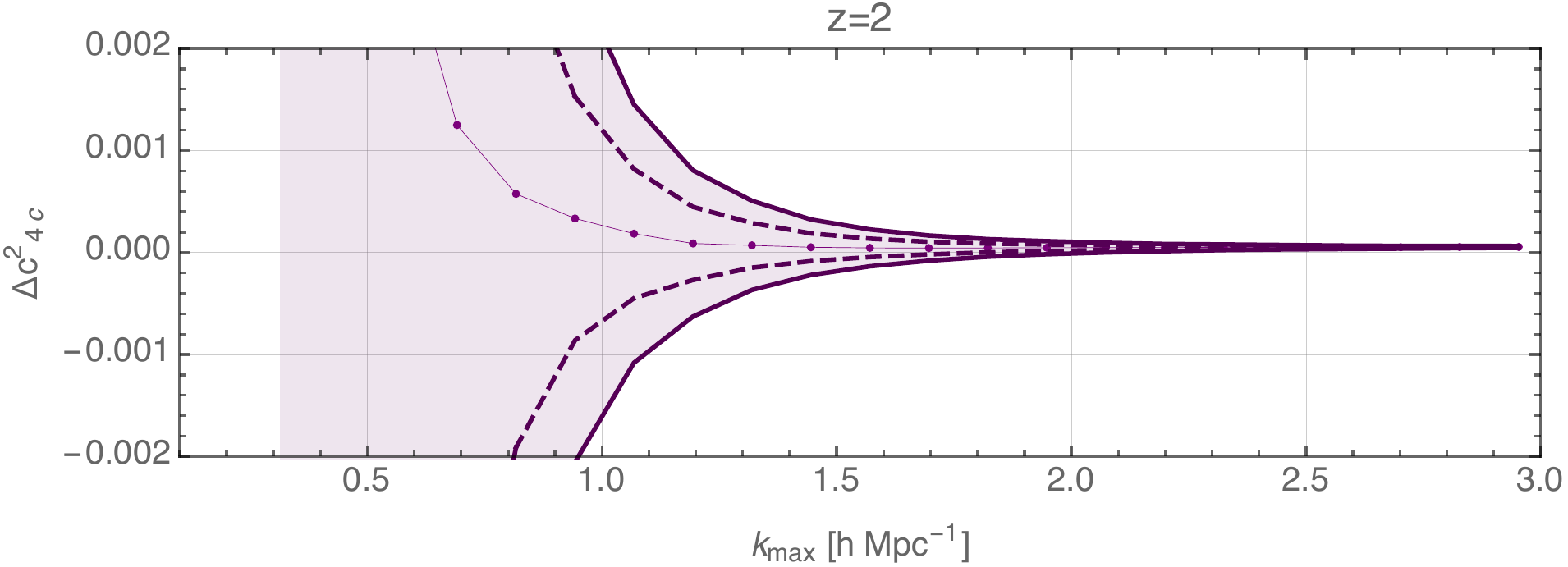}  \\  \includegraphics[width=12cm]{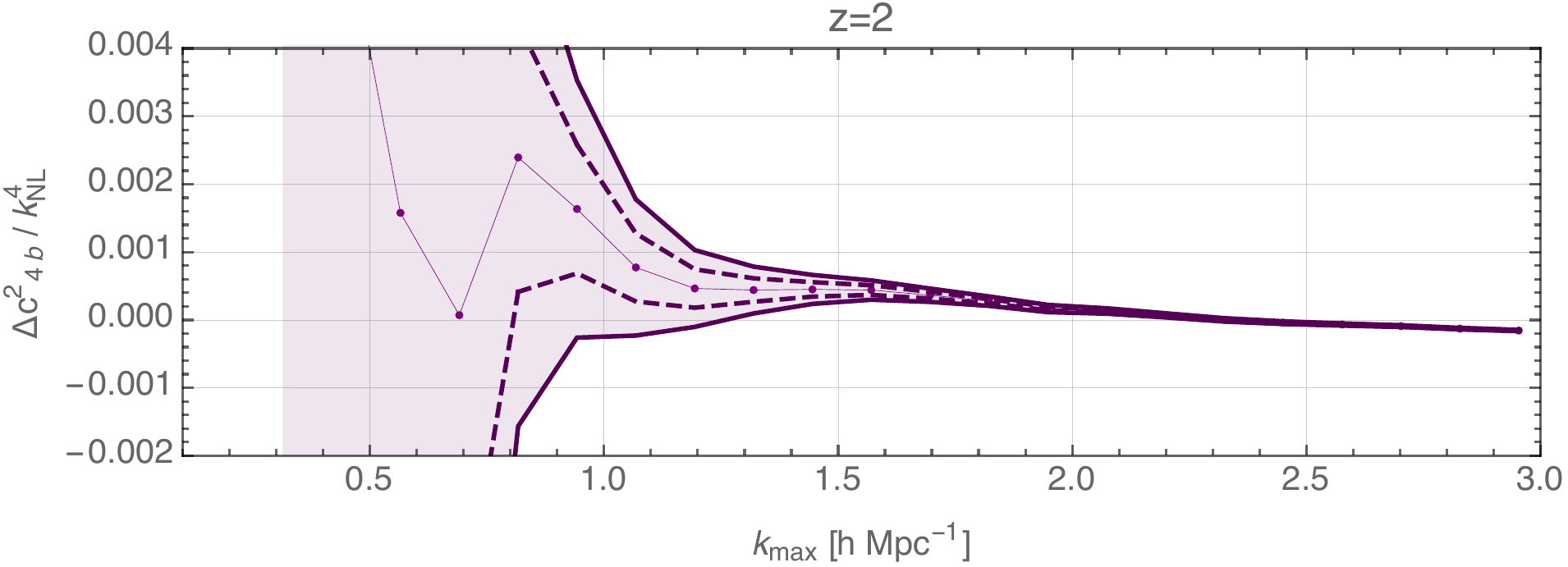}
            \end{tabular} 
    \end{center}
    \caption{ \footnotesize Determination of $\Delta c^2_{4\sigma}$ and $k_{\rm fit}^R$.  }
   \label{triangleplots3}
\end{figure}

  \begin{table}[h]
\centering
\begin{tabular}{c | c |  c | c  | c }
\hline \hline 
$z$ & $\Delta c_{c(1)}^2 / \knl^2$ &  $\Delta c_{b(1)}^2/ \knl^2$ &  $ c_{s(1)}^2/ \knl^2$ & $$  \\ 
\hline
$  0 $   &   $-0.26 \pm 0.32   $    &    $  4.3 \pm 0.24  $   &   $12  \pm 6.6 $  &   $ \times 10^{-2}$    \\ \hline
$  0.125 $   &   $0.034 \pm 0.23     $    &    $  4.7 \pm 0.17   $   &   $ 9.7 \pm6.2 $    &   $ \times 10^{-2}$   \\ \hline
$ 0.25  $   &   $ 0.022 \pm 0.23  $    &    $  4.8 \pm0 .17   $   &   $8.0 \pm 5.7   $   &   $ \times 10^{-2}$    \\  \hline
$ 0.375  $   &   $  7.6 \times 10^{-5} \pm 0.24 $    &    $ 4.6 \pm 0.17   $   &   $ 6.4 \pm 5.0  $    &   $ \times 10^{-2}$    \\  \hline
$ 0.5  $   &   $-0.016 \pm 0.24   $    &    $   4.4 \pm 0.17   $   &   $ 4.2 \pm 4.1 $     &   $ \times 10^{-2}$  \\  \hline
$ 0.75  $   &   $  -0.036  \pm 0.24  $    &    $   3.9 \pm 0.18     $   &   $  2.4 \pm 3.0  $     &   $ \times 10^{-2}$ \\ \hline
$ 1  $   &   $-0.038 \pm 0.24   $    &    $  3.2 \pm 0.18  $   &   $  3.2 \pm 1.9  $  &   $ \times 10^{-2}$      \\  \hline
$ 1.25  $   &   $ -0.040 \pm 0.24    $    &    $  2.4 \pm 0.18   $   &   $ -0.027 \pm 1.4 $   &   $ \times 10^{-2}$    \\  \hline
$ 1.5  $   &   $ -0.032 \pm 0.11   $    &    $   2.1 \pm 0.081   $   &   $ 1.9 \pm 0.97  $    &   $ \times 10^{-2}$    \\  \hline
$ 1.75  $   &   $  -0.028 \pm 0.098  $    &    $  1.6 \pm 0.073  $   &   $ 0.13 \pm  0.70  $    &   $ \times 10^{-2}$   \\  \hline
$ 2  $   &   $ -0.026 \pm 0.099    $    &    $  1.1 \pm 0.075   $   &   $ 1.2 \pm 0.50   $   &   $ \times 10^{-2}$    \\  \hline
$ 2.25 $   &   $ -0.21 \pm 0.89   $    &    $  8.1 \pm 0.68   $   &   $ 3.9 \pm 3.8   $  &   $ \times 10^{-3}$      \\  \hline
$ 2.75 $   &   $ -0.11 \pm 0.52  $    &    $  4.2 \pm 0.40    $   &   $  7.0 \pm 2.2 $    &   $ \times 10^{-3}$        \\  \hline
$ 3  $   &   $  -0.052 \pm 0.40  $    &    $  3.0 \pm 0.30  $   &   $  11 \pm 1.7 $ &   $ \times 10^{-3}$         \\  \hline
$ 3.25 $   &   $ -0.047 \pm 0.34   $    &    $ 2.1 \pm 0.27   $   &   $  2.8 \pm 1.4 $   &   $ \times 10^{-3}$        \\  \hline
$ 3.5  $   &   $ -0.011 \pm 0.27    $    &    $  1.5 \pm 0.22  $   &   $ 1.2 \pm 1.1    $    &   $ \times 10^{-3}$       \\  \hline
$ 3.75 $   &   $0.0025 \pm 0.24    $    &    $ 1.1 \pm 0.19   $   &   $  2.8 \pm 0.89  $    &   $ \times 10^{-3}$        \\  \hline
$ 4 $   &   $  0.013 \pm 0.22   $    &    $  0.73 \pm 0.18 $   &   $  7.0 \pm 0.72  $   &   $ \times 10^{-3}$        \\
\hline
\end{tabular}
\caption{\footnotesize Best fit and $1\sigma$ errors for EFT parameters.  Since the parameters are unitless, the columns have units given by the explicit factors of $\knl$.  All values of measured parameters and errors should be multiplied by the factor in the last column.}
\label{cs1table}
\end{table}

  \begin{table}[h]
\centering
\begin{tabular}{c | c |  c | c  | c }
\hline \hline 
$z$ & $\Delta c_{1c}^2 / \knl^2$ &  $\Delta c_{1b}^2/ \knl^2$ &  $ c_{1s}^2/ \knl^2$ & $$  \\ 
\hline
$  0 $   &   $ 0.53 \pm 0.68   $    &    $ -1.4 \pm 0.50    $   &   $ -29 \pm 77 $  &   $ \times 10^{-2}$    \\ \hline
$  0.125 $   &   $ -0.0049 \pm 0.44     $    &    $ -1.9 \pm 0.32    $   &   $ -18 \pm 75  $    &   $ \times 10^{-2}$   \\ \hline
$ 0.25  $   &   $ 0.019 \pm 0.49   $    &    $  -2.2 \pm 0.35   $   &   $  -11 \pm 73  $   &   $ \times 10^{-2}$    \\  \hline
$ 0.375  $   &   $ 0.074 \pm 0.54   $    &    $-2.2 \pm 0.39    $   &   $  -0.15 \pm 63  $    &   $ \times 10^{-2}$    \\  \hline
$ 0.5  $   &   $ 0.12 \pm 0.60  $    &    $ -2.1 \pm 0.43     $   &   $  19 \pm 50  $     &   $ \times 10^{-2}$  \\  \hline
$ 0.75  $   &   $  0.18 \pm 0.74   $    &    $ -1.8 \pm 0.54      $   &   $  27 \pm 35   $     &   $ \times 10^{-2}$ \\ \hline
$ 1  $   &   $ 0.20 \pm 0.90  $    &    $ -1.1 \pm 0.67   $   &   $  13 \pm 21   $  &   $ \times 10^{-2}$      \\  \hline
$ 1.25  $   &   $  0.22 \pm 1.1    $    &    $ -0.29 \pm 0.82    $   &   $ 27 \pm 14   $   &   $ \times 10^{-2}$    \\  \hline
$ 1.5  $   &   $ 0.18 \pm0.37   $    &    $ -0.73 \pm 0.26      $   &   $  9.7  \pm 9.5  $    &   $ \times 10^{-2}$    \\  \hline
$ 1.75  $   &   $  0.16 \pm 0.36   $    &    $ -0.15 \pm 0.26   $   &   $ 16  \pm 6.7  $    &   $ \times 10^{-2}$   \\  \hline
$ 2  $   &   $  0.15 \pm 0.43     $    &    $  0.24 \pm 0.32   $   &   $  6.7 \pm 4.5    $   &   $ \times 10^{-2}$    \\  \hline
$ 2.25 $   &   $1.2 \pm 4.2     $    &    $  2.0 \pm 3.2   $   &   $ 80 \pm 33  $  &   $ \times 10^{-3}$      \\  \hline
$ 2.75 $   &   $ 0.76 \pm 2.4    $    &    $  1.8 \pm 1.8     $   &   $  10 \pm 19  $    &   $ \times 10^{-3}$        \\  \hline
$ 3  $   &   $  0.50 \pm 1.7   $    &    $ 1.6 \pm 1.3    $   &   $ -33 \pm 15  $ &   $ \times 10^{-3}$         \\  \hline
$ 3.25 $   &   $0.45 \pm 1.6     $    &    $ 1.2 \pm 1.2    $   &   $ 4.0 \pm 12   $   &   $ \times 10^{-3}$        \\  \hline
$ 3.5  $   &   $ 0.28 \pm 1.2    $    &    $  1.1 \pm 0.97   $   &   $  3.9 \pm 9.3    $    &   $ \times 10^{-3}$       \\  \hline
$ 3.75 $   &   $  0.21 \pm 1.2    $    &    $ 0.84 \pm 0.91   $   &   $  -11 \pm 7.5   $    &   $ \times 10^{-3}$        \\  \hline
$ 4 $   &   $  0.15 \pm 1.1   $    &    $ 0.73 \pm 0.88   $   &   $ -41 \pm 6.0    $   &   $ \times 10^{-3}$        \\
\hline
\end{tabular}
\caption{\footnotesize Best fit and $1\sigma$ errors for EFT parameters.  Since the parameters are unitless, the columns have units given by the explicit factors of $\knl$.  All values of measured parameters and errors should be multiplied by the factor in the last column.}
\label{c1stable}
\end{table}

  \begin{table}[h]
\centering
\begin{tabular}{c | c |  c | c  | c }
\hline \hline 
$z$ & $\Delta c_{4c}^2 / \knl^4$ &  $\Delta c_{4b}^2/ \knl^4$ &  $ c_{4s}^2/ \knl^4$ & $$  \\ 
\hline
$  0 $   &   $0.19 \pm 0.23   $    &    $  0.17 \pm 0.18  $   &   $ -40 \pm 56  $  &   $ \times 10^{-2}$    \\ \hline
$  0.125 $   &   $  0.030 \pm 0.11    $    &    $  0.0030 \pm 0.089   $   &   $ -27 \pm 48  $    &   $ \times 10^{-2}$   \\ \hline
$ 0.25  $   &   $ 0.033 \pm 0.11   $    &    $  -0.078 \pm 0.090   $   &   $ -19 \pm 41   $   &   $ \times 10^{-2}$    \\  \hline
$ 0.375  $   &   $ 0.038 \pm 0.12   $    &    $ -0.77 \pm 0.090   $   &   $  -9.9 \pm 31  $    &   $ \times 10^{-2}$    \\  \hline
$ 0.5  $   &   $ 0.040 \pm 0.12  $    &    $   -0.090 \pm 0.091   $   &   $  -0.62 \pm 20  $     &   $ \times 10^{-2}$  \\  \hline
$ 0.75  $   &   $ 0.036 \pm 0.12   $    &    $ -0.060 \pm 0.092      $   &   $ 4.0 \pm 10    $     &   $ \times 10^{-2}$ \\ \hline
$ 1  $   &   $ 0.027 \pm 0.12  $    &    $ 0.019 \pm 0.092   $   &   $ 1.4 \pm 4.2    $  &   $ \times 10^{-2}$      \\  \hline
$ 1.25  $   &   $   0.020 \pm 0.12    $    &    $ 0.046 \pm 0.094    $   &   $ 2.8 \pm 2.0   $   &   $ \times 10^{-2}$    \\  \hline
$ 1.5  $   &   $  0.012 \pm 0.22  $    &    $ 0.0093 \pm 0.017      $   &   $  0.95 \pm 0.98  $    &   $ \times 10^{-2}$    \\  \hline
$ 1.75  $   &   $ 0.0075 \pm 0.017   $    &    $ 0.018 \pm 0.014   $   &   $ 1.0 \pm 0.49  $    &   $ \times 10^{-2}$   \\  \hline
$ 2  $   &   $ 0.0045 \pm 0.017     $    &    $ 0.028 \pm 0.014     $   &   $ 0.46 \pm 0.25    $   &   $ \times 10^{-2}$    \\  \hline
$ 2.25 $   &   $ 0.028 \pm 0.14    $    &    $  0.15 \pm 0.11   $   &   $ 3.7 \pm 1.3  $  &   $ \times 10^{-3}$      \\  \hline
$ 2.75 $   &   $  0.010 \pm 0.043   $    &    $  0.073 \pm 0.035     $   &   $ 0.71 \pm 0.46   $    &   $ \times 10^{-3}$        \\  \hline
$ 3  $   &   $ 0.0051 \pm 0.023    $    &    $  0.051 \pm 0.018   $   &   $  -0.20 \pm 0.29 $ &   $ \times 10^{-3}$         \\  \hline
$ 3.25 $   &   $  0.0037 \pm 0.018   $    &    $ 0.027 \pm 0.014    $   &   $ 0.14 \pm 0.16   $   &   $ \times 10^{-3}$        \\  \hline
$ 3.5  $   &   $  0.0020 \pm 0.011   $    &    $ 0.017 \pm 0.0087     $   &   $ 0.029 \pm 0.10     $    &   $ \times 10^{-3}$       \\  \hline
$ 3.75 $   &   $ 0.0013 \pm 0.0083     $    &    $ 0.011 \pm 0.0067   $   &   $  -0.089 \pm 0.069   $    &   $ \times 10^{-3}$        \\  \hline
$ 4 $   &   $  9.1 \times 10^{-4} \pm 0.0064   $    &    $  0.0083 \pm 0.0052   $   &   $ -0.26 \pm 0.050    $   &   $ \times 10^{-3}$        \\
\hline
\end{tabular}
\caption{\footnotesize Best fit and $1\sigma$ errors for EFT parameters.  Since the parameters are unitless, the columns have units given by the explicit factors of $\knl$.  All values of measured parameters and errors should be multiplied by the factor in the last column.}
\label{c4stable}
\end{table}

  \begin{table}[h]
\centering
\begin{tabular}{c | c |  c  }
\hline \hline 
$z$ & $k_{\rm fit}^{P} \, [ \unitsk ]$ & $ k_{\rm fit}^R   \, [ \unitsk ] $  \\ 
\hline
$  0 $   &   $ 0.320  $    &    $ 0.80   $     \\ \hline
$  0.125 $   &   $ 0.333      $    &  $1.18  $    \\ \hline
$ 0.25  $   &   $ 0.349   $    &    $  1.21   $   \\  \hline
$ 0.375  $   &   $  0.369  $    &    $ 1.25   $      \\  \hline
$ 0.5  $   &   $ 0.392  $    &    $    1.25  $   \\  \hline
$ 0.75  $   &   $  0.450   $    &    $    1.25   $    \\ \hline
$ 1  $   &   $ 0.520  $    &    $  1.25  $         \\  \hline
$ 1.25  $   &   $ 0.602     $    &    $  1.25   $      \\  \hline
$ 1.5  $   &   $  0.695   $    &    $      1.7 $      \\  \hline
$ 1.75  $   &   $ 0.798   $    &    $ 1.8   $      \\  \hline
$ 2  $   &   $ 0.910     $    &    $   1.8  $       \\  \hline
$ 2.25 $   &   $  1.03   $    &    $   1.9  $     \\  \hline
$ 2.75 $   &   $ 1.30   $    &    $     2.4   $    \\  \hline
$ 3  $   &   $  1.44   $    &    $  2.7   $           \\  \hline
$ 3.25 $   &   $ 1.60    $    &    $  2.9  $        \\  \hline
$ 3.5  $   &   $1.76     $    &    $   3.2  $          \\  \hline
$ 3.75 $   &   $   1.93   $    &    $   3.4 $          \\  \hline
$ 4 $   &   $ 2.11     $    &    $ 3.6   $          \\
\hline
\end{tabular}
\caption{\footnotesize Values of $k_{\rm fit}$ used in our fits: $k_{\rm fit}^P$ is for the two-loop dark-matter-only power spectrum fit using \eqn{candbtwoloop}, and $k_{\rm fit}^R$ is for the ratio fit using \eqn{ratiofittingfns}.  }
\label{kreachtab}
\end{table}

\clearpage

%
%
%

\section{Relaxing $k_{\rm fit}^R$} \label{changekfitapp}

\begin{figure}[h!]
\centering
\includegraphics[height=6.5cm]{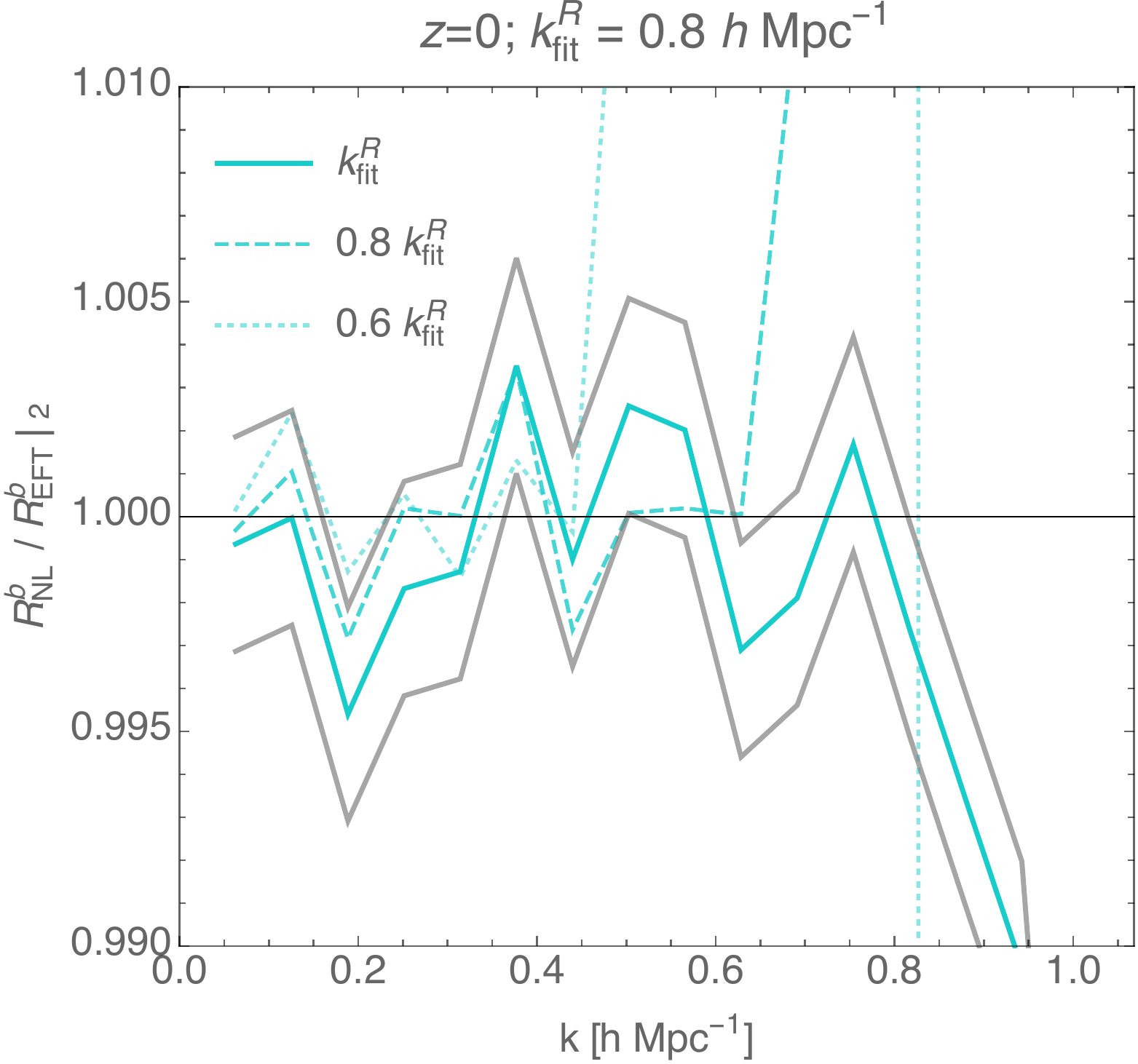} \hspace{.4in}
\includegraphics[height=6.5cm]{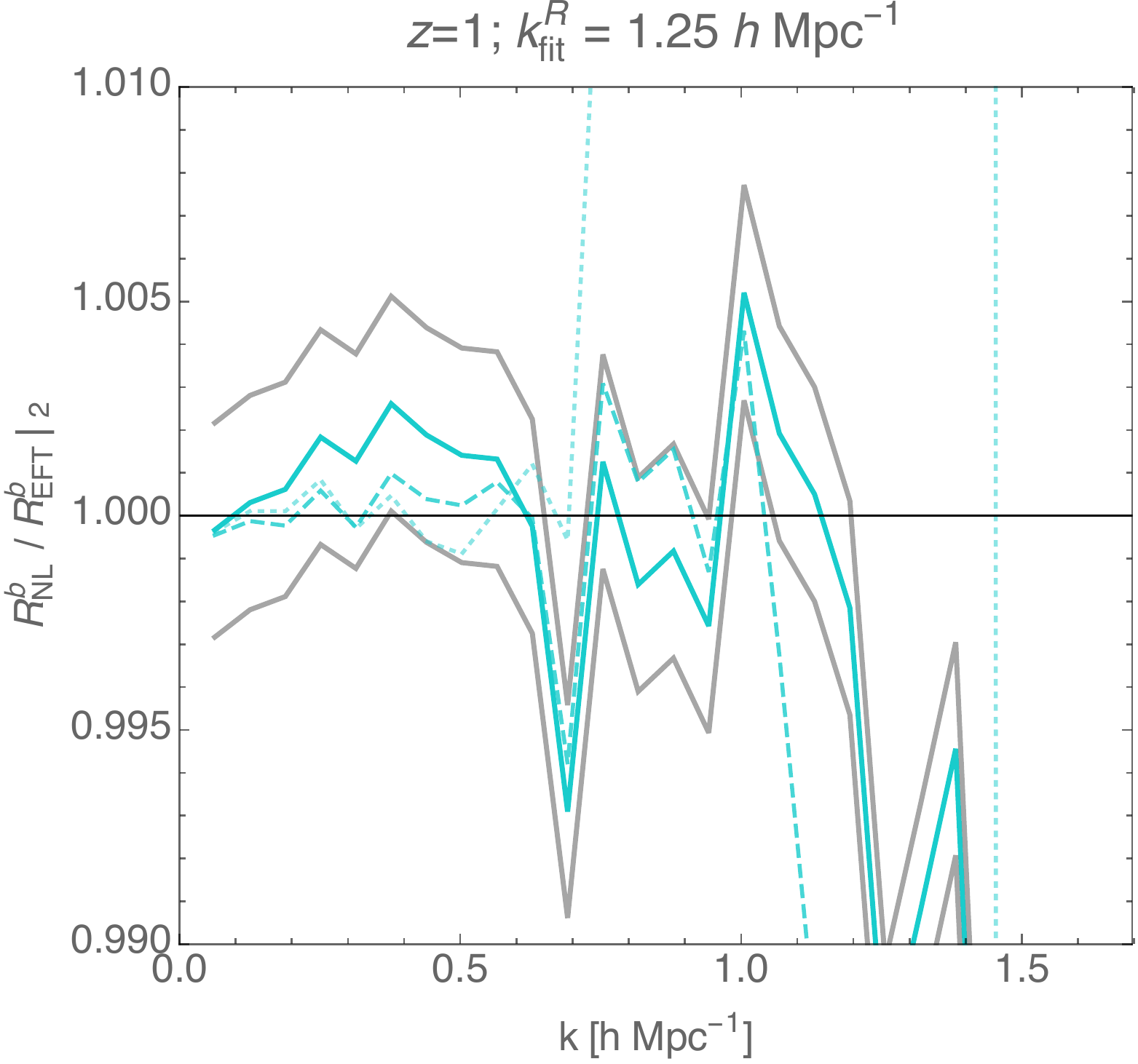} 
\caption{\footnotesize Residuals of baryon fits compared to non-linear data at $z = 0$ and $z=1$.  In gray we plot the error on the non-linear data that we used, $0.25\%$, which we see is justified by the scatter in the non-linear data.  In particular, we plot our results using $ k_{\rm fit}^R(a)$ (solid curve), $0.8 \, k_{\rm fit}^R(a)$ (dashed curve), and $0.6 \, k_{\rm fit}^R(a)$ (dotted curve).  We plot only the error associated with $ k_{\rm fit}^R(a)$ to avoid clutter.} \label{resb}
\end{figure}

\begin{figure}[h!]
\centering
\includegraphics[height=6.5cm]{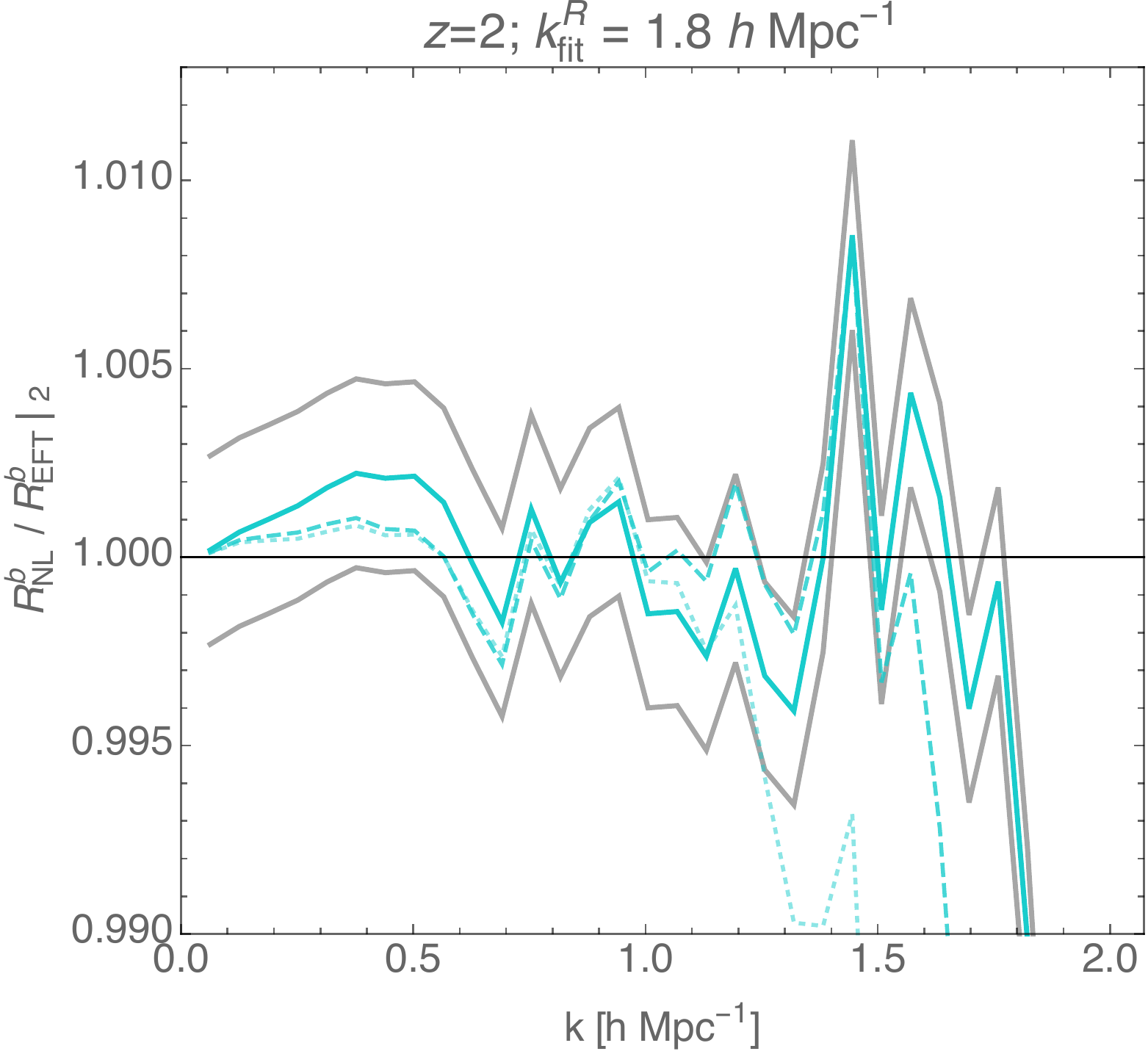} \hspace{.4in}
\includegraphics[height=6.5cm]{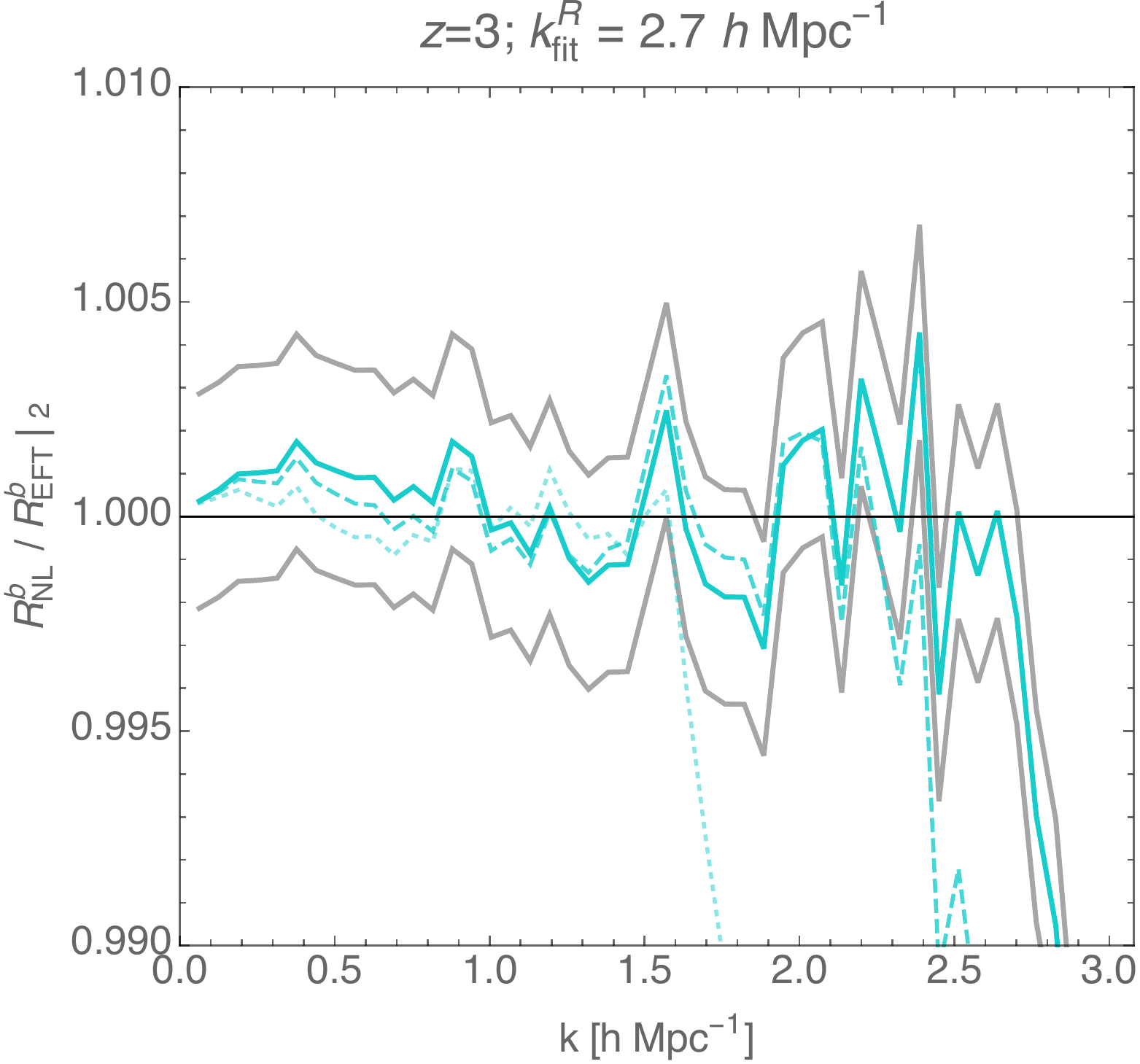}
\caption{\footnotesize Residuals of baryon fits compared to non-linear data at $z = 2$ and $z=3$.  In gray we plot the error on the non-linear data that we used, $0.25\%$, which we see is justified by the scatter in the non-linear data.  In particular, we plot our results using $ k_{\rm fit}^R(a)$ (solid curve), $0.8 \, k_{\rm fit}^R(a)$ (dashed curve), and $0.6 \, k_{\rm fit}^R(a)$ (dotted curve).  We plot only the error associated with $ k_{\rm fit}^R(a)$ to avoid clutter.} \label{resb2}
\end{figure}

\begin{figure}[h!]
\centering
\includegraphics[height=6.5cm]{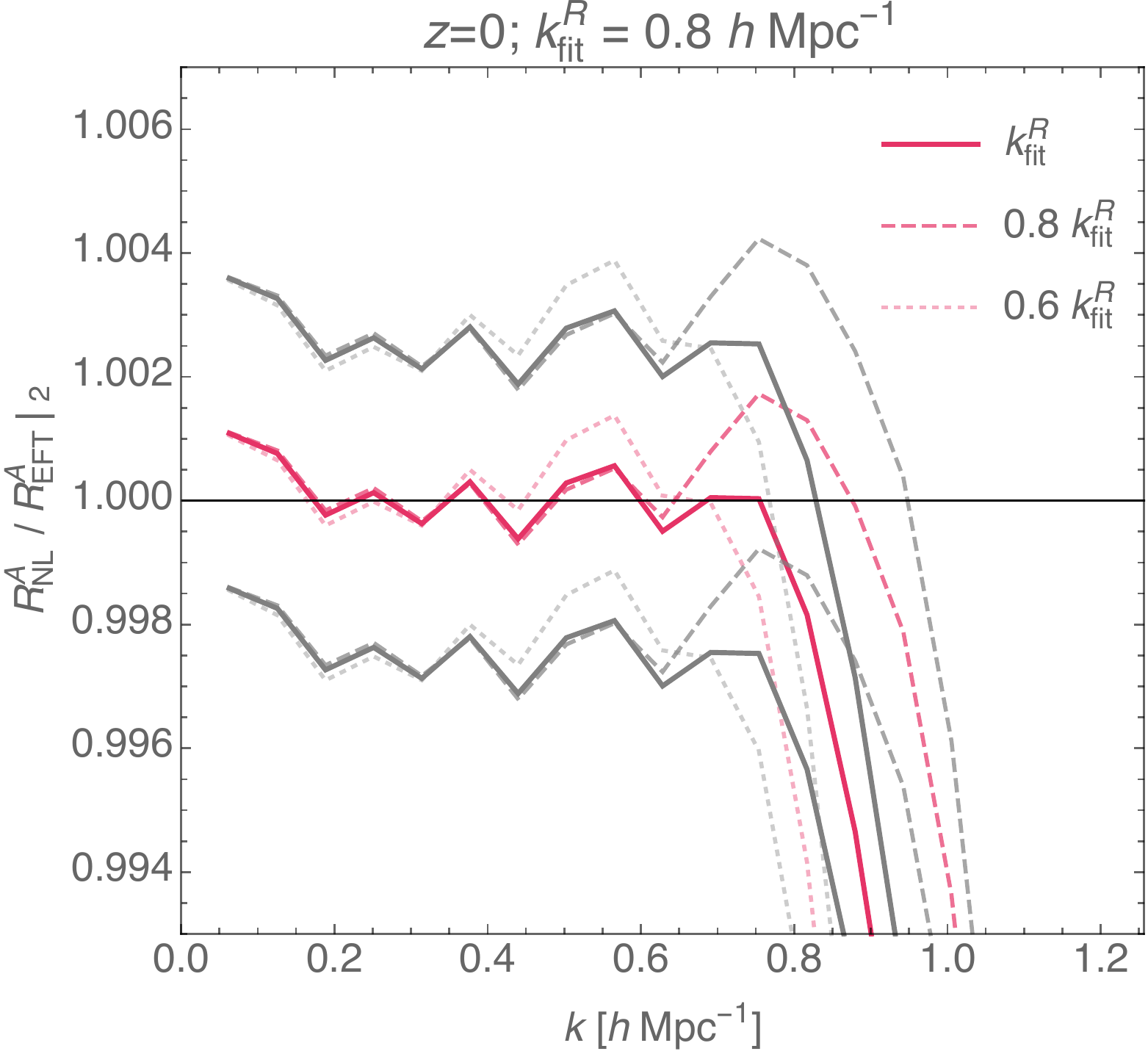} \hspace{.4in}
\includegraphics[height=6.5cm]{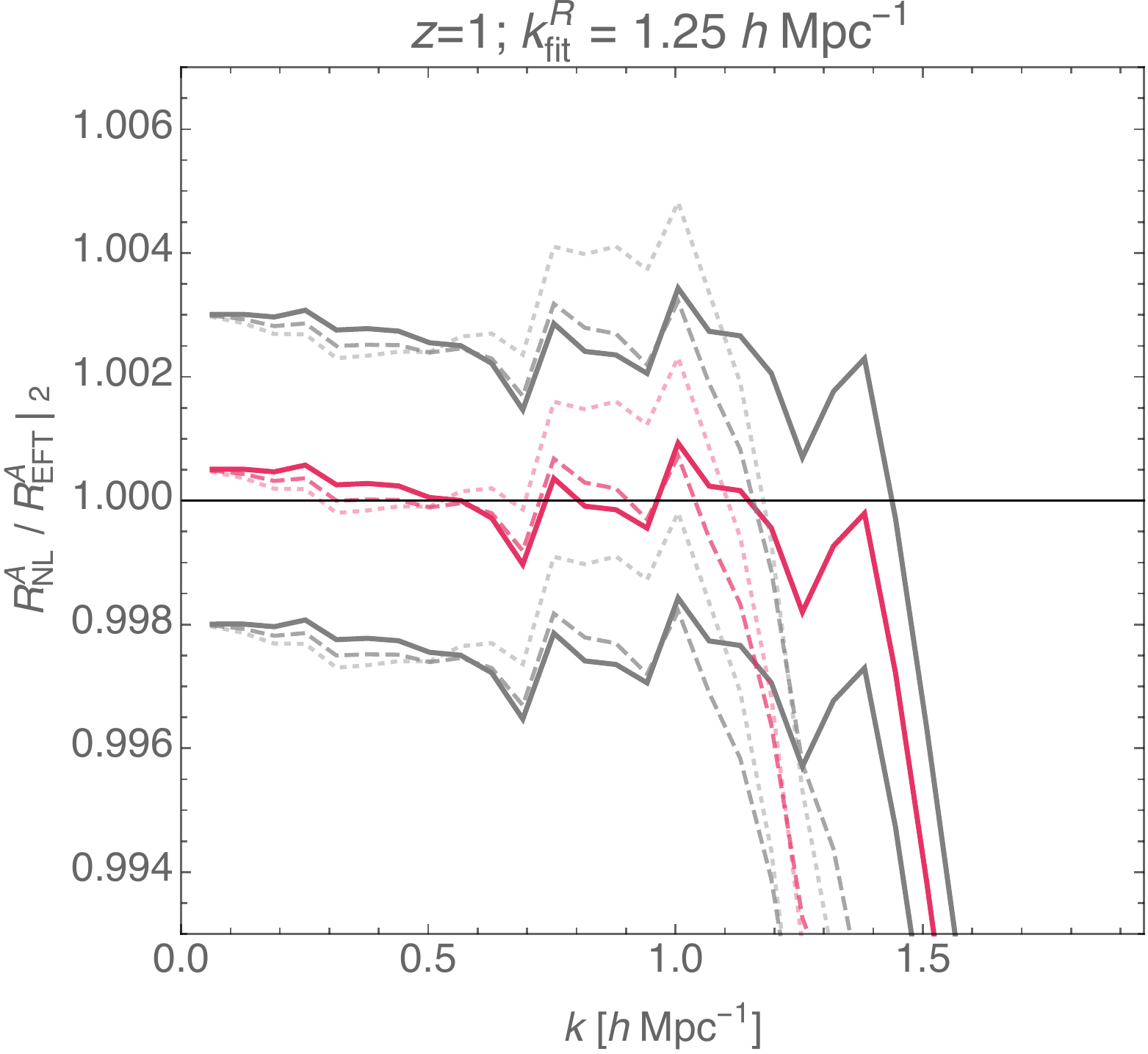} 
\caption{\footnotesize Residuals of adiabatic fits compared to non-linear data at $z = 0$ and $z=1$.  In gray we plot the error on the non-linear data that we used, $0.25\%$.  In particular, we plot our results using $ k_{\rm fit}^R(a)$ (solid curve), $0.8 \, k_{\rm fit}^R(a)$ (dashed curve), and $0.6 \, k_{\rm fit}^R(a)$ (dotted curve).  } \label{resA}
\end{figure}

\begin{figure}[h!]
\centering
\includegraphics[height=6.5cm]{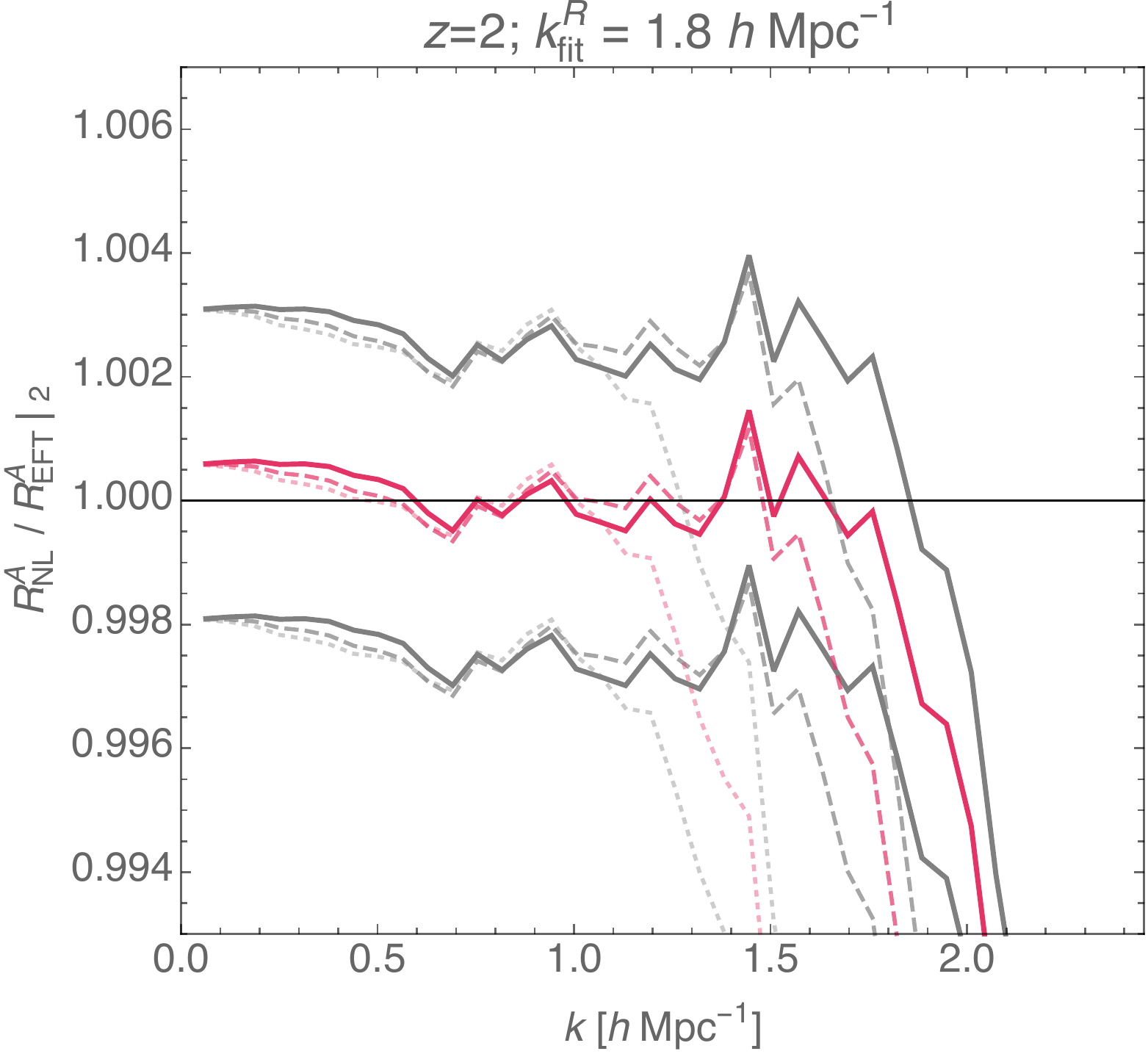} \hspace{.4in}
\includegraphics[height=6.5cm]{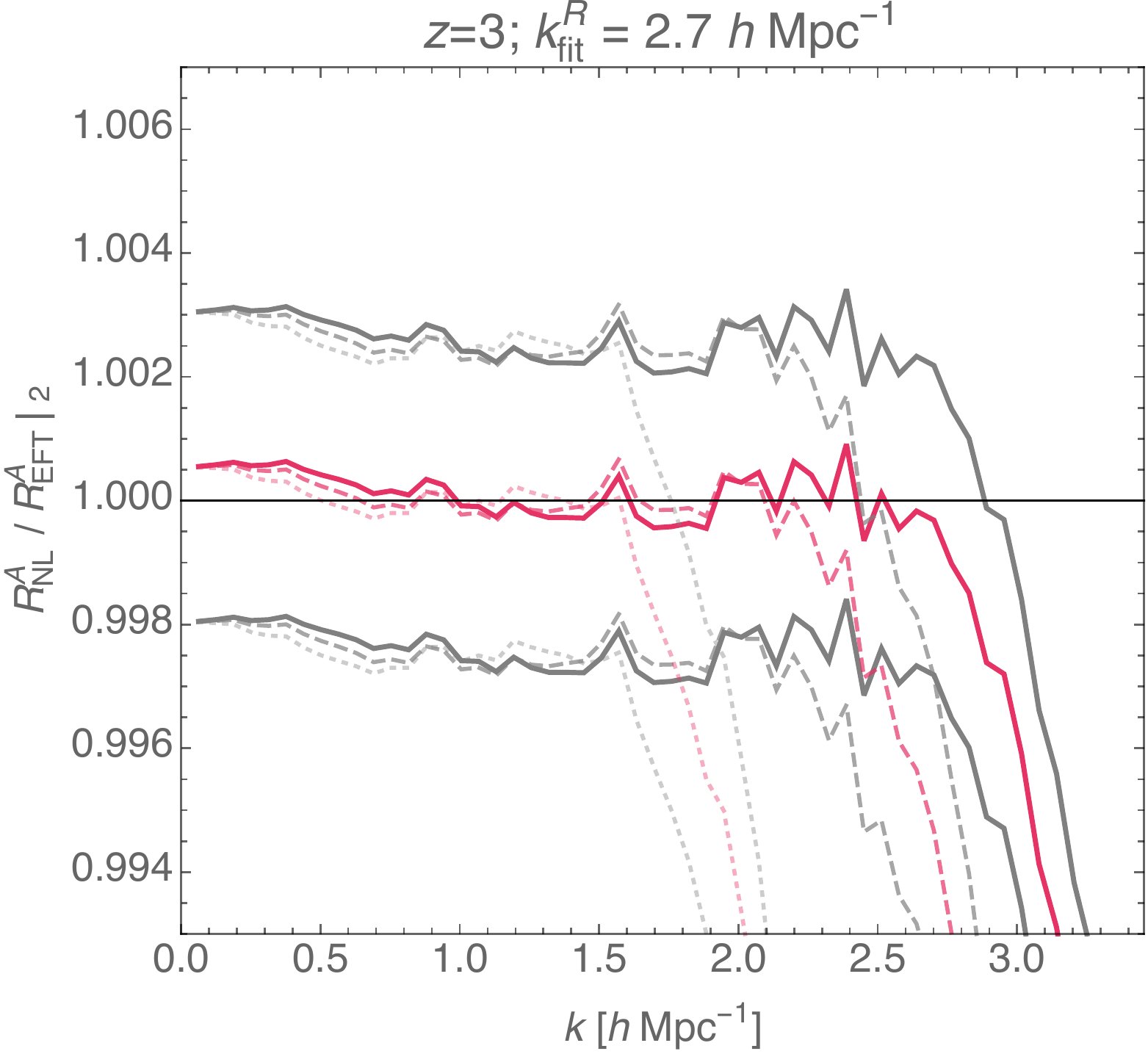}
\caption{\footnotesize Residuals of adiabatic fits compared to non-linear data at $z = 2$ and $z=3$.  In gray we plot the error on the non-linear data that we used, $0.25\%$.  In particular, we plot our results using $ k_{\rm fit}^R(a)$ (solid curve), $0.8 \, k_{\rm fit}^R(a)$ (dashed curve), and $0.6 \, k_{\rm fit}^R(a)$ (dotted curve).  } \label{resA2}
\end{figure}

Our procedure to determine $k_{\rm fit}^R(a)$, described in detail in \appref{tableapp}, has some ambiguity as to which exact  $k_{\rm fit}^R(a)$ we choose at each redshift. For example we choose the somewhat arbitrary, though reasonable, condition that the central value drifts outside of the $2\sigma$ regions of the lower values of~$k_{\rm max}$.  For that reason, we examine our choices of $k_{\rm fit}^R (a)$ in this Appendix.  In Figs. (\ref{resb}) - (\ref{resA2}), we plot our fits for the adiabatic and baryon ratios, $R^A$ and $R^b$, compared to the non-linear data.  We plot our fits using the $k_{\rm fit}^R(a)$ used in the rest of the paper (see \tabref{kreachtab}) and the smaller, more conservative, values $0.8 \, k_{\rm fit}^R(a)$ and $0.6 \, k_{\rm fit}^R(a)$.  

\begin{figure}[h!]
\centering
\begin{tabular}{cc}
\hspace{-.4in}\includegraphics[width=17cm]{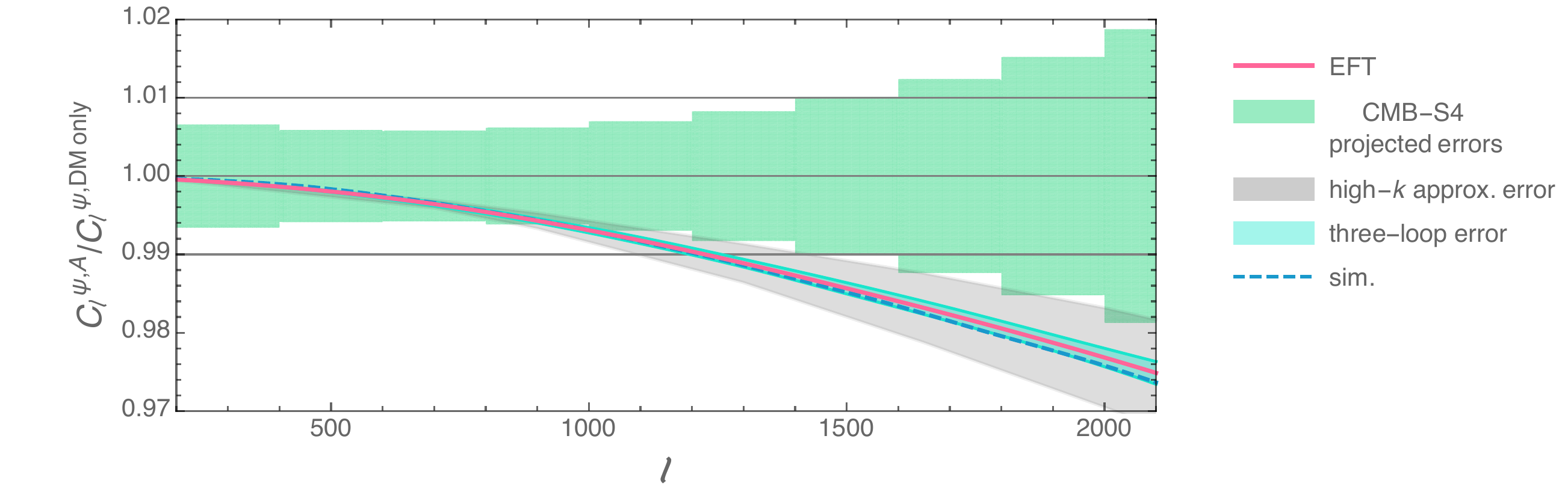} 
\end{tabular} 
\caption{\footnotesize Lensing calculation, same as \figref{lensing3}, using fits with $0.8 \, k_{\rm fit}^R(a)$.  The central value of our calculation remains virtually unchanged, although the theoretical error (gray band) increases slightly.  }
\label{lensing4}
\end{figure}

In the baryon fits, \figref{resb} and \figref{resb2}, we see that the fits with $k_{\rm fit}^R(a)$ have a slight residual bending with respect to the non-linear data, particularly at low redshift, which suggests a slight over-fitting~\footnote{Such a behavior of the fit is expected given that our way of determining $k_{\rm fit}^R(a)$ is such that the values of the counterterms we choose with $k_{\rm max}=k_{\rm fit}^R(a)$ begin to become slightly incompatible with the values at lower $k_{\rm max}$'s.}. However, one can see that the residual bend is absent when using $0.8 \, k_{\rm fit}^R(a)$, which justifies our ultimate choice, {\it i.e.} the overfit, if present, is very small.  On the other hand, we see that the adiabatic fits, \figref{resA} and \figref{resA2}, are very stable at low-$k$ and do not show any residual bending.  As a check, we reproduced our lensing calculation using the fits with $0.8 \, k_{\rm fit}^R(a)$ in \figref{lensing4}, where we see that the result is virtually unchanged, apart from a slight increase in the theoretical error, and  it is still significantly smaller than the expected cosmic variane  of CMB-S4 data.  Notice also that, as compared to the simulation result, our estimate of the theory error appears to be extremely conservative.

\clearpage

 \bibliographystyle{JHEP}
 \small
\bibliography{matt_master_bib}

 \end{document}